\documentclass[11pt,a4paper]{article}

\usepackage{jheppub}

\usepackage{amsmath}
\usepackage{amssymb}
\usepackage{mathtools}
\usepackage{physics}
\usepackage{slashed}
\usepackage{simpler-wick}

\usepackage{microtype}
\usepackage{dsfont}
\usepackage{eufrak}
\usepackage{bbold}

\usepackage{graphicx}
\usepackage{float}
\usepackage{subcaption}
\usepackage[export]{adjustbox}

\usepackage{tikz-feynman}
\tikzfeynmanset{compat=1.0.0}

\makeatletter
\renewcommand*\env@matrix[1][\arraystretch]{%
  \edef\arraystretch{#1}%
  \hskip -\arraycolsep
  \let\@ifnextchar\new@ifnextchar
  \array{*\c@MaxMatrixCols c}}
\makeatother

\newcommand{\bev}[1]{\expval{#1}_{\circ}}
\def\Vol{\operatorname{Vol}}

\title{
Fermions in Boundary Conformal Field Theory : Crossing Symmetry and $\epsilon$-Expansion 
}
\author{Christopher P. Herzog, Vladimir Schaub}
\emailAdd{christopher.herzog@kcl.ac.uk}
\emailAdd{vladimir.schaub@kcl.ac.uk}
\affiliation{Mathematics Department, King's College London, \\
The Strand, London,  WC2R 2LS, UK}
\abstract{
We use the equations of motion in combination with crossing symmetry to constrain the properties of
interacting fermionic boundary conformal field theories. 
This combination is an efficient way of determining operator product expansion
coefficients and anomalous dimensions at the first few orders of the $\epsilon$ expansion.    
Two necessary ingredients for this procedure are knowledge of the boundary and bulk spinor conformal blocks.  
The bulk spinor conformal blocks are derived here for the first time. We then consider a number of examples.
For $\phi$ a scalar field and $\psi$ a fermionic field,
we study the effects of a $\phi \bar \psi \psi$ coupling in $4- \epsilon$ dimensions, a $\phi^2 \bar \psi \psi$ coupling
in $3 -\epsilon$ dimensions, and a $(\bar \psi \psi)^2$ coupling in $2+\epsilon$ dimensions.  
We are able to compute some new anomalous dimensions for operators in these theories.  Finally, we relate the anomalous
dimension of a surface operator to the behavior of the charge density near the surface.
}

\makeatletter
\def\@fpheader{\vspace{0cm}}
\makeatother

\begin{document}
\maketitle

\section{Introduction}

One of quantum field theory's major successes is its ability to model the behavior of
physical systems near the point of a continuous phase transition
\cite{ZinnJustin}.  
Following a quantum field theory to a fixed point in the renormalization group flow,
anomalous dimensions of operators at the fixed point  can be used to calculate
critical exponents which in turn determine the scaling behavior of a large number of physical quantities such as specific heat, magnetization,
and correlation length
at or near the phase transition.  
A key player in this story is the $O(N)$ model: a set of $N$ interacting scalar fields with $O(N)$ symmetry.  This model has been
used successfully to explain the critical behavior of the liquid-gas transition in simple gases, the superfluid transition in liquid helium, and the 
Heisenberg transition in ferromagnets.  
Indeed, quantum field theory can model not only these thermal phase transitions but quantum phase transitions as well, driven by the tuning of a coupling strength
or external field at zero temperature \cite{sachdev1999quantum}.  The motivation for this work is to look at a more general class of quantum field theories that involve both relativistic fermions and boundaries in order to broaden the set of critical systems which can be successfully modeled.

Relativistic fermions have played an important role in the high energy community for decades as the quarks and leptons of
the standard model of particle physics.  Both in the development of the standard model and also in an attempt to understand better certain aspects of it, 
such as confinement and the existence of a Higgs particle, a number of auxiliary fermionic quantum field theories have been developed and studied
 over the years, named after their
inventors -- Gross, Neveu, Schwinger, Thirring, Yukawa.  While the emphasis in the condensed matter community has primarily been on 
non-relativistic fermions, the attitude has evolved
 in the last two decades with development
of new materials such as graphene and topological insulators. 
These so-called Dirac and Weyl materials \cite{Wehling:2014cla,Vafek:2013mpa} support half-filled 
nodal Fermi surfaces where the dispersion
relation for the fermionic degrees of freedom is linear.   
 Moreover, as a function of interaction strength, some of these materials may undergo continuous phase transitions \cite{gomes2012designer,gutierrez2016imaging,bao2021experimental,qu2022ubiquitous}.

Indeed, more precise conjectures have been formulated.  Take for example the Yukawa model which has
Lagrangian density
\begin{equation}
\label{GNY}
{\mathcal L} = -\frac{1}{2} \partial_\mu \phi \partial^\mu \phi -  \overline{\psi}_a \slashed{\partial} \psi_a  - g \phi \overline{\psi}_a \psi_a  - \frac{h}{4!}  \phi^4 \ .
\end{equation}
(See appendix \ref{app:conventions} for our conventions.)
Here $\phi$ is a real scalar field and $\psi_a$ is a Dirac fermion in $4-\epsilon$ dimensions with 
$a = 1, \ldots, N_f$ components.  
Using the relevant mass term to drive the phase transition, 
 $\phi$ gets an expectation value and gives $\psi_a$ a mass, breaking the chiral symmetry of the fermions.  
The case $N_f = 2$ has been proposed to describe a semi-metal insulator transition in graphene \cite{Herbut:2006cs}.
For $N_f = 1/4$, the model is conjectured to be supersymmetric at the quantum critical point, and could describe the boundary of a topological phase
\cite{Grover:2013rc}.
Disorder can also drive phase transitions.  In the context of the replica trick, the $N_f \to 0$ limit may describe the critical point between a relativistic semimetallic state and a diffusive metallic phase in 3D Weyl semimetals 
\cite{Louvet:2016zbb}.

While the studies above describe infinite systems, critical behavior can arise in the presence of a boundary as well \cite{Diehl:1996kd}.  
The critical exponents and scaling behavior associated with the surface  have been in some cases both measured and successfully compared with theoretical
predictions.  
For example, Alvarado et al.\
\cite{alvarado1982surface} measured the surface magnetization of a nickel ferromagnet close to the ordering transition. 
There is reasonable agreement with the corresponding surface magnetization critical exponent extracted from an epsilon expansion of 
the O(3) model \cite{diehl1986critical,diehl1980scaling,reeve1980critical}.
There are also measurements of surface critical exponents for an iron-aluminum alloy, a liquid mixture and a molecular solid
that agree with predictions from the Ising model in the presence of a boundary
\cite{mailander1990near, burandt1993near,sigl1986order}. 

To our knowledge, relatively little theoretical work has been done on critical fermionic systems with boundaries.  While a recent tour-de-force 
calculation \cite{Zerf:2017zqi} computes critical exponents in the Yukawa model at fourth order in the epsilon expansion in the
infinite system, 
the state of the art in the presence of a boundary is $O(\epsilon)$ \cite{Giombi:2021cnr}.  Refs.\ \cite{Giombi:2021cnr,Carmi:2018qzm} consider
also the large $N_f$ limit.
On the experimental side, the associated critical exponents have not yet been measured in the infinite system nor in the presence of a boundary, yet
much promising work has been done to engineer graphene and graphene like systems which undergo this type of chiral symmetry breaking
phase transition  
(see e.g.\
\cite{gomes2012designer,gutierrez2016imaging,bao2021experimental,qu2022ubiquitous}).

A major point of our work is to develop  tools to make the calculation of critical exponents for fermionic systems with boundaries more straightforward.
We are inspired by ref.\ \cite{Rychkov:2015naa}
 which pointed out the utility of the equations of motion in reducing the number of Feynman diagrams 
that need to be evaluated, and also by the revival of the conformal bootstrap program, especially as it pertains to systems with fermions
\cite{vanLoon:2017xlq,Giombi:2017rhm,Iliesiu:2015qra}
and boundaries \cite{Liendo:2012hy,Dey:2020jlc,Kaviraj:2018tfd,Mazac:2018biw}. In the context of the fermionic CFT, the equation of motion procedure has been implemented successfully without boundary in refs.\ \cite{Ghosh:2015opa,Raju:2015fza} to rederive anomalous dimensions. In ref.\ \cite{Giombi:2021cnr} the equation of motion technique was further extended to fermions in bCFT, where they calculated some spinorial boundary 
anomalous dimensions in the $\epsilon$-expansion for the first time.
Our work can be viewed as a variation of \cite{Giombi:2021cnr}, augmenting the equation of motion technique with a conformal bootstrap approach
instead of with Feynman diagram computations.

The key idea in the conformal bootstrap is the notion of crossing symmetry.
In systems with boundaries, this symmetry can be imposed on two-point correlation functions in the bulk.  One either brings the two bulk operators
close together or close to the boundary.  In the first case, the operator product decomposes into a sum over bulk operators.  As only scalar operators can
have nonzero one point functions in boundary conformal field theory, the sum reduces to a sum over these bulk scalar one point functions.
In the second case, the bulk operators each decompose as a sum over boundary operators.  The bulk two-point function is then expressed as a sum over boundary two-point functions.  The equivalence of these two procedures is the crossing constraint.

To employ crossing symmetry, however, one must have an efficient formalism to deal with the sums.  In conformal field theory, these operators arrange themselves into multiplets -- a conformal primary along with an infinite number of descendants obtained by acting with derivatives on the primary.  
In the context of crossing symmetry, 
these multiplets get resummed into conformal blocks.  While the boundary conformal blocks
for spinor two-point functions were known already \cite{Nishida:2018opl,Herzog:2019bom}, here we derive the bulk conformal blocks as well.

With this technical advance in hand, we proceed to analyze three different interacting fermionic theories in the $\epsilon$ expansion in the
presence of a boundary.  We find that the combination of crossing symmetry with the equations of motion allows us efficiently to put nontrivial constraints
on combinations of the surface and bulk critical exponents.  
Similar to \cite{Giombi:2021cnr}, we consider the Yukawa model (\ref{GNY}) in $4-\epsilon$ dimensions and also 
the Gross-Neveu model in $2+\epsilon$ dimensions:
\begin{equation}
\label{GN}
{\mathcal L} =-  \overline{\psi}_a \slashed{\partial} \psi_a  + \frac{g}{2} (\overline{\psi}_a \psi_a) (\overline{\psi}_b \psi_b)  \ .
\end{equation}
In these two cases, we can take advantage of the known bulk critical exponents to fix the boundary data.  

It is often argued that the phase transition of these two models in the limit $\epsilon = 1$ is in the same universality class.  Thus one can
gain further control over the $d=3$ dimensional theory by interpolating the results for the two theories at $\epsilon = 1$.
There is however a further wrinkle in the boundary case because of the the different choices of boundary condition for the scalar field $\phi$.  Ref.\ \cite{Giombi:2021cnr} argues that the Gross-Neveu model in $2+\epsilon$ dimensions matches onto the extraordinary boundary condition of the 
Yukawa model in $4-\epsilon$ dimensions.  The extraordinary phase transition corresponds to a bulk ordering phase transition at a point in 
the phase diagram where the surface is already ordered,  $\langle \phi \rangle \sim r^{-\Delta_{\phi}}$.
While we have some remarks to make about the extraordinary case, the emphasis in this work is on 
the so-called special and ordinary phase transitions of the Yukawa model, which in the classical limit become Dirichlet and Neumann boundary conditions for the scalar.\footnote{
 The boundary condition for the fermion is a projection condition $\frac{1}{2} (1 + \chi \gamma_n) \psi = 0$ where $\gamma_n$ is the gamma matrix perpendicular to the boundary and $\chi = \pm 1$.   This boundary condition corresponds to an ``armchair'' edge in graphene \cite{Biswas:2022dkg}.
}

Our third model is the little studied
\begin{equation}\label{eq:lag3dyuk}
{\mathcal L} = -\frac{1}{2} \partial_\mu \phi_a \partial^\mu \phi_a -  \overline{\psi}_b \slashed{\partial} \psi_b  - \frac{g}{2} \phi^2 \overline{\psi}_b \psi_b    - \frac{h}{6!} (\phi^2)^3
\end{equation}
in $d = 3 - \epsilon$ dimensions and the indices $a, b, \ldots$ run either from
one to $N_s$ or one to $N_f$, depending on whether they decorate a scalar or fermion.  
The results we obtain for this third model are new as far as we are aware, even in the absence of a boundary.\footnote{%
 See however related work on supersymmetric Chern-Simons theories in exactly $d=3$.  In these theories, kinetic terms
 involve gauge covariant derivatives, there is a Chern-Simons term, and there is a flavor mixing term
 $(\bar \psi \phi) (\phi^* \psi)$ since the $\phi$ and $\psi$ fields are part of the same chiral multiplet and are assumed
 to transform in the same representation.  We have in mind (1.3) of \cite{Jain:2013gza}.
}
One reason it has not been studied has to do with 
the behavior of the purely bosonic version of this theory.
The large $N_s$ limit  has both a UV and an IR fixed point which annihilate at $0<\epsilon_c < 1$ \cite{Pisarski:1982vz}; it is thus not possible to set
$\epsilon$ equal to one in order to learn something about a 2d strongly interacting fixed point.  It is nonetheless an interesting playground in which to explore the intersection of crossing symmetry and the equations of motion.
Similar to what happens in the purely bosonic theory with boundary \cite{diehi1987walks,eisenriegler1988surface}, the boundary critical exponents get nonanalytic corrections of $O(\sqrt{\epsilon})$.

\subsection*{Summary of Perturbative Results}

Our application of crossing symmetry together with the equations of motion fix a number of anomalous dimensions as well as an infinite number of OPE coefficients to the first few orders in the $\epsilon$-expansions.  Below we tabulate the anomalous dimensions.  The OPE coefficients can be found in the text.

\subsubsection*{Yukawa model}

At the RG fixed point, the coupling is of order $g^2 = O(\epsilon)$. For convenience, we express the results using a rescaled coupling $\kappa_{s}g^2 \equiv \lambda \epsilon$. The constant $\kappa_s^{-1} = (d-2) \Vol(S^{d-1})$ is the usual normalization of the free scalar two-point function in $d$ dimensions.
 Crossing symmetry and the equations of motion
enforce the following relations between the scaling dimensions of $\psi$ and $\phi$, and the boundary value of $\psi$ denoted here by $\rho$  
\begin{eqnarray}
\Delta_\psi &=& \frac{3}{2} - \frac{\epsilon}{2} + \frac{\lambda}{8}\epsilon + O(\epsilon^2) \ , \\
\Delta_\phi &=& 1 + \frac{\epsilon}{2} - \frac{3}{4} \lambda \epsilon + O(\epsilon^2)  \ , \\
\Delta_\rho &=& \frac{3}{2} -\frac{\epsilon}{2} + \frac{ (4 - \chi_s)}{8}\lambda \epsilon + O(\epsilon^2) \ .
\end{eqnarray}
The constant $\chi_s$ selects Neumann $\chi_s = 1$ or Dirichlet $\chi_s = -1$ boundary conditions for the scalar field $\phi$. Note the presence or absence of a boundary cannot affect the bulk anomalous dimensions and beta function calculation.  
Thus a standard beta function calculation in the system without boundary will fix the relation between $g$ and $\epsilon$ at the conformal
fixed point and thus the surface scaling dimension $\Delta_\rho$.  The results for the beta function can be looked up in many places,
for example \cite{Zerf:2017zqi}. Our results agree with the computation of \cite{Giombi:2021cnr} although our procedure requires less input from
Feynman diagram calculations.

\subsubsection*{Gross-Neveu model}

This model has the advantage that with the same amount of effort, 
we are able to go to one higher order in the $\epsilon$ expansion compared to the other two models. 
In this case $g\sim \epsilon$ while in the other two cases $g^2 \sim \epsilon$. The resulting expressions are simpler in the rescaled coupling $\lambda$ where $\kappa_f g \equiv \epsilon \lambda$. The constant $\kappa_f^{-1} = \Vol(S^{d-1})$ is the usual normalization of the free fermion two-point function in $d$ dimensions.
Furthermore, crossing symmetry and the equations of motion at $O(\epsilon^2)$ allow us to fix
\begin{equation}
g _*= \frac{\epsilon}{2(N-1)\kappa_f} + O(\epsilon^2)\ ,
\end{equation}
without needing to invoke results from the literature for the case without boundary.

We obtain the following scaling dimensions
\begin{eqnarray}
\Delta_\psi &=& \frac{1}{2} + \frac{\epsilon}{2} + \frac{N-\frac{1}{2} }{8(N-1)^2} \epsilon^2  + O(\epsilon^3) \ , \\
\Delta_{\psi^2} &=& 1 -\frac{1}{2(N-1)} \epsilon  + \gamma^{(2)}_{\psi^2} \epsilon^2 + O(\epsilon^3)\ , \\
\Delta_{\psi^4} &=& 2 + O(\epsilon^2) \ , \\
\Delta_{\rho} &=& \frac{1}{2} + \left( \frac{1}{2} + \frac{N-\frac{1}{2}}{2(N-1)} \right) \epsilon +\left( \frac{N-\frac{1}{2}}{4(N-1)^2}- \frac{ \gamma^{(2)}_{\psi^2}}{2} \right)\epsilon^2 + O(\epsilon^3) \ .
\end{eqnarray}
The result for $\Delta_\rho$ at $O(\epsilon)$ was presented already in \cite{Giombi:2021cnr}.
The result $\gamma^{(2)}_{\psi^2}$ is known from calculations in the literature for the case without boundary (see e.g.\ \cite{Gracey:2008mf}):
\begin{equation}
\gamma^{(2)}_{\psi^2} = - \frac{ N - \frac{1}{2}}{4(N-1)^2} \ ,
\end{equation}
which gives the boundary anomalous dimension
\begin{equation}
\gamma_\rho^{(2)} = \frac{3 \left( N - \frac{1}{2} \right)}{8 (N-1)^2} \ .
\end{equation}

\subsubsection*{$\phi^2 \bar \psi_a \psi_a $ model}

Similar to the 4d Yukawa model, the fixed point happens for $g^2 \sim \epsilon$. In contrast to the 4d Yukawa model, however, the surface anomalous
dimensions get contributions already at order $g \sim \sqrt{\epsilon}$.  The contribution occurs at leading order because the nonzero 
one point function for $\phi^2$
in the presence of a boundary mimics the effect of a conformal mass term for the fermion \cite{Herzog:2019bom}.  The one point function $\langle \phi \rangle$ in the Yukawa model, 
 which could play a similar role, vanishes for Dirichlet and Neumann boundary conditions.
 We work with the rescaled coupling $\lambda$ where
$\kappa_s g \equiv 2 \lambda \sqrt{\epsilon}$.  
 The full results are 
\begin{eqnarray}
\Delta_\psi &=&1 -\frac{\epsilon}{2}+ \frac{N_s\lambda^2}{3}\epsilon   +  O(\epsilon^2) \ , \\
\Delta_{\phi^2} &=& 1 + O(\epsilon) \ , \\
\Delta_{\psi^2} &=& 2 -\epsilon + \frac{8N_s}{3}\lambda^2 \epsilon + O(\epsilon^2) \ , \\
\Delta_{\psi \phi} &=& \frac{3}{2} + O(\epsilon) \ , \\
\label{Deltarho2d}
\Delta_{\rho} &=& 1 -\frac{\epsilon}{2}+\frac{\chi \chi_s N_s}{2} \lambda \sqrt{\epsilon}
-\lambda^2 N_s\left(\chi \chi_s(1+N_f)+\frac{4}{3}\right)\epsilon +  O(\epsilon^{3/2}) \ .
\end{eqnarray}

A standard beta function calculation in the theory without boundary, performed in appendix \ref{app:perturbative},  gives the value of the coupling $\lambda$ at the fixed point
\begin{eqnarray}
\lambda_*^2 &=& \frac{3 }{8(2N_f + N_s - 3)}  + O(\epsilon)\ .
\end{eqnarray}

\subsubsection*{Outline}

In section \ref{sec:correlators}, 
we set up the embedding formalism  that will allow us to linearize
the constraints of conformal symmetry.  We then use this embedding formalism
to review the form of scalar, vector and spinor correlation functions in the presence of a boundary.
Next in section \ref{sec:blocks}, we use the embedding formalism to derive the bulk and boundary
conformal partial waves (or blocks) that we will use to enforce the crossing symmetry constraint.
Section \ref{sec:crossing} is the heart of the paper.  We discuss how crossing symmetry is satisfied by
free fermions, make some comments about the extraordinary phase transition, and then embark on 
a solution of the crossing symmetry constraints for our three perturbative models.
Section \ref{sec:discussion} is a discussion where, among other things, we propose that a particular surface anomalous dimension 
may have observable consequences for the electron density of states near the boundary, at the point of a quantum 
phase transition, for a Dirac or Weyl material.

A large number of appendices contain auxiliary results.  Appendix \ref{app:conventions} spells out our conventions for dealing 
with fermions.  Appendix \ref{app:conformalintegral} presents some integrals in embedding space that were useful in deriving the conformal blocks.
Appendix \ref{app:OPEresummation} derives the conformal blocks in a different way, by explicitly summing the OPE.
Some conformal data for the 3d Yukawa model without boundary are derived perturbatively in an $\epsilon$-expansion in appendix \ref{app:perturbative}.
Finally, appendix \ref{app:EAdS} discusses the relationship between our correlation functions in flat space with boundary
and their Weyl transformed cousins in anti-de Sitter space. 

\section{Correlators of Scalars and Spinors}
\label{sec:correlators}

Conformal invariance places strong constraints on correlation functions, even in the presence of a boundary. 
Let us begin by writing down the set of correlation functions that will be important to us in this work along with the conformal invariance constraints
imposed on them.
As it requires the least formal development, we give these correlation functions first in coordinate space. 
Deriving the conformal invariance constraints, however, is more easily done in embedding space.  Moreover, embedding space will be important
in the next section when we discuss the conformal block decomposition of the correlation functions.  
Thus, after listing our main cast of characters, we embark on a discussion of embedding space.  

Please note, the only new aspects of our embedding space discussion concern spinors in the presence of a boundary.  
Earlier discussions of spinors in embedding space without a boundary can be found already in refs.\ \cite{Weinberg:2010ws,Isono:2017grm}.
Also, our real space results for free spinors with a boundary match the classic results  \cite{McAvity:1995tm}.  Even in the presence of a boundary,
our ideas are implicitly contained in a treatment of the embedding space for spinors in anti-de Sitter (AdS) and de Sitter (dS) space \cite{Nishida:2018opl,Pethybridge:2021rwf}.
The relation between  AdS, dS, and bCFT occurs because of the
relation between their metrics under Weyl rescaling and (in the case of dS) Wick rotation. We offer an embedding space perspective on this mapping in Appendix \ref{app:EAdS}.

In what follows, we have occasion to discuss two point functions of 
bulk spinors $\psi$, scalars $\phi$, and conserved currents $J^\mu$.  We will also require
boundary spinors $\rho$ and occasionally boundary scalars $\tau$ and vectors $v^i$.  
We divide the coordinates $x^\mu = (z, x^i)$ into a normal coordinate $z$ and parallel 
coordinates $x^i$.

In the case of bulk $\phi$ and boundary $\tau$ scalar fields, the tensor structures are trivial, and one finds
\begin{eqnarray}
\label{oneandtwophi}
\langle \phi(x)  \rangle &=& \frac{a_\phi}{z^{\Delta_\phi}} \ , \\
\label{phisigma}
\langle \phi(x) \tau(x') \rangle &=&  \frac{c_{\phi \tau}}{(2 z)^{ \Delta_\phi - \Delta_\tau} |x -x'|^{2 \Delta_\tau}} \ , \\
\label{phiphi}
\langle \phi(x) \phi'(x') \rangle &=& \frac{H(\xi)}{(2z)^{\Delta} (2z')^{\Delta'}} \ .
\end{eqnarray}
The constants $\Delta_\phi$ and $\Delta_\tau$ are the scaling dimensions of the corresponding operator insertions.
In bCFT, only one point functions of bulk scalars can get a nontrivial expectation value.  All other one-point functions will vanish because of Lorentz invariance.
Note that the bulk one-point function and bulk-boundary two-point function are fixed up to constant coefficients $a_\phi$ and $c_{\phi \sigma}$ because
of the absence of an appropriate cross ratio.  The bulk-two point function, however, can depend on an undetermined function $H(\xi)$ of the 
conformally invariant cross-ratio 
\begin{equation}
\xi = \frac{(x-x')^2}{4 z z'} \ .
\end{equation}

More generally, a correlation function of two bulk operators 
will be a sum over some number of tensor structures, whose form is fixed up to coefficients
which are functions of the  cross ratio $\xi$.  
Indeed, given two bulk points $x$ and $x'$, we can form exactly one conformally invariant cross ratio, 
although sometimes it is useful to work with the related quantity $v \equiv \xi / (1+\xi)$. 
The bulk coincident limit is given by $\xi \rightarrow 0$, the boundary limit by $\xi \rightarrow \infty$. 
 The $\xi$ variable also has an interpretation in terms of AdS geodesic distance.

For a conserved curent $J^\mu$, we have the correlation functions
\begin{eqnarray}
\label{Jtau}
\langle J^\mu(x) \tau(x') \rangle &=& \frac{c_{J \tau}}{|x-x'|^{2(d-1)}}  X^\mu \delta_{\Delta_\tau, d-1}\ , \\
\label{Jva}
\langle J^\mu(x) v^i (x') \rangle &=& \frac{c_{J v}}{(2z)^{d-1 - \Delta_v} |x-x'|^{2 \Delta_v}}I^{\mu i}(x-x')  \ , \\
\label{JJ}
\langle J^\mu (x) J^\nu (x') \rangle &=& \frac{1}{|x-x'|^{2(d-1)}} \left(Q(\xi) I_{\mu\nu}(x-x')  + (R(\xi) - Q(\xi) ) X_\mu X'_\nu  \right) \ .
\end{eqnarray}
Here the tensor structures are
\begin{eqnarray}
 I_{\mu\nu}(x) = \delta_{\mu\nu} - 2 \frac{x_\mu x_\nu}{x^2}  \ , \; \; \;
 X_\mu = z \frac{v}{\xi} \partial_\mu \xi \ , \; \; \; X'_\mu = z' \frac{v}{\xi} \partial'_\mu \xi \ .
\end{eqnarray}
Conservation imposes the relation $v \partial_v R = (d-1) (R-Q)$.  Furthermore, $\langle J^\mu \tau \rangle$ must vanish unless $\tau$ has conformal dimension $d-1$.  

Finally, for spinor insertions, we have
\begin{eqnarray}
\label{psirho}
\langle \psi(x) \overline{\rho}(x') \rangle &=&  \frac{\gamma_\mu(x-x')^\mu}
{(2z)^{\Delta_\psi - \Delta_\rho} |x-x'|^{2 \Delta_\rho+1}} (c_{\psi \rho}+ c'_{\psi\rho} \gamma_z)   \ , \\
\label{psipsi}
\langle \psi(x) \overline{\psi}'(x') \rangle &=& \frac{   \gamma_\mu (x-x')^\mu F(\xi) +  \gamma_z  \gamma_\mu (x'-\tilde x)^\mu G(\xi) }{(2z)^{\Delta+\frac{1}{2}} (2z')^{\Delta'+\frac{1}{2}}} , 
\end{eqnarray}
where $\tilde x = (-z, x^i)$ and the $\gamma^\mu$ are the usual gamma matrices.  
It is natural to treat $\rho$ as a spinor with respect to the smaller Lorentz group on the surface, and thus to take it as an eigenvector of the normal gamma matrix $\gamma_z \rho = \pm \rho$.  As a result, we can replace $c_{\psi \rho} - c'_{\psi \rho} \gamma_z$ in the
bulk-boundary two point function with $\frac{\mu_{\psi \rho}}{2} (1\mp \gamma_z)$, depending on the choice of eigenspace for $\rho$.  

\subsection{Embedding Space and Correlation Function in a Pure CFT}

The projective light-cone is the natural space 
in which to define CFT correlation functions 
because it linearizes the action of the conformal group \cite{Weinberg:2010ws,Costa:2011wa}. It is the quotient
\begin{align*}
	\flatfrac{\{P^{A}\in \mathbb{R}^{1,d+1} \mid P^2=0\}}{\{ \lambda \in \mathbb{R}^{+} \mid P^{A}\sim \lambda P^{A}\}} \ . 
\end{align*}
 
 Primary fields  are defined to have homogeneous scaling. In particular, for a scalar field of conformal dimension $\Delta$, one has 
\begin{align}
	\phi(\lambda P)=\lambda^{-\Delta}\phi(P) \ . 
\end{align}
The definition of primary generalizes to tensorial operators $F_{A_1 \cdots A_n}$.  To furnish an irreducible representation of the conformal group,
these tensor operators must be $P$-transverse and obey a ``gauge equivalence'' relation.  For a current $J_A$, these two conditions amount to the constraints
$P^A J_A = 0$ and $J_A \sim J_A + \alpha P_A$ for any $\alpha$.

There exists an index-free formalism for general tensor type which simplifies their manipulation \cite{Costa:2011wa,Costa:2014rya}. For symmetric tracelesss tensors (STT), one contracts all free indices with a polarisation vector $Z^{A}$ satisfying $Z^2=0$. We then require that $Z^{A}$ only appears in transverse structures, i.e.\ in objects invariant under $Z\rightarrow Z+\alpha P$. Because of the gauge equivalence relation, we can set $Z\cdot P =0$. Tensorial operators are then homogeneous, $P$-transverse polynomials in the polarisation vector $Z^{A}$. 

Projection to real space is done by choosing a section of the lightcone, specified through a parametrisation of the $P^{A}$. The projection is encoded through a set of replacement rules for the various quantities entering the correlation function. To be concrete, consider the Poincar\'e section, which maps to a CFT living on flat space,
\begin{align}
	P^{A} = \left(P^{+},P^{-},P^{\mu}\right) =\left(1,x^2,x^{\mu}\right) \ .
\end{align}
A solution for $Z^{A}$ compatible with the constraints $Z^2 = 0$ and $Z \cdot P = 0$ is $Z^A = (0, x \cdot z, z^\mu)$, provided $z^2 = 0$.
The projection to flat space can then be encoded in the replacement rules 
\begin{equation}
\label{reprules}
\begin{aligned}
	P_{ij}=-2P_{i}\cdot P_j &\rightarrow \abs{x_i-x_j}^{2}=\abs{x_{ij}}^2\ ,  \\
	P_{i}\cdot Z_{j}&\rightarrow (x_i-x_j)\cdot z_j=x_{ij}\cdot z_j \ , \\
	Z_{ij} = Z_{i} \cdot Z_{j} &\rightarrow z_i \cdot z_j 
	\ , 
\end{aligned}
\end{equation}
and the resulting expressions describe the correlation function of operators in real space.

We will write correlation functions evaluated without a boundary using the symbol $\expval{\ldots}_{\circ}$. Using the constraints previously considered, one can classify correlation functions of scalar and vector operators 
\begin{align}
	\bev{\phi(P_1)}= 0 \ , \; \; \;
\bev{\phi(P_1) \phi(P_2)} = 
\frac{1}{(P_{12})^{\Delta} }
\ , \\
\bev{Z_1 \cdot J(P_1) Z_2 \cdot J(P_2)} = c_{JJ} \frac{{\mathcal S}_1}{(P_1 \cdot P_2)^{\Delta_J}}
\ ,
\end{align}
where we have defined the transverse tensor structure
\begin{equation}
{\mathcal S}_1 \equiv Z_1 \cdot Z_2 - \frac{(Z_1 \cdot P_2) (Z_2 \cdot P_1)}{P_1 \cdot P_2} \ .
\end{equation}
If the current is furthermore conserved, then $\Delta_J = d-1$.  Note the transversality condition fixes the relative normalization between the two terms in ${\mathcal S}_1$.  Using the replacement rules (\ref{reprules}), this tensor structure becomes $I_{\mu\nu}$ in real space.

Spinors are dealt with similarly \cite{Weinberg:2010ws,Iliesiu:2015qra,Isono:2017grm}. A primary spinor field $\Psi$ obeys a scaling and irreducibility relation 
\begin{align}
	\Psi(\lambda P)&=\lambda^{-\Delta-\frac{1}{2}}\Psi(X) \ , & P^{A}\Gamma_A \Psi(P)&= 0 \, .
\end{align}
An extra complication is the relation between the embedding $\Psi$ and real space $\psi$ spinor representations.
Generically, we look at insertions of Dirac spinors $\psi$ of $\mathbb{R}^{d}$. These are uplifted to Dirac spinors $\Psi$ of $\mathbb{R}^{1,d+1}$, which are  twice as big. Correspondingly, if the Clifford algebra in real space is given through 
\begin{align}
	\{\gamma_\mu,\gamma_\nu\}=2\delta_{\mu\nu} \ ,
\end{align}
then one can construct the embedding space algebra through the tensor product 
\begin{align}
	\Gamma_{\mu} &= \gamma_\mu \otimes \sigma_3 \ , & \Gamma_{0} &= \mathbb{1} \otimes i\sigma_2 \ , & \Gamma_{d+1} &=\mathbb{1}\otimes \sigma_1 \ .
\end{align}
If required, a ``gamma-five'' $\Gamma_\star$ matrix can be included as well. 
Multiplying a given spinorial structure by $\Gamma_\star$ gives a second allowed structure, but with the same conformal block.  As the examples
we consider in this work do not require the $\Gamma_\star$ structures, we ignore this added complication.

It is convenient to work with scalar objects. 
The index-free formalism generalises to spinors \cite{Iliesiu:2015qra,Isono:2017grm}. One contracts the spinor field $\Psi$ with a (Grassman even) polarisation conjugate spinor $\overline{S}$, which incorporates the constraints, hence also the projection to real space. To be explicit, in the Poincar\'e section, one can take 
\begin{align}
	\overline{S} &= \overline{s}\begin{pmatrix}
		\slashed{x} & 1
	\end{pmatrix} \ , & \quad S &= \begin{pmatrix}
		1 \\ -\slashed{x}
	\end{pmatrix}s \ .
\end{align}
Using these expressions, one can then find as before a set of replacement rules encoding the real space limit. Allowed spinorial structures in correlation functions are then the non-vanishing Lorentz-scalars one can build using the elements introduced by the insertions. Structures are efficiently written in a bra-ket notation, e.g.\ $\bra{S_1}P_3\ket{S_2}=\bra{1}P_3\ket{2}=\overline{S}_1\slashed{P}_3S_2$. We will abuse the bra-ket notation to designate both  the sandwiching of the matrices between the embedding spinor polarisations, $S$, and the real-space polarisations, $s$.  We will also write $\bra{1}x_{12} \ket{2}=\overline{s}_1\gamma \cdot x_{12} s_2$. Which is meant should hopefully be clear from context. Note that with the conventions spelled out previously, this structure is real under conjugation and simultaneous exchange of 1 and 2.

Using this notation, we can efficiently constrain correlation functions of spinors 
\begin{align}
	\label{psipsiembedding}
	\bev{\psi(P_1,\overline{S}_1)\overline{\psi}(P_2,S_2)}&=\frac{\braket{1}{2}}{(P_{12})^{\Delta +\frac{1}{2}}} \ , \\
	\bev{\psi(P_1,\overline{S}_1)\overline{\psi}(P_2,S_2)\phi(P_3)}&= \lambda^{(1)}_{\phi}\frac{\braket{1}{2}}{P_{13}^{\frac{\Delta_1+\Delta_3-\Delta_2}{2}}P_{23}^{\frac{\Delta_2+\Delta_3-\Delta_1}{2}}P_{12}^{\frac{\Delta_1+\Delta_2-\Delta_3+1}{2}}}\\
	&+ \lambda^{(2)}_{\phi}\frac{\bra{1}P_3\ket{2}}{P_{13}^{\frac{\Delta_1+\Delta_3+1-\Delta_2}{2}}P_{23}^{\frac{\Delta_2+\Delta_3+1-\Delta_1}{2}}P_{12}^{\frac{\Delta_1+\Delta_2-\Delta_3}{2}}} \ . \nonumber
\end{align}
These equations set our conventions for the CFT data coefficients, and will reappear when we consider the bulk OPE of operators with a boundary. Note 
the $\lambda^{(i)}$ are real numbers, provided one uses Lorentzian conjugation to define reality. From the conventions previously given, real-space expressions are given using replacement rules (see also Appendix \ref{app:conventions}) 
\begin{equation}
	\begin{aligned}
	\braket{1}{2} &\rightarrow \bra{1}x_{12}\ket{2} = \overline{s}_1\, \gamma\cdot x_{12} \, s_2  \ ,\\
	\bra{1}P_3\ket{2} &\rightarrow \bra{1}x_{13} x_{32} \ket{2} = \overline{s}_1 \, \gamma\cdot  x_{13} \, \gamma \cdot x_{32}  \, s_2 \ .
\end{aligned}
\end{equation}
These rules yield from (\ref{psipsiembedding}) the familiar two point function of a Dirac spinor primary 
\begin{align}
	\bev{\psi(x)\overline{\psi}(x')}&=\frac{\bra{1}(x-x')\ket{2}}{\abs{x-x'}^{2\Delta+ 1}} \ .
\end{align}
One can proceed similarly to construct correlation functions involving more complicated correlation functions, but at this point we have what we need
before proceeding to the more elaborate case of a boundary.

\subsection{Correlator in the Presence of a Boundary}

In a boundary conformal field theory, one has to deal with an explicit breaking of the conformal group $SO(1,d+1)\rightarrow SO(1,p+1)$, $p=d-1$. In real space, this breaking is induced by a preferred direction specified through a unit normal $n^{\mu}$. In the embedding picture, this breaking is associated to a unit vector $V^{A}$, $V^2=1$ \cite{Liendo:2012hy}. It follows that one should distinguish two types of insertions, bulk $P\cdot V \neq 0\neq x\cdot n$, and boundary, $P\cdot V = 0= x\cdot n $. 

It is useful at times to consider an invariant inner product, $A \bullet B = A \cdot B - A \cdot V B\cdot V$. While bulk insertions are defined as in the pure CFT case, with fields transforming as irreducible representations of $SO(1,d+1)$, boundary ones have more structure, as bulk representations are generically reducible with respect to the boundary symmetry group. More precisely, one can consistently separate the normal and parallel components of tensor operators, which then transform separately.

Boundary operators in embedding space can be introduced as before, but now with some extra constraints.  
They cannot depend on the normal direction, and hence $P \cdot V = 0$.  Furthermore, if we contract a symmetric traceless boundary tensor 
with a polarization vector $W_A$ such that $W^2 = 0$, then we should choose $W$ such that $W \cdot V = 0$.

Correlation functions can then be constructed as in the pure CFT case: by building the most general Lorentz invariant object in the embedding given the objects and constraints at hand, and then projecting them to real space.  
The conformally invariant cross ratio can be written in embedding space:
\begin{align}
	\xi = \frac{-2P_1\cdot P_2}{(-2P_1\cdot V)(-2P_2 \cdot V)} \ .
\end{align}
 For the bulk scalar and vector, we have almost immediately 
\begin{align}
\expval{\phi(P)}&= \frac{a_{\phi}}{(2P\cdot V)^{\Delta}} \ , \\
\expval{\phi(P_1)\phi(P_2)}&=\frac{H(\xi)}{(2P_1\cdot V)^{\Delta_1}(2P_2\cdot V)^{\Delta_2}} \ , \\
\expval{Z_1 \cdot J(P_1) Z_2 \cdot J(P_2)} &= \frac{1}{(-2 P_1 \cdot P_2)^{\Delta_J}} ( Q(\xi) {\mathcal S}_1 + (R(\xi) - Q(\xi)) {\mathcal S}_2 ) 
\end{align}
where ${\mathcal S}_1$ is the same as in the case without boundary,
but now there is a new structure:
\begin{equation}
{\mathcal S}_2 \equiv \left( Z_1 \cdot V - \frac{(Z_1 \cdot P_2) (P_1 \cdot V)}{P_1 \cdot P_2} \right) \left(Z_2 \cdot V - \frac{(Z_2 \cdot P_1) (P_2 \cdot V)}{P_1 \cdot P_2} \right) \ .
\end{equation}
These expressions project down to the real space results (\ref{oneandtwophi}), (\ref{phiphi}), and (\ref{JJ}).  As they are anyway well known, 
we will leave the embedding space form of (\ref{phisigma}), (\ref{Jtau}), and (\ref{Jva})  as exercises.

In this work, we are concerned with insertions of fields transforming as spinors under the rotation group, both as  bulk and boundary insertions. The initial question is how to generalise the analysis of boundary tensors to that of spinors. 

On the boundary, the preserved generators of the conformal group will commute with the matrices $\slashed{V}$ in the embedding, respectively $\slashed{n}$ in real space. Hence, when we consider generic boundary spinor insertions $\rho(x)$, $x\cdot n =0$, we can impose irreducibility by requiring 
\begin{align}
	\slashed{n}\rho(x)& =\pm \rho(x) \ , & \slashed{V}\rho(P)&=\pm\rho(P) \ .
\end{align}
The boundary primaries are then ``chiral spinors'' $\rho_{\pm}$. In odd dimensions $d=2k+1$, a Dirac spinor of $Spin(d)$ is also a Dirac spinor of $Spin(p)$. Then, the $\rho_{\pm}$ can be thought of as Weyl fermions under the stabiliser group of the boundary, i.e.\ intrinsically to the boundary theory, they are simply Weyl spinors. When $d=2k+2$, the analysis is more intricate; it turns out that $\rho_{\pm}$ transform under the two unitarily inequivalent Dirac spinor representations of $Spin(p)$. Intrinsically, we then have a doublet of Dirac spinors, which are exchanged under ``parity". These are however technical details, useful when working intrinsically, with only boundary insertions.

When we consider a mix of bulk and boundary insertions, it is most convenient, as for tensors, to encode the boundary objects using constrained bulk representations, as we will do from now on. This is efficiently done by using the projectors $\Pi_{\pm}=\frac{1}{2}\left(\mathds{1}\pm \slashed{V} \right)$, and $\mathbb{P}_{\pm}=\frac{1}{2}\left(\mathds{1}\pm \slashed{n}\right)$. 
As before, one can work in an index-free formalism, by contracting all objects with polarisation spinors, but now the polarisations for boundary insertions have a definite helicity. 

One should however be cautious when comparing real space and embedding expressions though, as the change in signature makes it such that $\overline{\rho}_{\pm}(x)\slashed{n}=\mp \overline{\rho}_{\pm}(x)$, while $\overline{\rho}_{\pm}(P)\slashed{V}=\pm \overline{\rho}_{\pm}(P)$, i.e.\ Dirac conjugation exchanges helicity eigenvalue in real space, but not in the embedding. For this reason, embedding space becomes useful to derive conformally invariant structures, but those are simpler to analyse in real space.  For completeness, one can define polarisation spinors
\begin{align}
	\overline{S}^{(\pm)}\slashed{V}& = \pm \overline{S}^{(\pm)} \ ,  & \slashed{V}S^{\pm}&=\pm S^{(\pm)} \ , \\
	\overline{s}_{\pm}\slashed{n}&=\pm \overline{s}_{\pm} \ , & \slashed{n}s_{\pm}&=\mp s_{\pm} \ .
\end{align}

These subtleties can create confusion. For example, if we consider the bra-ket notation, with an added label for helicity $\bra{1,\alpha}$, this leads to projections from embedding space to real space where we have  schematically
\begin{align}
	\bra{1,\alpha} &\rightarrow \bra{1,\alpha}(\ldots) \ ,  \\
	\ket{2,\alpha} &\rightarrow (\ldots)\ket{2,-\alpha} \ .
\end{align}
Although the rules may appear intricate, they are simple to deploy. Consider for example the two point function of the boundary spinor 
\begin{align}
	\expval{\rho_{\alpha}(x)\overline{\rho}_{\beta}(y)}=\frac{\bra{1,\alpha}(x-y)\ket{2,-\beta}}{\abs{x-y}^{2\Delta+1}}=\overline{s}_1\mathbb{P}_{\alpha}\frac{\gamma\cdot(x-y)}{\abs{x-y}^{2\Delta+1}}\mathbb{P}_{-\beta}s_2 \ .
\end{align}
Clearly $\alpha=\beta$ for this correlator not to vanish. Note again that under complex conjugation as well as exchange of $(1,x) \leftrightarrow (2,y)$, this correlator is mapped to itself, since the kets exchange helicity. The embedding space spinor structure in a pure CFT is $\bra{1}\ket{2}$, which is diagonal; hence we need to project $\ket{2,+}$ in the embedding into something proportional to $\ket{2,-}$ in real space, 
\begin{align}
	\expval{\rho_{\alpha}(P_1)\overline{\rho}_{\beta}(P_2)}=\frac{\bra{1,\alpha}\ket{2,\beta}}{(P_{12})^{2\Delta+1}} \ .
\end{align}
Being even more explicit, we can view the real-space and embedding space expressions as the same ones, under the following set of supplementary replacement rules 
\begin{equation}
\begin{aligned}
	P_i\cdot V & \rightarrow x_i\cdot n \ ,  \\
	\bra{i}V\ket{j} & \rightarrow \bra{1} n (-\tilde x_i+ x_j)\ket{2} \ ,  \\ 
	\bra{i}\ket{j,\alpha} & \rightarrow \bra{i}(x_i-x_j)\ket{j,-\alpha} \ ,  \\
	\bra{i,\alpha}\ket{j} & \rightarrow \bra{i,\alpha}(x_i-x_j)\ket{j}  \ , \\
	\bra{i,\alpha}\ket{j,\beta} & \rightarrow \bra{i,\alpha}(x_i-x_j)\ket{j,-\beta} \ ,
\end{aligned}
\end{equation} 
and we define $\tilde{x}^{\mu}= x^{\mu}-2n^{\mu} x\cdot n$, the boundary reflection of $x^{\mu}$.

With these details spelled out, we can turn to the classification of correlation functions. The simplest spinorial correlator  
is the bulk-boundary correlator of $\psi$ and $\rho$
\begin{align}
	\expval{\psi(P_1)\overline{\rho}_{\alpha}(P_2)}&=\mu_{\Delta_2} \frac{\braket{1}{2,\alpha}}{(2P_1\cdot V)^{\Delta_1-\Delta_2}(-2P_1\cdot P_2)^{\Delta_2+\frac{1}{2}}} \ .
\end{align}
The real-space result for the bulk-boundary two-point function of spinors (\ref{psirho})  follows straightforwardly by projection.  It is useful to spell out the $\langle \rho \bar \psi \rangle$ correlation function as well:
\begin{align}
	\expval{\rho_{\alpha}(x_1,s_1)\overline{\psi}(x_2,s_2)}&=\mu_{\Delta_1}^{\dagger}\frac{\bra{1,\alpha} x_{12}\ket{2}}{(2z_2 )^{\Delta_2-\Delta_1}\abs{x_{12}}^{2\Delta_1+1}} \ ,
\end{align}
which follows by taking the complex conjugate of (\ref{psirho}). One can in general perform a phase redefinition of $\rho_{\alpha}$ to make $\mu_{\Delta}$ real valued, as we now assume. 

The natural next step in complexity is to consider spinor bulk-bulk correlators. There are two allowed spinorial structures.  For spinors in embedding space, we have
\begin{align}
	\expval{\psi(P_1,\overline{S}_1)\overline{\psi}(P_2,S_{2})}&=\frac{F(\xi)\braket{1}{2}+G(\xi)\bra{1}V\ket{2}}{(2P_1\cdot V)^{\Delta_1+\frac{1}{2}}(2P_2\cdot V)^{\Delta_2+\frac{1}{2}}} \ .
\end{align}
The spinor two point function and its two structures are the central object of study of this paper. More intuition for these structures can be found from the real space expression (\ref{psipsi}), where the second structure is associated to the reflection of the first one by the boundary. These structures match 
the structures that appear for free fermions \cite{McAvity:1993ul,McAvity:1995tm}, but we see from this analysis that no additional structures can appear
and be consistent with conformal invariance.\footnote{%
 We are ignoring the possibility of $\Gamma_\star$ structures here.
}

In the rest of this paper, we investigate in detail these functions $F(\xi)$ and $G(\xi)$. We decompose them in a distinguished basis, the conformal blocks, and constrain them using crossing symmetry, in a few perturbative models. The derivation of the bulk conformal blocks and perturbative analysis of our three models form the core of our results.

The kinematic classification for more complicated insertions, for example tensors and tensor-spinors,  is done more straightforwardly in the real space picture. The tactic is to use the different tensor and spinorial structures already found, and to multiply them together before taking out all $\gamma$-traces. 

\section{The Operator Product Expansion and Conformal Blocks}
\label{sec:blocks}

We determined that the correlator of interest, $\langle \psi(x) \overline \psi(x') \rangle$, is fixed up to two functions of a cross-ratio, $F(\xi)$ and $G(\xi)$.
These functions, however, are further constrained by the existence of two different operator product expansions (OPEs). The first one is the boundary OPE, where we take a bulk field and Taylor-expand it near the boundary, yielding a sum over boundary primary operators and their descendants. Meanwhile, one can bring two bulk insertions close together, and merge them into a sum of bulk operators; this is the bulk OPE. 
Grouping the sums into conformal multiplets, each of which consists of a primary operator and all its descendants, 
 we find the conformal block decomposition of the correlation function. 

In this section, we derive the conformal blocks.  The boundary conformal blocks were first derived in \cite{Nishida:2018opl,Herzog:2019bom} from AdS; while 
our methods are different, the results are the same.  Our results for the bulk conformal blocks, however, are new.
We compute the blocks by using a representation of them as conformal integrals.  We are then able to check the computation by explicitly resumming the OPE, the details of which are presented in appendix \ref{app:OPEresummation}.

\subsection{Conformal Blocks}

The conformal blocks depend on the type (bulk, boundary), and the properties of the exchanged operator: 
\begin{align}
	\expval{\psi(x)\overline{\psi}(y)}&= \sum_{\Delta, i=1,2} a_{\Delta}\lambda^{(i)}_{\Delta}\mathcal{W}^{}_{\Delta,i}(x,y)  \\
	&= \sum_{\widehat{\Delta}, \alpha = \pm} \mu_{\widehat{\Delta},\alpha}^{2}\widehat{\mathcal{W}}_{\widehat{\Delta},\alpha}(x,y) \ ,
\end{align}
where the hat is used to indicate the boundary.
Equivalently, this is an expansion for the invariant functions $F(\xi)$ and $G(\xi)$:
\begin{align}
	F(\xi)&=\sum_{\Delta,i=1,2}  a_{\Delta}\lambda^{(i)}_{\Delta} f_{\Delta,i}(\xi) & G(\xi)&=\sum_{\Delta,i=1,2} a_{\Delta}\lambda^{(i)}_{\Delta}g_{\Delta,i}(\xi) \\
	&= \sum_{\widehat{\Delta}, \alpha = \pm} \mu_{\widehat{\Delta},\alpha}^{2} \widehat{f}_{\widehat{\Delta},\alpha}(\xi) \ , &  
	&= \sum_{\widehat{\Delta}, \alpha = \pm} \mu_{\widehat{\Delta},\alpha}^{2} \widehat{g}_{\widehat{\Delta},\alpha}(\xi) \ .
\end{align}
This basis of blocks is easy to work with.
With the reality structure we picked, the blocks and coefficients multiplying them are manifestly real. 
It turns out that $\widehat{f}_{\widehat{\Delta},\alpha}$ is $\alpha$ independent, while $\widehat{g}_{\widehat{\Delta},\alpha}$ is proportional to $\alpha$. 
Moreover, $f_{\Delta, 2} = 0 = g_{\Delta, 1}$.

We distinguish between the $\mathcal{W}^{}_{\Delta,i}(x,y)$ and $\widehat{\mathcal{W}}_{\widehat{\Delta},\alpha}(x,y)$, which we call conformal partial waves (CPWs) and  the $f_{\Delta,i}(\xi)$, $g_{\Delta,i}(\xi)$, $\widehat{f}_{\Delta,\alpha}(\xi)$, and $\widehat{g}_{\Delta,\alpha}(\xi)$ 
we call conformal blocks \cite{Dolan:2004up}.  
There are two distinct methods for computing these objects: 
solving a differential equation and computing a sum.  
The differential equation route, involves identifying the CPWs as 
eigenfunctions for the bulk or boundary Casimir operators.
The sum route involves resumming the OPE \cite{Dolan:2001wg,Billo:2016vm}. The OPE relations are best given in real space. For the bulk OPE one has
\begin{align}
\label{bulkpsiOPE}
	 \psi_{1}(x)\overline{\psi}_{2}(y)&= \sum_{\Delta,i=1,2} \frac{\lambda_{\Delta}^{(i)}\mathfrak{C}^{(i)}_{(\Delta_1,\Delta_2,\Delta)}[x-y,\partial_y]\phi_{\Delta}(y)}{\abs{x-y}^{\Delta_1+\Delta_2+1-\Delta}} +(\ell>0)\\
	 &=\sum_{\Delta}\frac{\left(\lambda_{\Delta}^{(1)}\gamma\cdot(x-y)+\lambda_{\Delta}^{(2)}\mathbb{1}\abs{x-y}\right)}{\abs{x-y}^{\Delta_1+\Delta_2+1-\Delta}}\phi_{\Delta}\left(y\right)+(\ell>0)+desc. \nonumber
\end{align}
While for the boundary OPE we find :
\begin{align}
\label{boundarypsiOPE}
	\psi_{\Delta}(x,z)&= \sum_{\widehat{\Delta},\alpha}\frac{\mu_{\widehat{\Delta},\alpha}}{(2z)^{\Delta_1-\Delta}}\mathfrak{D}_{(\Delta,\widehat{\Delta})}^{(\alpha)}(z,x,\partial_x)\rho_{\widehat{\Delta},\alpha}\left(x\right) \\
	&= \sum_{\widehat{\Delta},\alpha}\frac{\mu_{\widehat{\Delta},\alpha}}{(2z)^{\Delta_1-\Delta}}\rho_{\widehat{\Delta},\alpha}\left(x\right)+desc.  \nonumber
\end{align}
In appendix \ref{app:OPEresummation}, we systematically determine the form of the sum over descendants, and then perform the resummation to obtain the conformal blocks.  
The resummation is technical, and we present the differential equation route here instead, 
computing the blocks through the shadow formalism as conformal integrals \cite{Ferrara:1972uy,Ferrara:1973yt}.  

The CPWs are uniquely fixed by their transformation properties under the conformal group and matching with the OPE leading terms. More specifically, they inherit the transformation properties of the whole correlator, but form an eigenfunction of the conformal Casimir \cite{Dolan:2012wt}. As such, any good ansatz satisfying these conditions will give a representation of the CPW. We represent the CPWs as conformal integrals -- as projections of the conformally-invariant sewing of lower point functions. These integrals are most easily evaluated in the embedding space \cite{Simmons-Duffin:2014wb}. The conformal integral is given symbolically by  
\begin{align}
	\int D^{d}X
\end{align}
and can integrate weight $-d$ functions in  the variable $X^{A}$. Its defining property is that it forms a conformally invariant integration measure over the point $X^{A}$. We also encounter the reduced integral with $d\rightarrow p = d-1$, which is invariant under the conformal group preserved by the boundary. Any such conformal integral can generically be reduced to a fundamental case, which is fixed from dimensional analysis
\begin{equation*}
	\int D^{d}X (-2 Y\cdot X)^{-d} \propto (- Y \cdot Y)^{-\frac{d}{2}} \ .
\end{equation*}
The partial waves then follow from sewing points of simpler correlators using this integral. One must further take into account a projection $\mathcal{P}$, and a normalisation $\mathcal{N}$. These are fixed by matching with the bulk or boundary OPE. The conformal integral yields a mixture of the block corresponding to the exchange of operator with dimension $\Delta$, as well as its shadow, which corresponds to the exchange of the unphysical operator with dimension $d-\Delta$. This second contribution must be projected out. The normalisation is fixed by matching to the OPE leading term. 

One encounters integrals which are a product of power laws with typical insertions in the numerator of the form $(Z\cdot X)^{l}$. These can be reduced to the fundamental case given earlier using standard methods. In appendix \ref{app:conformalintegral}, we explicitly perform this analysis in full generality to compute all possible conformal integrals for the partial wave type considered in this work.

\subsection{Conformal Integral for Boundary Block}

The boundary blocks are eigenvectors of the boundary conformal Casimir acting on either of the spinor insertions.  We construct an ansatz by sewing together two bulk to boundary spinor correlators
\begin{align}
	\widehat{\mathcal{W}}_{\Delta,\alpha}(P_1,P_2,\overline{S}_1,S_2) =\frac{\mathcal{P}}{\mathcal{N}}\int D^{p}X\expval{\psi(P_1,\overline{S}_1)\overline{\rho}_{\Delta}(X)}\slashed{X}\Pi_{\alpha}\expval{\rho_{p-\Delta}(X)\overline{\psi}(P_2,S_2)} \ . 
\end{align}
By writing $\rho_{\Delta}(X)$ in place of $\rho_{\Delta, \alpha}(X,S)$, we have implicitly removed the $S$ dependence from the correlation function.
We claim this object has the correct properties to be a CPW.  It is an eigenvector of the boundary Casimir operator.  The extra power of $X_A$ is required
by dimensional analysis, and the $\Pi_\alpha$ picks out a conformal block of a particular helicity.

Once this integral is computed, we extract the putative blocks, and from there, fix the normalisations. These follow from the matching with the OPE in some limit of the cross-ratio. In the case of the boundary blocks, we have the normalisation conditions 
\begin{align}
	\lim_{\xi \rightarrow \infty} \xi^{\Delta+\frac{1}{2}}\hat{f}_{\Delta,\alpha}(\xi)&=\frac{1}{2}  \ , & \lim_{\xi\rightarrow \infty} \xi^{\Delta+\frac{1}{2}}\hat{g}_{\Delta,\alpha}(\xi)&=\frac{-\alpha}{2} \ .
\end{align}
The integral is given by 
\begin{equation}
	\begin{aligned}
	&= \frac{\overline{S}_1\Gamma_{A}\Pi_{\alpha}S_2}{(2P_1\cdot V)^{\Delta_1-\Delta}(2P_2\cdot V)^{\Delta_2-\tilde{\Delta}}} \frac{\mathcal{P}}{N}\int D^{p}X\frac{X^{A}}{(-2P_1\bullet X)^{\Delta+\frac{1}{2}}(-2P_2\bullet X)^{\tilde{\Delta}+\frac{1}{2}}} \ ,
	\end{aligned}
\end{equation}
where $\tilde \Delta = p  - \Delta$. 
It is now a straightforward exercise to use the formulas given in appendix \ref{app:conformalintegral}, plus the normalization conditions, to obtain a closed form expansion for the boundary blocks. After the dust settles, one finds the remarkably simple expression for the partial wave
\begin{equation}
\label{boundaryCPW}
	\begin{aligned}
	\widehat{\mathcal{W}}_{\Delta,\alpha} &=\frac{1}{2\xi^{\Delta+\frac{1}{2}}}\frac{1}{(2P_1\cdot V)^{\Delta_1+\frac{1}{2}}(2P_2\cdot V)^{\Delta_2+\frac{1}{2}}}\\
	&\times  \overline{S}_1 \Bigg(\, _2F_1\left(\Delta +\frac{1}{2},\Delta -\frac{p}{2};1-p+2 \Delta;-\frac{1}{\xi}\right) \\
	&-\alpha \slashed{V} \, _2F_1\left(\Delta +\frac{1}{2},1+\Delta-\frac{p}{2};1-p+2\Delta;-\frac{1}{\xi}\right) \Bigg)S_2 \ ,
	\end{aligned}
\end{equation}
and the expression for the blocks follows:
\begin{align}
	\widehat{f}_{\widehat{\Delta},\alpha}(u)&= \frac{1}{2}
	\xi^{-\widehat \Delta -\frac{1}{2}}
	\,_2 F_1\left(\widehat{\Delta}+\frac{1}{2} ,\widehat{\Delta}-\frac{p}{2} ; 2\widehat{\Delta}-p+1; -\frac{1}{\xi}\right) \ , \\
	\widehat{g}_{\widehat{\Delta},\alpha}(u)&= \frac{-\alpha}{2} 
	\xi^{-\widehat \Delta -\frac{1}{2}}
	\,_2 F_1\left( \widehat{\Delta} +\frac{1}{2},1+\widehat{\Delta}-\frac{p}{2} ;2\widehat{\Delta}-p+1 ; -\frac{1}{\xi}\right) \ .
\end{align}
Closely related expressions appear in  \cite{Nishida:2018opl,Herzog:2019bom,Kawano:1999au}.  

Because of the simple $\alpha$ dependance of these blocks, it is convenient to write the block decomposition using the linear combination that appears in the first and second structure:
\begin{equation}
\label{boundaryblocks}
\begin{aligned}
	\sum_{\alpha}\mu_{\widehat{\Delta},\alpha}^2\widehat{f}_{\widehat{\Delta},\alpha}(\xi)&=\nu_{\widehat{\Delta}}^{(1)}\widehat{f}_{\widehat{\Delta}}(\xi) \ , &  \nu_{\widehat{\Delta}}^{(1)}&=\frac{\mu_{\widehat{\Delta},+}^2+\mu_{\widehat{\Delta},-}^2}{2}  \ , \\
	\sum_{\alpha}\mu_{\widehat{\Delta},\alpha}^2\widehat{g}_{\widehat{\Delta},\alpha}(\xi)&=\nu_{\widehat{\Delta}}^{(2)}\widehat{g}_{\widehat{\Delta}}(\xi)  \ , & \nu_{\widehat{\Delta}}^{(2)}&=\frac{\mu_{\widehat{\Delta},-}^2-\mu_{\widehat{\Delta},+}^2}{2} \ .
\end{aligned}
\end{equation}

\subsection{Bulk Block From Conformal Integrals}

Consider now the bulk CPW. It is by definition an eigenvector of the bulk conformal Casimir acting on the two spinor insertions. It then follows that it can be computed through the ansatz
\begin{align}
	\mathcal{W}_{\Delta,i}(P_1,P_2,\overline{S}_1,S_2) =\frac{\mathcal{P}}{\mathcal{N}}\int D^{d}X \bev{\psi(P_1,\overline{S}_1)\overline{\psi}(P_2,S_2)\phi_{\Delta}(X)}^{(i)} \expval{\phi_{d-\Delta}(X)} \ ,
\end{align}
with $\bev{\ldots}^{(i)}$ enumerating the independent structures in the correlator. Indeed, provided the integration is invariant under the whole conformal group, as well as the projection, then this object manifestly satisfies the properties to be a CPW, since the pure CFT three point functions are eigenvectors of the bulk Casimir operator. One must further fix the normalization. The CPW asymptotics are constrained by the bulk OPE, which translate into the conditions  
\begin{equation}
\begin{aligned}
	\lim_{\xi\rightarrow 0} \xi^{\frac{\Delta_1+\Delta_2+1-\Delta}{2}}f_{\Delta,1}(\xi)&=1 \ ,  & f_{\Delta,2}(\xi)&= 0  \ , \\
	g_{\Delta,1}(\xi) &= 0 \ , & \lim_{\xi\rightarrow 0} \xi^{\frac{\Delta_1+\Delta_2-\Delta}{2}}g_{\Delta,2}(\xi)&=1 \ .
\end{aligned}
\end{equation}
With these elements fixed, one can proceed with the evaluation of the integrals to find the blocks. We now have two different cases to consider. The first one is $\mathcal{W}_{\Delta,1}$, which is just a straightforward conformal integral with no insertion
\begin{equation}
\label{bulkCPWone}
	\begin{aligned}
		\mathcal{W}_{\Delta,1}(P_1,P_2,\overline{S}_1,S_2)&=\frac{\overline{S}_1 S_2}{(2P_1\cdot V)^{\Delta_1+\frac{1}{2}}(2P_2\cdot V)^{\Delta_2+\frac{1}{2}}}  \xi^{\frac{\Delta-\Delta_1-\Delta_2-1}{2}} \\ 
	&\times  \, _2F_1\left(\frac{\Delta +\Delta_1-\Delta_2}{2},\frac{\Delta -\Delta_1+\Delta_2}{2};\Delta-\frac{d}{2}+1;-\xi\right) \ .
	\end{aligned}
\end{equation}
For the computation of the second partial wave, $\mathcal{W}_{\Delta,2}$, we encounter an integral with one insertion in the numerator, whose explicit form simplifies to give 
\begin{equation}
\label{bulkCPWtwo}
	\begin{aligned}
	\mathcal{W}_{\Delta,2}(P_1,P_2,\overline{S}_1,S_2)&=\frac{\overline{S}_1 \slashed{V} S_2}{(2P_1\cdot V)^{\Delta_1+\frac{1}{2}}(2P_2\cdot V)^{\Delta_2+\frac{1}{2}}} \xi^{\frac{\Delta-\Delta_1-\Delta_2}{2}} \\ 
	&\times  \, _2F_1\left(\frac{\Delta+1 +\Delta_1-\Delta_2}{2},\frac{\Delta+1 -\Delta_1+\Delta_2}{2};\Delta-\frac{d}{2}+1;-\xi\right) \ .
	\end{aligned}
\end{equation}

\section{Solutions to Crossing Symmetry}
\label{sec:crossing}

We have derived the block decomposition of the two point function of bulk spinor fields. The constraint of crossing symmetry is the equality
\begin{align}
	\sum_{\Delta} \begin{pmatrix}[2]
		\lambda a_{\Delta}^{(1)}\ f_{\Delta}^{(\Delta_1,\Delta_2)}(\xi) \\
		\lambda a_{\Delta}^{(2)}\ g_{\Delta}^{(\Delta_1,\Delta_2)}(\xi)
	\end{pmatrix} = \sum_{\widehat{\Delta}}
	\begin{pmatrix}[2]
		\nu^{(1)}_{\widehat{\Delta}}\ \widehat{f}_{\widehat{\Delta}}(\xi)\\
		\nu^{(2)}_{\widehat{\Delta}}\ \widehat{g}_{\widehat{\Delta}}(\xi)
	\end{pmatrix}
\end{align}
with functions given by 
\begin{equation}
\begin{aligned}
	f_{\Delta}^{(\Delta_1,\Delta_2)}(\xi)&=\sqrt{\xi}^{\Delta-\Delta_1-\Delta_2-1}\,_2 F_1\left(\frac{\Delta+\Delta_1-\Delta_2}{2} , \frac{\Delta-\Delta_1+\Delta_2}{2} ;\Delta+\frac{1-p}{2}; -\xi\right)  \ , \\
	g_{\Delta}^{(\Delta_1,\Delta_2)}(u)&= \sqrt{\xi}^{\Delta-\Delta_1-\Delta_2} \,_2 F_1\left(\frac{\Delta+\Delta_1-\Delta_2+1}{2} ,\frac{\Delta-\Delta_1+\Delta_2+1}{2} ; \Delta+\frac{1-p}{2} ; -\xi\right)   \ , \\
	\widehat{f}_{\widehat{\Delta}}(\xi)&=\frac{1}{\sqrt{\xi}^{2\widehat{\Delta}+1}}\,_2 F_1\left(\widehat{\Delta}+\frac{1}{2} ,\widehat{\Delta}-\frac{p}{2} ; 2\widehat{\Delta}-p+1; -\frac{1}{\xi}\right) \ , \\
	\widehat{g}_{\widehat{\Delta}}(\xi)&= \frac{1}{\sqrt{\xi}^{2\widehat{\Delta}+1}}\,_2 F_1\left( \widehat{\Delta} +\frac{1}{2},1+\widehat{\Delta}-\frac{p}{2} ;2\widehat{\Delta}-p+1 ; -\frac{1}{\xi}\right) \ .
\end{aligned}
\end{equation}
In this section, we explore this constraint in the context of the $\epsilon$-expansion of interacting QFT. We consider situations with $\Delta_1= \Delta_2=\Delta_\psi$. To simplify the notation, we drop redundant information, abbreviating $f^{(\Delta_\psi,\Delta_\psi)}_{\Delta}(\xi)\equiv f_{\Delta}$ and
$g^{(\Delta_\psi,\Delta_\psi)}_{\Delta}(\xi)\equiv g_{\Delta}$. 
Moreover, in this special case $\Delta_1= \Delta_2$, we can use (\ref{boundaryblocks}) to write the $\nu_{\widehat \Delta}^{(i)}$ in terms of the manifestly
non-negative quantities $\mu_{\widehat \Delta, \alpha}^2$.  

The starting point of our analysis is a free spinor. 
We follow with the generalized free field (GFF) version of the spinor, i.e.\ a spinorial field with a generalized dimension but whose correlation
functions still follow from Wick's Theorem.
We then consider three different interacting models generated by classically marginal deformations near dimensions $4$, $3$ and $2$:
\begin{itemize}
	\item The Yukawa model, $\mathcal{L}_{\rm int}= -g \phi \overline{\psi}\psi$, which describes the interaction of a fermion with a scalar field, which we study in $d=4-\epsilon$. At first order in $\epsilon$, the addition  of the field $\phi$ to the bulk OPE induces a reorganisation of the field content through a fake-primary effect. 
	\item The Gross-Neveu model, $\mathcal{L}_{\rm int}=\frac{g}{2} (\overline{\psi}_{a}\psi^{a})^2$, of an interacting $O(N)$ vector of fermions in $d=2+\epsilon$. In this model at the fixed point one has $g\sim \epsilon$, which allows us to fix the correlation function to order $\epsilon^2$. Surprisingly, this decomposition also fixes the value of the zero of the beta function, although it leaves unfixed one bulk anomalous dimension.
	\item A Yukawa-type model, $\mathcal{L}_{\rm int}= -\frac{g}{2} \phi_{a}\phi^{a} \overline{\psi}\psi$, which describes a fermion interacting with an $O(N_s)$-vector of scalar fields. In $d=3-\epsilon$, this model remarkably generates an  expansion in $\sqrt{\epsilon}$ rather than $\epsilon$ for boundary data.  Similar behavior was remarked for the $\phi^6$ scalar field theory in $3-\epsilon$ dimensions \cite{diehi1987walks,eisenriegler1988surface}. More surprisingly still, crossing symmetry of $\expval{\psi\overline{\psi}}$ does not fully fix the boundary anomalous dimension at the next order. This ambiguity is resolved by bootstrapping the leading order deformation of the scalar-scalar bulk correlator.
\end{itemize}

In each case, we reproduce data concerning the bulk critical point from this boundary bootstrap. We employ the same methodology for each interacting model. Starting from the equation of motion applied to the correlation function, we find differential equations for $F(\xi)$, $G(\xi)$. We solve them  perturbatively in $\epsilon$, up to integration constants. This ansatz is then decomposed on the bulk and boundary block basis. Compatibility fixes the integration constants and allows us to read off the anomalous dimensions and field content of the theory.

\subsection{(Gaussian) Free Theory}

Let us first consider the free theory
\begin{align}
	\expval{\psi(x)\overline{\psi}(x')}&=\frac{\bra{1}(x-x')\ket{2}}{\abs{x-x'}^{\frac{p+1}{2}}}+\chi\frac{\bra{1}n(x'-\tilde{x})\ket{2}}{\abs{\tilde{x}-x'}^{\frac{p+1}{2}}} \\
	&= \mathcal{W}_{0}^{(0)}+\chi \mathcal{W}_{p}^{(1)}= (1-\chi)\widehat{\mathcal{W}}_{\frac{p}{2},+}+(1+\chi)\widehat{\mathcal{W}}_{\frac{p}{2},-} \nonumber
	\ ,
\end{align}
where the boundary condition on the spinor is imposed through a choice of $\chi$.  
The second line solves the crossing symmetry constraint.  
In terms of the functions $F(\xi)$ and $G(\xi)$, we have more precisely
 \begin{align}
	\begin{pmatrix}[2]
		F(\xi)\\
		G(\xi)
	\end{pmatrix}&= \begin{pmatrix}[2]
		\xi^{-\frac{p+1}{2}} \\
		\chi(1+\xi)^{-\frac{p+1}{2}}
	\end{pmatrix}=\begin{pmatrix}[2]
		 f_{0} \\
		\chi \, g_{p}
	\end{pmatrix}=\begin{pmatrix}[2]
		 \widehat{f}_{\frac{p}{2}} \\
		\chi  \, \widehat{g}_{\frac{p}{2}}
	\end{pmatrix} \ .
\end{align}
The positivity of the boundary blocks constrains $\chi$, 
\begin{align}
	\abs{\mu_{\frac{p}{2},\alpha}}^2 \geq 0 \Rightarrow -1 \leq \chi\leq1 \ .
\end{align}
Indeed, a very similar constraint appears when looking at the boundary blocks for a free scalar. 
In both cases, the choices $\chi = \pm 1$ play a special role.  In the scalar case, they select for either Dirichlet or Neumann boundary 
conditions.  In the spinor case, decomposing the boundary value of $\psi$ into boundary spinors of plus or minus chirality 
with respect to $\gamma_z$, they set one of the boundary spinors $\rho_\pm$ to zero.

We can use Wick contraction to explain the operators that show up in the bulk decomposition:
\begin{align}
	\psi(x)\overline{\psi}(x') &= \wick{ \c\psi(x)\overline{\c \psi}(x')}+{:}\psi(x)\overline{\psi}(x'){:} \\
	&= \frac{\gamma\cdot(x-x')}{\abs{x-x'}^{p+1}}-\frac{\mathbb{1}}{\Tr(\mathbb{1})}{:}\overline{\psi}(x)\psi(x){:}-\frac{\gamma^\mu}{p+1}{:}\overline{\psi}(x)\gamma_\mu\psi(x){:}+\ldots 
	\nonumber
\end{align}
where we used the generalized Fierz identity on the spinor bilinear and Taylor expanded the second insertion to bring it to a coincident point. 
The ellipsis denotes terms depending on derivatives and possibly higher powers of the gamma matrices.
Because we are in a bCFT, the one point functions of all operators with nonzero spin vanish.  Indeed only the 
first two terms survive after taking an expectation value.  As there is a unique dimension zero operator, the identity, in a CFT, we already knew
the origin of $\mathcal{W}_{0}^{(0)}$.  However, now we see very clearly that $\mathcal{W}_{p}^{(1)}$ corresponds to the mass operator
${:} \overline \psi \psi {:}$.  

The richer case of a  GFF spinor follows straightforwardly. The two point function is taken to be instead
\begin{align}
	\expval{\psi(x)\overline{\psi}(x')}&=\frac{\bra{1}(x-x')\ket{2}}{\abs{x-x'}^{2\Delta+1}}+\chi\frac{\bra{1}n(x'-\tilde{x})\ket{2}}{\abs{\tilde{x}-x'}^{2\Delta+1}} \ ,
\end{align}
corresponding to the bulk decomposition 
\begin{align}
\label{FGFFbulk}
	F(\xi) &=  f_{0}  \ ,  \\
\label{GGFFbulk}
	G(\xi) &= \chi \sum_{k=0}^{\infty}\frac{(-1)^k}{k!}\frac{\left(\Delta +\frac{1}{2}\right)_k \left(\Delta -\frac{p}{2}\right)_k}{\left(2 \Delta -\frac{p+1}{2}+k\right)_k}\,  g_{2\Delta+2k} \ .
\end{align}
The bulk blocks in the second structure correspond naturally to the exchange of double trace operators of the schematic form
 ${:}\overline{\psi}\Box^k \psi{:}$. In the limit $\Delta \rightarrow \frac{p}{2}$, we retrieve the free theory result of a single contribution from the spinor bilinear, while the remaining operators become null.
 This example clearly show the non-positivity of the coefficients entering the bulk block expansion. 

In the boundary channel, we find a sum over an infinite tower of single trace operators, with schematic form ${:}\partial_n^{k}\rho_{\pm}{:}$
\begin{align}
\label{FGFFbry}
	F(\xi) &= \sum_{k=0}^{\infty} \frac{1}{k!}\frac{\left(\Delta +\frac{1}{2}\right)_k (2 \Delta -p)_k}{4^k\left(\Delta -\frac{p-1}{2}\right)_k}\,  \widehat{f}_{\Delta+k}  \ , \\
	\label{GGFFbry}
	G(\xi)&= \chi\sum_{k=0}^{\infty} \frac{(-1)^k}{k!}\frac{\left(\Delta +\frac{1}{2}\right)_k (2 \Delta -p)_k}{4^k\left(\Delta -\frac{p-1}{2}\right)_k}\,  \widehat{g}_{\Delta+k} \ ,
\end{align}
where now, in the free field limit, the Dirac equation imposes $\partial_n \rho_+ = - \gamma^{i}\partial_i \rho_-$, making all fields, except the leading ones, descendant. As previously, reality of the expansion coefficients imposes $\chi^2\leq 1$.

Starting from free theory, one can try to perform a slight perturbation to the operator spectrum of the theory and its CFT data, to explore a conformal manifold. Generically, it is not so easy to find such manifolds. In bCFT, the parameter $\chi$ seems to give us an example of a marginal parameter. However, one expects that $\chi$ is fixed to $\pm 1$ in a free theory from a stress-tensor Ward Identity \cite{Herzog:2021spv}. To find continuous deformations away from mean-field theory which satisfy crossing symmetry, it is then convenient to perturb more than the field content. One path to a non-trivial deformation is the $\epsilon$-expansion, where one analytically continues the dimension $d$, or $p=d-1$, to non-integer values. This deformation gives us a parameter with which to explore points away from these Gaussian systems. Our goal for the remainder of this section is to explore some examples of solutions to crossing symmetry in the $\epsilon$-expansion. 

\subsection{The Yukawa Model at $1$-loop}

We now turn to a more interesting system, the Yukawa Model with a boundary. Our starting point is to assume there exists a perturbative fixed point, $g^2\sim \epsilon$, and draw the consequences from the equations of motions at leading order in $\epsilon$. This is analogous to the computation of \cite{Giombi:2021cnr}. The generic form of the two point function is as before
\begin{equation}
\begin{aligned}
	\expval{\psi(x)\overline{\psi}(x')}&=\frac{F(\xi)\bra{1}(x-x')\ket{2}+G(\xi)\bra{1}n(x'-\tilde{x})\ket{2}}{(4x\cdot n \, x'\cdot n)^{\Delta_\psi+\frac{1}{2}}} \ , \\
	\slashed{\partial}_x\expval{\psi(x)\overline{\psi}(x')}\overset{\leftarrow}{\slashed{\partial}}_{x'}&=\frac{\mathcal{D}_{F}\bra{1}(x-x')\ket{2}+\mathcal{D}_{G}\bra{1}n(x'-\tilde{x})\ket{2}}{(4x\cdot n \, x'\cdot n)^{\Delta_\psi+\frac{1}{2}}} \ ,
\end{aligned}
\end{equation}
where the differentiated blocks are given by 
\begin{equation}
\begin{aligned}
	\mathcal{D}_{G}&= -4\left( \xi  (1+\xi) F''(\xi )+ \left(\frac{p+3}{2}+\left(p+2\right)\xi\right) F'(\xi )+\left(\Delta_\psi+\frac{1}{2}\right)\left(p-\Delta_\psi+\frac{1}{2}\right)F(\xi )\right) \ , \\
\mathcal{D}_{G}&= -4 \left( \xi (1+\xi ) G''(\xi )+\left(\frac{p+1}{2}+(p+2)\xi\right) G'(\xi)+\left(\Delta_\psi+\frac{1}{2}\right)\left(p-\Delta_\psi+\frac{1}{2}\right)G(\xi)\right) \ .
\end{aligned}
\end{equation}
Replacing the explicit form of the Yukawa interaction inside the correlator, and evaluating the resulting four point function using Wick contractions, we find
\begin{equation}
\begin{aligned}
	\slashed{\partial}_x\expval{\psi(x)\overline{\psi}(x')}\overset{\leftarrow}{\slashed{\partial}}_{x'}&=-g^2\expval{\phi(x)\phi(x')}\expval{\psi(x)\overline{\psi}(x')}\\
	&= -\frac{g^2 H(\xi)F(\xi)\bra{1}(x-x')\ket{2}+g^2 H(\xi)G(\xi) \bra{1}n(x'-\tilde{x})\ket{2}}{(4x\cdot n \, x'\cdot n)^{\Delta_\psi+\frac{1}{2}+\Delta_\phi}} \ .
\end{aligned}
\end{equation}
We will take $\kappa_s g^2\equiv \lambda \epsilon$, and use the free field propagator for $H(\xi)$.
The anomalous dimension one finds for the spinor will enter only through the explicit factors of $\Delta_\psi$ in $\mathcal{D}_{F/G}$. The scalar field two point function depends on a reflection coefficient $\chi_{s}$,
\begin{equation}
	H(\xi)=\frac{1}{\xi}+\frac{\chi_s}{1+\xi} \ .
\end{equation}
The equations of motion simplify to two ordinary differential equations:
\begin{equation}
	\begin{aligned}
	\mathcal{D}_{F}&=-\epsilon\lambda H(\xi)F(\xi) \ ,  \\
	\mathcal{D}_{G}&=-\epsilon\lambda H(\xi)G(\xi) \ ,
	\end{aligned}
\end{equation}
which we solve perturbatively in $\epsilon$, setting $\Delta_\psi = \frac{p}{2}+ \gamma_\psi^{(1)}\epsilon+ O(\epsilon^2)$. 
At each order in $\epsilon$, we need to specify two boundary conditions for $F$ and two for $G$.
At zeroth order, the boundary conditions mean setting the correlation function to be the free theory result.  To $O(\epsilon^2)$, the solution is 
\begin{equation}
\label{FGsol}
	\begin{aligned}
		F(\xi)&=\frac{1}{\xi^{2}}+\frac{\epsilon}{\xi^2}\bigg(\delta \kappa -\frac{1}{2}\left(\frac{1}{1+\xi}+\log(1+\frac{1}{\xi})\right)+c_1\left(\frac{1}{1+\xi}+\log(1+\xi)\right)  \\
	&\quad \quad \quad \quad \quad \quad +\frac{\lambda}{8}\left(\frac{1+2\chi_s}{1+\xi} +\log(1+\frac{1}{\xi})\right) \bigg) + O(\epsilon^2) \ , \\
	G(\xi)&=\frac{\chi}{(1+\xi)^2}+\frac{\epsilon\chi}{(1+\xi)^2}\bigg(\delta \chi + \frac{1}{2}  \left(\frac{1}{\xi}+\log(1+\frac{1}{\xi})\right) - c_2\left(\frac{1}{\xi}-\log(\xi)\right)  +  \\
	& \quad \quad \quad \quad \quad \quad \quad \quad \quad \quad \quad +\frac{\lambda}{8}\left(\frac{2+\chi_s}{\xi}+\chi_s\log(1+\frac{1}{\xi})\right) \bigg) + O(\epsilon^2) \ .
	\end{aligned}
\end{equation}
with integration constants  $\delta \kappa$, $\delta \chi$, $c_1$, and $c_2$.  
The constant $\delta \kappa$ renormalizes the contribution from the identity block, while the difference between $\delta \chi$ and $\delta \kappa$ controls shifts of the boundary condition.

The task at hand is now to fix these integration constants by comparison with the boundary and bulk OPE's of the spinor correlation function.
These OPEs depend on a number of undetermined coefficients as well: the anomalous dimensions and OPE coefficients.  Nevertheless,
we will see that the comparison fixes all but a few of these unknown quantities.  The remaining quantities in the cases we study can then
be fixed by studying the corresponding CFT without boundary.  For example, here we are unable to determine $\lambda$ through the crossing constraint.
The value of $\lambda$ is the same whether or not there is a boundary however, and hence this value
 is easily looked up in the literature \cite{Fei:2016sgs, Zerf:2017zqi}.
 
We  start with the boundary channel and the function $F(\xi)$. The interaction gives an anomalous dimension to the fields present in the free theory.  Moreover the anomalous dimension of the bulk field  converts it essentially to a GFF at this order in the $\epsilon$-expansion, 
which means we expect to find at order $\epsilon$ a tower (\ref{FGFFbry})
of spinors on the boundary. One can explicitly verify the matching 
\begin{align}
\label{FYukawadecomp}
	F(\xi)= (1+\delta \kappa \, \epsilon)\, \widehat{f}_{\frac{p}{2}+\gamma_\rho}+
	\frac{\epsilon \lambda}{4} \left( \chi_s\, \widehat{f}_{\frac{3}{2}+1}+
	\sum_{k=1}^{\infty} \frac{\Gamma (k) \Gamma (2+k)}{ \Gamma (1+2k)} \, \widehat{f}_{\frac{3}{2}+k} \right) \ , 
\end{align}
which allows us to fix $c_1 = \frac{1}{2}-\gamma_\rho^{(1)}$. 
At a technical level, we are fixing the anomalous dimension for $\rho$ by matching the coefficient of a logarithm that appears in (\ref{FGsol}) with a logarithm
that appears in the boundary OPE decomposition. 
The matching requires expanding a hypergeometric function in its indices.  A simple approach which works here is to expand the hypergeometric first in a near boundary limit and then to expand the Taylor series coefficients as functions of the indices of the hypergeometric.  (Later for the bulk blocks, one can replace a 
near boundary limit with a coincident limit.)

We choose to write the decomposition (\ref{FYukawadecomp}) this way to bring some attention to the final sum. 
As in the GFF model, the single trace operators are of the form $\partial_n^k\rho$.
At level $k=1$, however, we also have the operator ${:}\phi \rho{:}$. The  primary field exchanged is made from a mixing of the two. This is how we understand this decomposition of the $k=1$ single-trace operator data into two contributions, one of which is coming from the $k=1$ term in the sum. 

To compare the tower with the GFF model (\ref{FGFFbry}) we considered above, take $\Delta = \frac{p}{2}+\gamma_\psi^{(1)} \epsilon$ in the decomposition. At leading order in epsilon we find
 \begin{align}
 	\lim_{\Delta \to  \frac{p}{2}+\gamma_\psi^{(1)} \epsilon}F_{GFF}(\xi)&= \widehat{f}_{\Delta_\psi}+2\gamma_\psi^{(1)}\epsilon \sum_{k=1}^{\infty} \frac{\Gamma (k) \Gamma (2+k)}{ \Gamma (1+2 k)}\times \widehat{f}_{\frac{p}{2}+k}+\mathcal{O}(\epsilon^2) \ , 
 \end{align}
from which it is clear that this tower of operators is that expected perturbatively from the GFF result. 

As we obtained on the boundary the infinite sum (\ref{FGFFbry}) of a GFF, we expect to find something like (\ref{FGFFbulk}) and (\ref{GGFFbulk}) in the bulk as well. For $F$, the GFF decomposition in the bulk is extremely simple, as there is only the identity, whose dimension is protected. We fix its normalisation to $1$, which in turn fixes $\delta \kappa = \gamma_{\rho}^{(1)}-\frac{\lambda\left(1+2 \chi _{\phi }\right)}{8}$. We are able to use a finite number of manually adjusted blocks. One has 
\begin{align}
	F(\xi)&=1\times f_{0}+\epsilon\bigg(\underbrace{-\frac{\lambda \chi_s}{4}f_{2}}_{:\phi^2:}+\underbrace{\frac{\lambda+8 \gamma_{\rho}^{(1)}}{16} f_{4}}_{:\phi\bar{\psi}\psi:} \bigg) \ . 
\end{align}
The matching of the logarithm from the ansatz and the block imposes the condition
\begin{align}
\label{gammapsi}
	\gamma_\psi^{(1)} = \frac{\lambda}{8} \ ,
\end{align}
our first anomalous dimension!  This result matches the literature \cite{Fei:2016sgs,Zerf:2017zqi}, but we have determined it in a new way.

 We can now turn to the decomposition on the boundary of the $G(\xi)$ function. The procedure is analogous to the previous one
\begin{align}
	G(\xi)&=\chi(1+\delta \chi \, \epsilon)\, \widehat{g}_{\frac{p}{2}+\gamma_\rho^{(1)}\epsilon}+\frac{\epsilon\chi  \lambda}{4} \left(
	\widehat{g}_{\frac{3}{2}+1}-\chi_{s} \sum_{k=1}^{\infty}(-1)^k\frac{\Gamma (k) \Gamma (2+k)}{ \Gamma (1+2 k)}\, \widehat{g}_{\frac{3}{2}+k} \right)
	 \ , 
\end{align}
which fixes an integration constant $c_2=\frac{1}{2}-\gamma_\rho^{(1)}$. Requiring that the boundary condition of the fermion is preserved as we crank up the interaction fixes $\delta \kappa = \delta \chi$ by comparing the coefficients multiplying the leading boundary spinor blocks in each structure. As previously, the coefficients of the tower are easily understood by comparing with the GFF limit
\begin{align}
	\lim_{\Delta_\psi \rightarrow \frac{p}{2}+\gamma_\psi^{(1)}\epsilon}G_{GFF}(\xi)=\chi \, \widehat{g}_{\Delta_\psi}+  \chi 2\gamma_\psi^{(1)} \epsilon  \sum_{k=1}^{\infty} (-1)^k\frac{\Gamma (k) \Gamma (2+k)}{\Gamma (1+2 k)}\,  \widehat{g}_{\frac{3}{2}+k} \ .
\end{align}

Finally, we can consider the bulk decomposition of $G(\xi)$. This last matching is more delicate than the others. 
Taylor expanding $G(\xi)$ around $\xi=0$, one encounters a term of order $\xi^{-1}$, which would correspond to a block
with $\Delta = 1$, the free scalar $\phi$.
However, such a block naively diverges, because it has a simple pole associated to the exchange of the null descendant $\Box\phi$ \cite{Penedones:2015aga}. More explicitly, by considering the form of the block and using a hypergeometric identity we find 
\begin{equation}
\label{fakehyperrel}
\begin{aligned}
	g_{\Delta}(\xi)&=\frac{(\Delta+1)^2}{4\left(\frac{p-3}{2}-\Delta \right)}\times \frac{g_{\Delta+2}(\xi)}{\left(\Delta -\frac{p-1}{2}\right)}
	\\
	&+\xi^{\frac{\Delta}{2}-\Delta_\psi}\, {}_2F_1 \left(\frac{\Delta+1}{2},\frac{\Delta+1}{2};1+\Delta-\frac{p-1}{2}; -\xi \right) \ .
\end{aligned}
\end{equation}
In the free-field limit, $\Delta = \frac{p-1}{2}$, the first term gives a singular contribution, while the second term is finite. From the equations of motions however, we expect $\Box \phi \sim \overline{\psi}\psi$; hence $\phi$ will gain an anomalous dimension, 
$\Delta_\phi=\frac{p-1}{2}+\gamma_\phi^{(1)}\epsilon$,  and $\bar{\psi}\psi$ becomes a descendant \cite{Fei:2016sgs}. 
The wrong thing to do would be to assume $\overline{\psi} \psi$ remains a primary of dimension nearly three.
This situation is often called the ``fake primary effect'',   
where a primary field at threshold signals itself through a block which corresponds to the exchange of its null-descendant. This effect was noticed and explained in applications of the numerical bootstrap \cite{Karateev:2019pvw}.

The diverging first term of (\ref{fakehyperrel}), in the limit $\epsilon \rightarrow 0$, 
reproduces the contribution from a primary with dimension $3$ instead of $1$. 
The logarithmic contributions of (\ref{FGsol}) can now be matched to the bulk OPE expansion, and
one finds
\begin{align}
\label{Gxibulk}
	G(\xi)&= \chi \left( -\gamma_\phi^{(1)} \epsilon + \nu_\phi \epsilon^2 \right) \,  g_{\Delta_\phi}
	-  \frac{\epsilon\lambda\chi\chi_s}{8} \sum_{k=1}^{\infty}(-1)^k\frac{\Gamma (k) \Gamma (k+2)}{\Gamma (1+2 k)}\, g_{3+2k} 
\end{align}
with the identifications
\begin{align}
\label{gammarhogammaphi}
	\gamma_\rho^{(1)} &=-\frac{1}{2}+\frac{\lambda(4-\chi _{\phi})}{8} \ , \;\;\;  \gamma_\phi^{(1)} = \frac{1}{2} -\frac{3}{4}\lambda \ , \\
	\nu_\phi &= -\gamma_\phi^{(2)} + \frac{1}{2} - \frac{1}{32} (3 \lambda-2) (3 \lambda(1+\chi_s) -4)  \ .
\end{align}
We have computed two more anomalous dimensions as a function of the coupling $\lambda$!
The result for $\gamma_\phi^{(1)}$ matches the literature \cite{Fei:2016sgs,Zerf:2017zqi}, while the result for $\gamma_\rho^{(1)}$ matches
the more recent \cite{Giombi:2021cnr}, but our methods are different. In their work, the bulk anomalous coefficients are used to extrapolate the boundary data, while we effectively derive both at the same time. 

Note that the tower of double trace operators in (\ref{Gxibulk}) furthermore has a structure that matches (\ref{GGFFbulk}) in an $\epsilon$-expansion
with $\Delta_\psi = \frac{p}{2} + \epsilon \gamma_\psi^{(1)}$.  

\subsection{Coupling to a scalar undergoing an extraordinary phase transition}

We have studied the coupling of fermions to a scalar field. Notoriously, scalar fields in the presence of a boundary admits multiple interesting phases, one of which is the extraordinary transition. In this configuration, the scalar field develops an expectation value  \cite{Diehl:1996kd}, contrasting starkly with the Dirichlet and Neumann boundary conditions that we just considered. 
Let us investigate the effect of the extraordinary transition for $\phi$ in the Yukawa model. 
Treating the expectation value of the scalar as a background field for the fermion, 
the spinor equation of motion is effectively modified to 
\begin{equation}
	\left( \slashed{\partial}\psi = -g \phi  \psi \right) \rightarrow  \left( \slashed{\partial} \psi =- \frac{\mu}{x\cdot n} \psi  \right) \ ,
\end{equation}
where $\mu = \frac{g a_{\phi}}{2}$ is related to the expectation value of the scalar $\phi$. We obtain an effective position dependent
 mass term for the fermion.  If we were to try to write an action for such a fermion with a position dependent mass (in absence of the scalar),
 we would find that the fermion is sensitive to the breaking of the bulk conformal invariance everywhere and not just through its boundary 
 condition at $z=0$.  By an abuse of language, we might call this sensitivity spontaneous breaking of local bulk conformal invariance.
 We call this breaking spontaneous because the full theory, including the scalar, remains sensitive to the breaking of bulk conformal invariance
 only through the boundary condition.  We call the invariance local to emphasize that the boundary condition anyway breaks it.
 In the AdS picture, Weyl rescaling changes the position dependent mass to a constant mass, 
and the situation is reminiscent of the Higgs mechanism  \cite{Carmi:2018qzm,Herzog:2019bom}.
Indeed,
generically for QFT in AdS, there is no bulk conformal invariance.

Indeed, bulk conformal invariance is an extra assumption. The natural form of conformal invariance in the presence of a boundary is $SO(1,p+1)$ invariance. That the bulk itself is invariant under the full $SO(1,d+1)$ far away from the boundary is a supplementary assumption. This is obvious in the AdS picture, where a generic QFT in AdS generates correlation functions satisfying boundary conformal invariance, although the bulk itself will not be a CFT. 

Given this modified equation of motion, we would like to determine the boundary OPE of the spinor field $\psi$ through analyzing
its correlation function with boundary operators
\begin{align}
	\left(\gamma^\nu\pdv{}{x^{\nu}}+\frac{\mu}{z}\right)\expval{\psi(x,z)\overline{\rho}_{\alpha}(y)}=0 \Rightarrow \begin{cases}
		\Delta_\psi &= \frac{p}{2} \\
		\Delta_{\rho,\alpha}&=\frac{p}{2}-\alpha \mu 
	\end{cases} \ ,
\end{align} 
The bulk field has a free-field dimension, as needed for its descendant to be a (null) primary field. The boundary spinors $\rho_{\pm}$ now have different dimensions, but they are still shadow dual in the sense that their dimensions sum to $p$. (The scalar field case is very similar \cite{Herzog:2019bom,Behan:2020vy}.) Noting that there is a unitarity bound at $\widehat{\Delta}\geq \frac{p-1}{2}$, we find the usual bound of $\abs{\mu}\leq\frac{1}{2}$, where alternate quantisation of the spinor in AdS is allowed. For $\abs{\mu}>\frac{1}{2}$, there will be a single quantisation scheme. One can then use the block decomposition to find the two point function this field content induces
\begin{align}
	\expval{\psi(x) \overline \psi(x')}=\nu_{+}\widehat{\mathcal{W}}_{\frac{p}{2}-\mu,+}+\nu_{-}\widehat{\mathcal{W}}_{\frac{p}{2}+\mu,-} \ .
\end{align}
Up to a power law prefactor, the same expression appears in computing the propagator of a free massive spinor field in AdS
(see appendix \ref{app:EAdS}). If we further impose boundary conditions on the spinor field, we can restrict to only one helicity. 
Normalizing the coincident limit, one fixes 
\begin{align}\label{eq:extra}
	\expval{\psi(x)\overline \psi(y)}=\frac{2\Gamma (\frac{p+1}{2}-\mu \alpha) \Gamma (1-\mu  \alpha )}{\Gamma(\frac{p+1}{2})\Gamma (1-2 \mu  \alpha)}\widehat{\mathcal{W}}_{\frac{p}{2}-\alpha\mu,\alpha} \ .
\end{align}
Unitarity places constraints on when this boundary conformal partial wave can be expressed as a sum over bulk conformal partial waves. 
If we look at the boundary $\hat g_{\frac{p}{2}-\alpha \mu}(\xi)$ conformal block in the coincident limit, we find a sum of two power series in $\xi$, 
the first with leading term $\xi^{\frac{1}{2}-\frac{p}{2}}$ and the second with constant leading term.  The more singular power series with
the $\xi^{\frac{1}{2}-\frac{p}{2}}$ behaviour is absent precisely
in the case $\mu=0$.  Furthermore, the more singular power series 
requires the existence of a bulk operator of dimension one in order to be expressed as a sum over bulk conformal blocks.  However, this dimension
is only consistent with the bulk unitarity bound when $d \leq 4$.  In fact, in the case $d=4$, this bulk scalar must be free and thus it cannot appear
in the bulk OPE of the two spinor operators.  By this argument, crossing symmetry for such a spinor with a position dependent mass term 
can only be consistent with unitarity in the case $d<4$.  (A similar argument looking at $\hat f_{\frac{p}{2}-\alpha \mu}(\xi)$ gives a weaker bound, that
$d<5$.)

\subsection{The Gross-Neveu Model at 2-loops}

We now focus the Gross-Neveu Model in the $\epsilon$ expansion $d=2+\epsilon$, up to order $\epsilon^2$. We will consider a model of $N$ fermions, with flavour index $a,b$.

The starting point of our analysis is the equation of motion
\begin{align}
	\slashed{\partial}\psi_a(x)= g\left(\sum_b \bar{\psi}_b\psi_b \right)\psi_{a}(x) \ .
\end{align}
At the fixed point, we anticipate $g = \frac{\lambda \epsilon}{\kappa_f}+\mathcal{O}(\epsilon^2)$. 
Acting on the two point function at leading order, we obtain the system of equations 
\begin{align}
	\slashed{\partial}\expval{\psi_a(x)\overline{\psi}_{b}(x')}&=-\epsilon\left(2N-1\right)\frac{\lambda\chi}{2x\cdot n}\expval{\psi_{a}(x)\overline{\psi}_{b}(x')} \ .
\end{align}
 In terms of the functions $F(\xi)$ and $G(\xi)$, the equation of motion translates into the differential equations
\begin{align}
	2 \xi F'(\xi) + \left(1+p \right)F(\xi) +\epsilon \lambda\chi\left(2N-1\right) G(\xi)&=0 \ , \\
	2 (1+\xi) G'(\xi) + \left(1+p\right)G(\xi) + \epsilon \lambda\chi\left(2N-1\right) F(\xi)&=0 \ .
\end{align}
In order to be able to write the two-point function purely in terms of $F(\xi)$ and $G(\xi)$ (and hence preserve conformal invariance), 
we find at this stage that we must set $\Delta_\psi = \frac{p}{2} + O(\epsilon^2)$.  
We solve this system in an $\epsilon$-expansion, and find
\begin{align}
	F(\xi) &= \frac{1}{\xi}+\frac{\epsilon}{\xi}\left( \delta \kappa+ \frac{\lambda(1-2n)}{2}\log (1+\xi)-\frac{1}{2}\log (\xi )\right) \ , \\
	G(\xi) &= \frac{\chi}{(1+\xi)}+\frac{\epsilon}{(1+\xi)}\left(\delta \chi+\chi\left(\frac{\lambda(1-2n)}{2} \log (\xi)-\frac{1}{2}\log (1+\xi)\right)\right) \ .
\end{align}
There are two integration constants, $\delta \kappa$ associated with the normalization of the coincident limit, and $\delta \chi$ associated with the choice
of boundary condition for the spinor.

These functions admit an expansion in terms of bulk and boundary blocks. In fact, one only needs a finite number of  blocks in each channel to reproduce $F(\xi)$ and $G(\xi)$ at $O(\epsilon)$
\begin{align}
	\begin{pmatrix}
		F(\xi) \\
		G(\xi)
	\end{pmatrix}=\begin{pmatrix}
		(1+\delta \kappa \, \epsilon) f_{0}+ \epsilon \lambda \frac{(1-2 n)}{2} f_{2} \\
		(\chi + \delta \chi \epsilon) g_{p+\gamma_\sigma^{(1)}\epsilon} 
	\end{pmatrix} =\begin{pmatrix}
		(1+\delta \kappa \,\epsilon)\widehat{f}_{\frac{p}{2}+\gamma_\rho^{(1)}\epsilon} \\
		(\chi + \delta \chi \epsilon)\widehat{g}_{\frac{p}{2}+\gamma_\rho^{(1)}\epsilon}
	\end{pmatrix} \ .
\end{align}
We denote by $\sigma \equiv \overline \psi \psi$ the mass operator in this theory, which has dimension nearly one.
The bulk field of dimension two is then naturally interpreted as $\sigma^2 \equiv (\overline \psi \psi)^2$.  
The matching works provided one takes
\begin{align}
	\chi^2 & =1 \ , & \gamma_\psi^{(1)} &= 0 \ ,   &  \gamma_\rho^{(1)} &= \frac{(2n-1)\lambda}{2} \ , & \gamma_\sigma^{(1)} &= -(2n-1)\lambda  \ .
\end{align}
We fix $\delta \kappa=0$ by normalising the identity block, and we absorb $\delta \chi$ into $\chi$. Crossing symmetry imposes at leading order $\chi=\pm 1$, and we will now anticipate and require that this condition is preserved by the interaction.  

To push to the next order in $\epsilon$, we use the equation of motion twice to constrain 
$\langle \slashed{\partial} \psi(x) \overline \psi(x')\overset{\leftarrow}{\slashed{\partial}} \rangle$, similar to what we did for the Yukawa model above.
The equation of motion produces a factor of $g^2 \sim \epsilon^2$, allowing us to evaluate the right hand side using free fields and Wick contractions.
 
One can then use as an input our formulas for $F$ and $G$ to order $\epsilon$ in order 
to find an equation for the $\epsilon^2$ components. Proceeding as described, we find 
\begin{equation}
\begin{aligned}
	\slashed{\partial}_x\expval{\psi_a(x)\overline{\psi}_b(y)}\overset{\leftarrow}{\slashed{\partial}}_{y}=& \ -g^2\sum_{m,n}\expval{ \bar{\psi}_{m}(x)\psi_{m}(x)\psi_a(x)\overline{\psi}_b(y)\bar{\psi}_{n}(y)\psi_{n}(y)} \ .
\end{aligned}
\end{equation}
Note the equations of motion for $\overline \psi_a$ and $\psi_a$ have opposite sign.
To evaluate these products of two point functions, it is convenient to specialise to points $x^{\mu}=zn^{\mu}$, $y^{\mu}=w n^{\mu}$. When evaluating objects such as $\expval{\psi(x)\overline{\psi}(x)}$, one must consider the regular part of the expression, as usual in a normal-ordered correlator. The final expression simplifies greatly and can be encapsulated as in the Yukawa case by a set of eigenvalue equations
\begin{align}
 	\mathcal{D}_{F}&=-\epsilon^2\lambda^2 H(\xi)F(\xi)  \ , \\
	\mathcal{D}_{G}&=-\epsilon^2\lambda^2 H(\xi)G(\xi) \ , 
\end{align}
with a function $H(\xi)$ encoding the Wick contractions given by 
\begin{align}
H(\xi)=\left(2N-1\right) \left(2N-1+ \frac{1}{\xi}-\frac{1}{(1+\xi)} \right) \ ,
\end{align}
where we have used that $\chi^2 = 1$.
The two point functions at order $\epsilon^2$ can then be solved for 
\begin{align}
	F(\xi)= \ldots &+ \frac{\epsilon^2}{\xi}\bigg(\frac{\log(\xi)^2}{8}+c_1 \log(1+\xi)+\lambda^2\frac{(2N-1)}{2}\big(\text{coth}^{-1}\left(1+2\xi\right)+(N-1)\log(\xi)\log(1+\xi)\big) \nonumber \\
	&+ \frac{\lambda(\lambda+1)(2N-1)}{8}\log(1+\xi)^2-  \lambda (1-2 (N-1)\lambda) \frac{(2 N-1)}{4} \text{Li}_2(-\xi )\bigg)  \ , \\
	G(\xi)=\ldots &+\frac{\epsilon^2\chi}{(1+\xi)}\bigg(\frac{\log(1+\xi)^2}{8}+c_2\log(\xi)+ \lambda \frac{2N-1}{4}\left(\log(\xi)\log(1+\xi)-2\lambda \text{coth}^{-1} \left(1+2\xi\right)\right) \nonumber \\
	&+\frac{\lambda(\lambda+1)(2N-1)}{8}\log(\xi)^2+\lambda(1-2 (N-1)\lambda)\frac{(2 N-1)}{4} \text{Li}_2(-\xi )\bigg) \ .
\end{align}
The variables $c_1$ and $c_2$ are new integration constants. 
We have fixed two integration constants -- the $O(\epsilon^2)$ analogs of $\delta \kappa$ and $\delta \chi$ -- from the field normalisation as well as the definition of $\chi$.  

One can then expand the relevant pieces in blocks. 
The bulk expansion of the $F(\xi)$ function is 
\begin{align}
	F(\xi) =&  f_{0}+\left(-\frac{\lambda(2N-1)}{2}\epsilon+\left(c_1-\lambda(\lambda(2N-3)-1)\frac{(2N-1)}{4}\right)\epsilon^2\right)f_{\Delta_{\sigma^2}} \nonumber \\
	&+\epsilon^2\frac{\lambda^2(2N-1)}{4}\sum_{k=1}^{\infty} \frac{(2N-1+(-1)^{k+1})\sqrt{\pi}\Gamma(k)}{4^k(1+k)\Gamma(\frac{1}{2}+k)}f_{2+2k} \ .
\end{align}
From the matching of the logarithms, one fixes the anomalous dimension  of the leading operators  $\psi$ and $\sigma^2$
\begin{align}
	\Delta_\psi &= \frac{1+\epsilon}{2}+\lambda^2\frac{(2N-1)}{4}\epsilon^2 +\ldots  \ , \\
	\Delta_{\sigma^2} &= 2+\epsilon - 2\lambda(N-1)\epsilon +\ldots  \ .
\end{align}
These agree with results computed for the theory without boundary \cite{ZinnJustin,Fei:2016sgs}.  

The boundary decomposition for $F(\xi)$ in contrast is given by
\begin{align}
\label{FbryGN}
	F(\xi)=& \left(1+\lambda\pi^2(1-2(N-1)\lambda)\frac{(2N-1)}{24}\epsilon^2\right)\widehat{f}_{\Delta_\rho}\nonumber \\
	&+\epsilon^2\frac{\lambda^2(2N-1)}{8}\sum_{k=1}^{\infty}\frac{(k-(-1)^{k}+1)\sqrt{\pi}\Gamma(1+k)}{4^k(k+1)\Gamma(\frac{3}{2}+k)}\widehat{f}_{\frac{3}{2}+k} \ .
\end{align}
Matching logarithms constrains anomalous dimensions.  Expanding the dimension of the leading boundary spinor as
\begin{equation*}
	\Delta_{\rho} = \frac{p}{2}+ \gamma_{\rho}^{(1)}\epsilon+\gamma_{\rho}^{(2)}\epsilon^2 \ ,
\end{equation*}
we find  $c_1 = -\gamma_{\rho}^{(2)}$.

We can now turn to the second function $G(\xi)$ and its bulk expansion. From the matching of the logarithms, one encounters a quadratic constraint 
which fixes the value of $\lambda$ to that needed for the vanishing of the $\beta$-function, $\lambda = \frac{1}{2(N-1)}$ or 
$\lambda =0$.  Using the non-trivial choice, one finds
\begin{align}
	G(\xi)&= \chi\times g_{\Delta_{\sigma}}+\epsilon^2\frac{\lambda^2(2N-1)}{2}\sum_{k=1}^{\infty}\frac{(2N-1+(-1)^{k}k)\sqrt{\pi}\Gamma(k)}{4^k k\Gamma(\frac{1}{2}+k)}g_{1+2k} \ .
\end{align}
This matching fixes the integration constant
\begin{align}
c_2 &=\frac{\gamma_{\sigma}^{(2)}}{2}-\frac{(2 N-1) }{8 (N-1)^2}  \ .
\end{align}
We have further expanded the scaling dimension of the mass operator as
\begin{align}
	\Delta_{\sigma}&=p+\gamma_\sigma\epsilon+\gamma_{\sigma}^{(2)}\epsilon^2+\ldots \nonumber \\
	&= p-(2N-1)\lambda\epsilon+ \gamma_{\sigma}^{(2)}\epsilon^2 +\ldots  \ . 
\end{align}

Finally, one can expand $G(\xi)$  in the basis of boundary blocks, yielding the decomposition
\begin{align}
\label{GbryGN}
	G(\xi)&= \chi\left(1- \epsilon^2\lambda\pi^2(1-2 (N-1) \lambda)\frac{(2 N-1)}{24}\right)\widehat{g}_{\Delta_{\rho}}
	\nonumber  \\
	&-\epsilon^2 \frac{\lambda^2(2N-1)}{8}\sum_{k=1}^{\infty}\frac{(-1)^{k}(k-(-1)^{k}+1)\sqrt{\pi}\Gamma(1+k)}{4^k(k+1)\Gamma(\frac{3}{2}+k)}\widehat{g}_{\frac{3}{2}+k} \ , 
\end{align}
which further requires $c_1 = -\gamma_\rho^{(2)}$. 
Comparing the coefficients of the boundary blocks $\widehat f_{\Delta_\rho}$ and $\widehat g_{\Delta_\rho}$ in (\ref{FbryGN}) and (\ref{GbryGN}), 
we see that the boundary condition is preserved at the fixed point

At this order in the $\epsilon$ expansion, we lack a constraint to fully fix our system based on crossing symmetry alone. 
From the $\expval{\sigma\sigma}$ correlator at $O(\epsilon^2)$ for the theory without boundary, 
one can derive or look up $\gamma_{\sigma}^{(2)}$  \cite{Fei:2016sgs},
\begin{align}
	\gamma_{\sigma}^{(2)}&=-\frac{(2N-1)}{2}\lambda^2 = -\frac{(2N-1)}{8(N-1)^2} \ ,
\end{align}
which further fixes
\begin{align}
	\gamma_{\rho}^{(2)}&=3\frac{(2N-1)}{4}\lambda^2 \ .
\end{align}
Thus the simple input of $\gamma_{\sigma}^{(2)}$ into the constraint of crossing symmetry fully specifies the 2-loop correction to this system.  

\subsection{The 3D Yukawa Model}

It is natural to look for a model of interacting fermions in $d=3-\epsilon$ dimensions. From power counting, one can construct a classically marginal interaction term of the form $\phi^2 \overline{\psi}\psi$. This model is quite similar to the Yukawa model studied before. One has an equation of motion 
\begin{align}
	\slashed{\partial}\psi = -\frac{g}{2} \phi^2 \psi \ ,
\end{align}
with $\kappa_s g \equiv 2 \lambda \sqrt{\epsilon}$ at the fixed point, for some order one $\lambda$ determined by the zero of the beta function. 
The main difference is that the CFT data admits an expansion in powers of $\sqrt{\epsilon}$ rather than $\epsilon$. 
From a perturbative perspective, the change comes about because the operator ${:} \phi(x)^2 {:}$ has a nonzero expectation value in
the theory with a boundary even when $g=0$; $\langle  \phi^2  \rangle = N_s \chi_s / (2 x \cdot n)$ where $\chi_s = \pm 1$
selects the boundary conditions, Dirichlet or Neumann, for the scalar field, and the $N_s$ is the number of scalar fields. This makes the leading order result analogous to the conformally massive fermion previously discussed.\footnote{%
 This ``conformal mass'' argument can also be invoked to show that a purely $\phi^6$ bosonic model will involve an expansion in $\sqrt{\epsilon}$:
 the equation of motion for the scalar will take the schematic form $\Box \phi \sim \langle \phi^4 \rangle \phi$.  Interestingly,
 refs.\ \cite{diehi1987walks,eisenriegler1988surface} derive this $\sqrt{\epsilon}$ behavior in a different way, through mixing in the RG flow
 of a bulk $\phi^6$ coupling and an induced boundary $\phi^4$ coupling.
} 

To analyse the system, we proceed as in the Gross-Neveu model, and first act once with the equation of motion for $\psi$, 
\begin{align}
	\slashed{\partial}\expval{\psi_a(x)\overline{\psi}_{b}(x')}
	&=-\lambda \sqrt{\epsilon}\frac{N_s \chi_s}{2x\cdot n}\expval{\psi_{a}(x)\overline{\psi}_{b}(x')} \ .
\end{align}
The equation of motion gives an expansion for the two point function
\begin{align}
	F(\xi) &=  \frac{1}{\xi ^{3/2}}+ \frac{\lambda\sqrt{\epsilon}}{\xi ^{3/2}} N_s \chi \chi _{\phi } \left(\sqrt{\frac{\xi }{1+\xi}}-\text{sinh}^{-1}\sqrt{\xi }\right) \ ,\\
	G(\xi) &= \frac{\chi }{(1+\xi)^{3/2}} + \frac{\lambda\sqrt{\epsilon}}{(1+\xi )^{3/2}} N_s \chi _{\phi } \left(\sqrt{\frac{1+\xi }{\xi}}-\text{sinh}^{-1}\sqrt{\xi }\right) \ .
\end{align}
We absorbed two integrations constants into the normalisation of the coincident limit and the definition of $\chi$.  
The matching to the blocks imposes $\chi^2 = 1 + \mathcal{O}(\epsilon)$. In fact, we will from now require that this is preserved also at higher order. These functions can be written as
\begin{align}
	\begin{pmatrix}
		F(\xi) \\
		G(\xi)
	\end{pmatrix}=\begin{pmatrix}
	 f_{0}-  \lambda\sqrt{\epsilon} \frac{N_s \chi \chi_s}{3} \, f_{3} \\
		\chi \, g_{p} + \lambda\sqrt{\epsilon } N_s  \chi_s  \, g_{1}
	\end{pmatrix} =\begin{pmatrix}
		(1+N_s \chi \chi _{\phi}  (1-\log (2)) \lambda \sqrt{\epsilon}) \, \widehat{f}_{\frac{p}{2}+\gamma_\rho\sqrt{\epsilon}} \\
		\chi(1+N_s \chi \chi _{\phi}  (1-\log (2)) \lambda \sqrt{\epsilon})\,  \widehat{g}_{\frac{p}{2}+\gamma_\rho \sqrt{\epsilon}}
	\end{pmatrix} \ .
\end{align}
This matching is contingent on the modification at $O(\sqrt{\epsilon})$
of the dimension of the boundary field
\begin{align}
\Delta_\rho &= \frac{p}{2}+ \frac{\lambda  N_s  \chi\chi_{\phi}}{2}   \sqrt{\epsilon}+O(\epsilon) \ .
\end{align}
The deformation of the conformal block decomposition at $O(\sqrt{\epsilon})$ is particularly simple. That is because the anomalous dimensions of the bulk fields
are untouched at this order in the expansion, while the boundary spectrum is highly constrained by the conformal mass structure.

To go to the next order, we mimic the Gross-Neveu discussion above and act with the equations of motions on both $\psi$ and $\overline \psi$ in the 
spinor two-point function. We find the same differential operators studied in the Yukawa model. The source for the system of differential equations is given by the Wick contraction of four scalar insertions. 
We find for $F(\xi)$ and $G(\xi)$ at $O(\epsilon)$ that  
\begin{align}
	F(\xi) &= \ldots + \frac{\epsilon }{\xi ^{3/2}}\bigg( \frac{\log (\xi )}{2} + 2c_1\left(\text{sinh}^{-1}\left(\sqrt{\xi }\right) -  \sqrt{\frac{\xi }{1+\xi}}\right)-2 \lambda^2 N_s  \chi _{\phi } \sqrt{\frac{\xi}{1+\xi}}\\
	& -\lambda^2 N_s \left( \frac{\log (\xi)}{3}+\log(1+\xi)\right)+\frac{\lambda^2  N_s^2}{2} \left(\text{sinh}^{-1}\left(\sqrt{\xi }\right)^2-2 \sqrt{\frac{\xi }{\xi +1}} \text{sinh}^{-1}\left(\sqrt{\xi }\right)\right) \bigg) \ , \nonumber \\
	G(\xi) &= \ldots +\frac{\epsilon }{(1+\xi) ^{3/2}}\bigg(
	\delta \chi^{(2)} + 
	\chi\frac{\log(1+\xi)}{2} +2c_2\left(\text{sinh}^{-1}\left(\sqrt{\xi }\right) -  \sqrt{\frac{1+\xi }{\xi}}\right) \\
	&+\lambda^2 N_s \chi\left(\frac{\log(1+\xi)}{3}+\log(\xi) \right) +\frac{\lambda^2 \chi N_s^2}{2} \left(\text{sinh}^{-1}\left(\sqrt{\xi }\right)^2-2 \sqrt{\frac{1+\xi }{\xi}} \text{sinh}^{-1}\left(\sqrt{\xi }\right)\right) \nonumber \\
	&+\lambda^2 \chi N_s^2+ 2 \lambda^2 N_s \chi \chi_\psi \sqrt{\frac{1+\xi}{\xi}} \bigg)  \ .  \nonumber
\end{align}

We expand these functions using the boundary and bulk conformal blocks.  For $F(\xi)$, we find in the bulk 
\begin{align}
	F(\xi)&=  f_{0}+ \left(-\frac{\lambda \sqrt{\epsilon}N_s \chi \chi_s}{3}+\epsilon\frac{2 c_1}{3} \right)\,  f_{3}- 2N_s\chi_s \epsilon \lambda^2 \,  f_{1}  \nonumber \\ 
	&+\epsilon\frac{\lambda^2 N_s}{2} \sum_{k=0}^{\infty} \frac{(-1)^k \Gamma(k+\frac{1}{2})\Gamma(k-\frac{1}{2})+2^{-2k+1} k N_s \pi \Gamma(2k-1)}{(1+k)\sqrt{\pi}\Gamma(\frac{1}{2}+2k)} \,  f_{2+2k} \ .
\end{align}
Matching the logarithms, one finds further the bulk anomalous dimension of the fermion
\begin{align}
\label{Deltapsid3}
	\Delta_\psi &= \frac{p}{2}+\frac{\lambda^2 N_s }{3}\epsilon  + O(\epsilon^2) \ .
\end{align}
The boundary expansion of $F(\xi)$ is instead
\begin{align}
	F(\xi)&= (1+ N_s \chi \chi_s (1-\log(2))\lambda\sqrt{\epsilon}+\delta \nu_F \lambda^2 \epsilon)\, \widehat{f}_{\Delta_\rho} \\ 
	&+ \frac{2N_s \lambda^2}{3}\epsilon\sum_{k>0}\frac{(4k^2+6k \chi_s-3(-1)^k-1)}{4^k (2k-1)k}\, \widehat{f}_{1+k} \ . \nonumber
\end{align}
Matching the logarithms partially constrains the anomalous dimension of $\rho$, 
\begin{equation}
\label{c1gammarhorel}
c_1=\frac{4N_s\lambda^2}{3}-\gamma_{\rho}^{(2)} \ , 
\end{equation}
where we write  $\Delta_\rho = \frac{p}{2}+\gamma_\rho^{(1)} \sqrt{\epsilon}+\gamma_{\rho}^{(2)}\epsilon + \ldots$.
The parameter $\delta \nu_F$ is complicated 
\begin{align}
\delta \nu_F \lambda^2 &= 2 c_1 ( \log 2 - 1) + \frac{N_s \lambda^2}{2} (-4 \chi_s + N_s (\log 2 - 2) \log 2)  \ .
\end{align}

We turn now to $G(\xi)$. The bulk expansion is 
\begin{align}
	G(\xi)&= (\chi + \epsilon \delta \chi^{(2)}) g_{\Delta_\sigma}+ (N_s \chi_s \lambda\sqrt{\epsilon}+2(N_s \chi\chi_s\lambda^2-c_2) \epsilon)\, g_{1}  \\
	& +\chi \frac{\lambda^2 N_s}{2}\epsilon\sum_{k=1}^{\infty} \frac{4^{-k} N_s\pi \Gamma(2k)+\frac{1}{3} (-1)^{k+1}\Gamma(k+\frac{1}{2})\Gamma(k+\frac{3}{2}) }{k \sqrt{\pi}\Gamma(2k+\frac{1}{2})}\, g_{2k+2}  \nonumber \\
	& +\chi \chi_s  \lambda^2 N_s \epsilon  \sum_{k=1}^\infty \frac{\sqrt{\pi} \Gamma\left(k - \frac{1}{2} \right)^2}{\Gamma \left( 2k- \frac{1}{2} \right)} g_{2k+1} \ . \nonumber
\end{align}
Matching the logarithms, the dimension of the bulk mass term is 
\begin{align}
\label{Deltasigmad3}
	\Delta_\sigma =p +\frac{8 N_s}{3}\lambda^2 \epsilon + O(\epsilon^2) \ .
\end{align}

Finally, one can expand $G(\xi)$ in boundary blocks, giving
\begin{align}
	G(\xi) &= \chi(1+N_s\chi \chi_s(1-\log(2))\lambda\sqrt{\epsilon}+\delta \nu_G \lambda^2 \epsilon)\widehat{g}_{\Delta_\rho}\\
	&+ \chi\frac{2N_s \lambda^2 \epsilon}{3}\sum_{k>0} \frac{3+(-1)^k(1-4k^2-6k \chi_s)}{4^k(2k-1)k}\widehat{g}_{1+k}  \ .\nonumber
\end{align}
One finds an additional constraint on the anomalous dimension of the boundary spinor, $c_2 = -\frac{4}{3} N_s \lambda^2  - \gamma_\rho^{(2)}$, 
which allows us to relate $c_1$ and $c_2$, using a similar constraint (\ref{c1gammarhorel}) 
from the boundary expansion of $F(\xi)$.  The $O(\epsilon)$ contribution
to the boundary $\rho$ block takes the form
\begin{align}
\chi \delta \nu_G \lambda^2 &= 2 c_2 (\log 2 - 1) +  \frac{N_s \lambda^2}{2 \chi} (4 \chi_s + N_s ( (\log 2 - 2) \log 2 + 2) + \delta \chi^{(2)} \ . \nonumber
\end{align}
Insisting on the boundary conditions $\delta \nu_F = \delta \nu_G$ then fixes $\delta \chi^{(2)}$ in terms of $\lambda$.

Unfortunately, there is a degeneracy in the system of equations which does not allow us to specify both $c_1$ and $c_2$, and so we cannot determine
$\gamma_\rho^{(2)}$ either. This situation is unsatisfying but can be remediated. Crossing symmetry of $\expval{\psi(x) \bar{\psi}(x')}$ is not enough to fix $\gamma_\rho^{(2)}$, but we can look at $\expval{\phi(x)\phi(x')}$ as well. To see how this would help, consider the bulk expansion we found for the $G$ function 
\begin{align}
	 G(\xi) &= \ldots + \bigg(\left[\lambda a_{\phi^2}^{(2)}\right]_{(1)}\sqrt{\epsilon}+\left[\lambda a_{\phi^2}^{(2)}\right]_{(2)}\epsilon\bigg) g_{1} + \ldots \, .
\end{align}
The coefficient $\left[\lambda a_{\phi^2}^{(2)}\right]_{(2)}$ contains $c_2$, or equivalently $\gamma_\rho$, and is the only unknown left. This term contains a contribution from the expectation value $a_{\phi^2}$ at order $\sqrt{\epsilon}$, and from the coupling $\lambda_{\phi^2}^{(2)}$ at order $\epsilon$. Notice that the latter can be computed using perturbation theory in the theory without a boundary. In such a setup, each interaction vertex $g$ contributes one fermion propagator, while $\lambda_{\phi^2}^{(1)}$ and $\lambda_{\phi^2}^{(2)}$ multiply respectively structures with odd and even numbers of gamma matrices. From this diagrammatic argument, we gather that $\lambda_{\phi^{2}}^{(2)}$ receives contributions only for odd powers of the coupling, $g^{2n+1}\sim \epsilon^{n}\sqrt{\epsilon}$; hence the ratio of the two coefficients multiplying the block measures precisely the correction to $a_{\phi^2}$ at order $g$, 
\begin{align}
	\frac{\left[\lambda a_{\phi^2}^{(2)}\right]_{(2)}}{\left[\lambda a_{\phi^2}^{(2)}\right]_{(1)}}&=\frac{[a_{\phi^2}]_{(1)}}{[a_{\phi^2}]_{(0)}} \ .
\end{align}
Crucially, this ratio can be determined from an analysis of $\expval{\phi(x)\phi(x')}$ at order $\sqrt{\epsilon}$. The equation of motions for the scalar is 
\begin{align}
	\Box \phi_a = g \overline{\psi}_b\psi_b \phi_a +\frac{h}{5!}(\phi^{2})^2 \phi_a \ .
\end{align}
At the fixed point, at order $\sqrt{\epsilon}$, we can neglect the second term. Applying this equation once to the scalar correlator allows us to solve for $H(\xi)$, 

\begin{align}\label{eq:corrH1}
	H(\xi)&= \frac{1+\delta\kappa \sqrt{\epsilon}}{\sqrt{\xi}}+\frac{\chi_s+c_3\sqrt{\epsilon}}{\sqrt{1+\xi}}+2\lambda\sqrt{\epsilon}N_f \chi \left(\frac{\chi_s \sinh[-1](\sqrt{\xi})}{\sqrt{\xi}}+\frac{\coth[-1](\sqrt{\frac{1+\xi}{\xi}})-\chi_s}{\sqrt{1+\xi}}\right)\, . 
\end{align}
We still need to fix the two unknowns $c_3$ and $\delta \kappa$. From the normalisation of the coincident limit, we fix $\delta\kappa=0$. To fix $c_3$, we  pick a boundary condition through some $\chi_s=\pm$, and adjust $c_3$ to maintain compatibility of $H(\xi)$ with either Dirichlet or Neumann boundary conditions. This gives $c_3 = -2 \lambda N_f \chi \chi_s $. With $\expval{\phi(x)\phi(x')}$ fully fixed, one can look at its bulk decomposition. The bulk conformal block decomposition of the correlator has the generic form
\begin{align}\label{eq:corrH2}
	H(\xi) &= \sum_{\Delta}\lambda a_{\Delta} h_{\Delta}  \, , \\
	&=\sum_{\Delta}\lambda a_{\Delta}\xi^{\frac{\Delta-2\Delta_\phi}{2}}\, _2F_1\left(\frac{\Delta}{2},\frac{\Delta}{2};\Delta-\frac{p-1}{2};-\xi \right) \, ,
	\nonumber
\end{align}
where we have used the explicit expressions for the bulk conformal blocks \cite{McAvity:1995tm}. The equations of motion are compatible with conformal invariance only for $\Delta_\phi=\frac{1}{2}+\mathcal{O}(\epsilon)$.
For $\epsilon=0$, the only two exchanged bulk operators are the identity and $\phi^2$. The identity operator contribution receives no correction beyond the field rescaling which we have fixed. One can systematically match expressions \eqref{eq:corrH1} and \eqref{eq:corrH2}. Focusing on the block of dimension $1$, we find 
\begin{align}
	H(\xi) &= \ldots + (\chi_s+c_3\sqrt{\epsilon})h_{1}+\ldots  \, ,  \nonumber \\
	&= \ldots + ([\lambda a_{\phi^2}]_{(0)}+[\lambda a_{\phi^2}]_{(1)}\sqrt{\epsilon})g_1 + \ldots 
\end{align}

We can now repeat the argument made previously for the fermion. $\lambda$ measures the coupling of $\bev{\phi(x)\phi(y):\phi^2(z):}$, whose first correction must come from one $h$ or two $g$ vertices, at order $\epsilon$. Hence, the ratio of the two coefficients must measure precisely the relative shift in the expectation value of $\phi^2$. This now fixes the fermion correlator, and gives a prediction for the boundary anomalous dimension of the leading spinor
\begin{align}
	\gamma_\rho^{(2)}&=-\lambda^2 N_s\left(\chi \chi_s(1+N_f)+\frac{4}{3}\right) \ .
\end{align} 

In other dimensions, we could compare our results with explicit perturbative results in the literature. Since we lacked those in this specific case, we went through their derivation to be able to showcase the agreement. The renormalisation group analysis of this system, 
given in Appendix \ref{app:perturbative}, indeed reproduces the bulk scaling dimensions $\Delta_\sigma$ and $\Delta_\psi$ we derived here. 

\section{Discussion}
\label{sec:discussion}

A central accomplishment of this work is the derivation of the bulk and boundary conformal blocks for spinor two point functions
in boundary CFT.  While the boundary blocks were known in some form previously \cite{Nishida:2018opl,Herzog:2019bom}, 
the bulk blocks are completely new, and we have put both
 to use for the first time in imposing crossing symmetry constraints on the spinor two point function.
In combination with the equations of motion, we used these constraints to derive a number of bulk and surface anomalous
dimensions as well as towers of OPE coefficients in three different interacting models, in the $\epsilon$-expansion.

Each of our models presented an interesting feature.  The Yukawa model in $4-\epsilon$ dimensions exhibited the fake
primary effect \cite{Karateev:2019pvw}, as we were cleanly able to demonstrate using our bulk conformal blocks.  Naively there should be a 
primary of $\Delta \approx 3$ corresponding to the operator $\overline \psi \psi$ when in fact, because of the equations of motion,
this operator is a descendant of $\phi$ and there is no $\Delta \approx 3$ primary in the bulk OPE of the spinor two-point function.

For the Gross-Neveu model in $2+\epsilon$ dimensions, for the same amount of work, 
we were able to get results to $O(\epsilon^2)$, in contrast to the other two models where our expansions terminate at $O(\epsilon)$.
As a result, we were able to compute the anomalous dimension for the leading fermion surface operator $\rho$ to one higher
order (\ref{Deltarho2d}) than had been computed heretofore \cite{Giombi:2021cnr}.

Finally, to our knowledge, our Yukawa model in $3-\epsilon$ dimensions with boundary has not been studied before, and thus our results
are new.  A quirk of this model is that it exhibits an expansion in $\sqrt{\epsilon}$ rather than $\epsilon$, similar
to what happens for the scalar $\phi^6$ model in $3-\epsilon$ dimensions in the presence of a boundary \cite{diehi1987walks,eisenriegler1988surface}.  
To perform the analysis on this little studied 3d model with boundary, we had to back up first and perform a standard perturbative
analysis on the theory without a boundary, in order to locate the zero of the beta functions. Interestingly, we found that the beta 
functions vanish in exactly $d=3$ for $N_s = N_f = 1$, which may indicate this case is supersymmetric. It would be interesting to 
look at this case more closely, and also to see if the zero of the beta function can be extended to larger $N_f = N_s \neq 1$ by
allowing for interactions between fermions and bosons that mix flavor. Another natural question is whether, in the large-$N$ limit, one finds that the 
two fixed points collide for an intermediate value of $\epsilon<1$, as is the case for the purely bosonic system \cite{Pisarski:1982vz}. On a tangential note, the tools we outlined allow for a generalisation to spinor-currents, from which one could discuss quantitatively supersymmetry in the presence of a boundary, at the level of correlators and blocks \cite{Erdmenger:2002ex,Drukker:2017dgn,Herzog:2018lqz}\footnote{After the completion of this work, we became aware of \cite{Gimenez-Grau:2020jvf}, which considers precisely supersymmetric correlation functions in the presence of a boundary using superspace techniques. This includes a derivation of the  superconformal blocks and allows them to study interactions in this setting.}.

One clear direction for further study is to explore how many anomalous dimensions can be calculated in this fashion and to what order in $\epsilon$.
While we got away with using free two-point functions on the right hand side of the equations of motion, 
to proceed further than we did requires considering interacting correlation functions on the right hand side.  Thus technically, the problem is more difficult
but not obviously insuperable. It would be interesting to have more analytical knowledge and control of this expansion, making contact with the works on analytical bootstrap in bCFT \cite{Bissi:2022mrs,Kaviraj:2018tfd,Mazac:2018biw,Bissi:2018mcq,Dey:2020jlc}. We have found that the crossing equation for the spinor is extremely similar to the one of the scalar.  We hope to present a more in depth analysis of the spinor crossing equations in the future. 

The surface anomalous dimension $\Delta_\rho$ is probably the most interesting quantity to come out this work.    
While we demonstrated the bulk anomalous dimensions can be computed efficiently using crossing symmetry
in boundary CFT, these bulk dimensions are unaffected by the presence of a boundary and were known previously 
in the 4d Yukawa model and 2d Gross-Neveu model. 

It would be interesting to find an experimentally measurable observable associated with $\Delta_\rho$.  
One possibility is angle resolved photo-emission spectroscopy (ARPES).  In such an experiment, a high energy photon
is used to eject electrons from the surface of the material.  The energy and momentum of the ejected electrons can then
be used to reconstruct the electron spectral function and by extension the fermionic Green's function.  In principle, then,
it should be possible to extract the surface anomalous dimension $\Delta_\rho$ from the fermionic Green's function near
the edge of a Dirac or Weyl material.
Whether such a measurement is possible in practice, we do not know.

A possibility for the measurement of a related surface anomalous dimension is to use electron transport.
Consider $J^\mu = \bar \psi \gamma^\mu \psi$, which we will 
assume to be conserved $\partial_\mu J^\mu=0$.  In the presence of a boundary at $z=0$, conservation requires also that the flux
of the current through the boundary vanish,
$J^z|_{z=0} = 0$.
 
Let us try to understand how this condition affects the structure of the boundary OPE for the current.
Conformal invariance, bulk conservation $\partial_\mu J^\mu = 0$, and unitarity together imply that $J^\mu$ can be decomposed
into a boundary scalar $\tau$ of dimension $\Delta = d-1$ (see (\ref{Jtau})) and a tower of boundary vectors $v^{i}$ of dimension $\Delta> d-2$ (see (\ref{Jva})) \cite{Liendo:2012hy}.\footnote{%
 A boundary conserved current, which saturates the unitarity bound, with $\Delta = d-2$,  
 must decouple from the bulk current, precisely because it is conserved.
}  

Consider the current-current two point function $\langle J_\mu(x) J_\nu(x') \rangle$ (\ref{JJ})
 for the particular configuration where
$x$ and $x'$ are on a line perpendicular to the boundary.  In this case, we can write
\begin{eqnarray}
\langle J_z(x) J_z(x') \rangle &=& \frac{R(v)}{|x-x'|^{2(d-1)}} \ , \;  \; \;
\langle J_i(x) J_j(x') \rangle = \frac{Q(v) \delta_{ij}}{|x-x'|^{2(d-1)}} \ .
\end{eqnarray}
The scalar of dimension $d-1$ that is allowed to contribute to this two point function has a $R(\xi) \sim v^{d-1} (v^{-1} + v)$ which is finite
in the boundary limit $v \to 1$.  Thus, in order for $J^\mu$ to be conserved, this scalar must be absent from the bOPE of $J^\mu$.
In contrast, the tower of boundary vectors have a $R(\xi) \sim (v^{-1} - v)^{\Delta - d+2}$ close to the boundary.  Thus, the unitarity bound
$\Delta > d-2$ is sufficient to guarantee current conservation for these operators.

Given this tower of boundary vectors and the unitarity bound, there will be one vector with smallest conformal dimension
which determines the near boundary behavior of $J^\mu$ through the bOPE.  
In a free fermion theory, this boundary vector is
just the boundary value of 
$\bar \psi \gamma^i \psi$, and has dimension $d-1$. More generally, we assume that this operator has dimension $\Delta_1 = d-1 + \gamma_1$ and 
that there is a gap in conformal dimension before the next boundary vector.  Note that $\gamma_1$ can be positive or negative but boundary unitarity enforces $\gamma_1 > -1$.  

What are the physical implications of this smallest boundary vector operator?
By dimensional analysis of the bOPE, it follows that $J^\mu$ has near boundary  behavior $J^i \sim z^{\gamma_1}$. 
 Indeed, the $z^{\gamma_1}$ scaling of $J^t$ 
should be a visible feature of the Fermi sea near the boundary for a material close to a phase transition.  The density of states should satisfy this power law behavior, either vanishing (for $\gamma_1 >0$) or 
diverging (for $\gamma_1 <0$) close to the boundary.  Moreover, this scaling should be reflected in the behavior of spatial currents $J^i$ near and parallel to the surface.
 
 To be more precise, we can isolate the contribution of this smallest boundary vector to the current-current two point function by looking at the near boundary limit:
\begin{eqnarray}
\langle J^z({x}) J^z({ x}') \rangle &\sim&  \frac{(z z')^{\gamma_1 +1}}{|{\bf x} - {\bf x}'|^{2\Delta_1+2}}  \ ,  \\
\langle J^i({ x}) J^j({ x}') \rangle &\sim&  \frac{(zz')^{\gamma_1}}{|{\bf x} - {\bf x'}|^{2\Delta_1}}  \frac{\Delta_1 - d+2}{8(d-1)} \left( \delta^{ij} - \frac{2 {\bf x}^i {\bf x}^j}{|{\bf x}|^2} \right)  \ .
\end{eqnarray}
Indeed $\langle J^z J^z \rangle$ will vanish by unitarity, consistent with conservation.  The $\langle J^i J^j \rangle$ contribution on the other
hand looks like the correlation function of a boundary vector with dimension $\Delta_1$, however with a power law dependent profile  as one
moves away from the boundary

For our fermionic theories, we determined $\gamma_\rho$, the anomalous dimension of the boundary value of the $\psi$ fields.
The anomalous dimension $\gamma_1$ will have a contribution from $2 \gamma_\rho$.  In general, however, it may also get 
contributions from connected loop diagrams that involve both $\rho$ fields.  It is an interesting project for the future to try to compute
these anomalous dimensions precisely, especially as they may have directly observable consequences.

\vskip 0.2in
\noindent
{\bf Erratum:} Regarding Appendix D, in previous versions, there was a sign mistake in (D.28), (D.33), and (D.34) which is now corrected.   Note also two $O(g^4)$ diagrams have been left out of the computation of (D.23).  Restoring these contributions, a real fixed point
still exists for generic values of $N_s$ and $N_f$.  These perturbative computations can be found in \cite{Jack:2016utw} which predates our work by several years.  We would like to thank L.~Fraser-Taliente, P.~Steudtner, and M.~\"Utrecht for bringing these issues to our attention.

\section*{Acknowledgments}
We would like to thank D.~Anninos, A.~Chalabi, P.~Kravchuk, E.~Lauria, S.~Parameswaran, S.~Simon, A.~Stergiou, and B.~van Rees for discussion. We would also like to thank H.~W.~Diehl for his comments on the draft. C.~H. would like to acknowledge the hospitality of the Oxford University Physics Department where part of this work was developed.
 C.~H. was supported in part by a Wolfson Fellowship from the Royal Society.
  This work was supported 
  by the U.K.\ Science \& Technology Facilities Council Grant ST/P000258/1.

\appendix

\section{Conventions}
\label{app:conventions}

The theories we consider are reflection-positive Euclidean CFTs, which under analytic continuation should map to unitary Lorentzian CFTs. We use Greek indices ranging from $1$ to $d$, contracted with a metric $ \delta_{\mu\nu}$. Intrinsic coordinates on the boundary are written using Latin indices $i,j,k = 0,\ldots, p-1$ with $p=d-1$. Embedding coordinates are written $A,B,C \ldots = 0,\ldots, d+1$, with one time-dimension. The (inward pointing) unit-normal to the boundary is taken to be $n^{\mu}=\delta^{d,\mu}$, without loss of generality. The induced metric on the boundary is $h_{\mu\nu}=\eta_{\mu\nu}-n_{\mu}n_{\nu}$, and intrinsically $\delta_{ij}$. Lightcone coordinates are given in the embedding using $P^{\pm}= P^{0}\pm P^{d+1}$, i.e.\ $\eta^{+-}=-2$, $\eta_{+-}=-\frac{1}{2}$.

Our convention for the Clifford algebra is that $\{\gamma_\mu,\gamma_\nu\}=2\delta_{\mu\nu}$.  We generically consider Dirac spinors unless otherwise specified. In real space, since we work in Euclidean signature, we take $\overline{\psi}$ and $\psi$ to be independent fields, upon which we impose a reality condition coming from Lorentzian signature. In the embedding, conjugation follows through in the same way, and so one must consider how Dirac conjugation works with two time dimensions. The index-free spinors are made up from spinor bilinears; hence their complex-conjugation properties are straightforward.

The embedding gamma-matrices are given by
\begin{align}
	\Gamma_{\mu} &=  \begin{pmatrix}
		1 & 0 \\ 0 & -1 
	\end{pmatrix} \otimes \gamma_\mu \ , & 
	\Gamma_{+} &= \begin{pmatrix}
		0 & 1 \\ 0 & 0 
	\end{pmatrix} \otimes \mathbb{1}  \ , & 
	\Gamma_{-} &=\begin{pmatrix}
		0 & 0 \\ -1 & 0 
	\end{pmatrix} \otimes \mathbb{1} \ .
\end{align}
For a generic point on the lightcone $P^{A}$ in the Poincar\'e patch, with real-space coordinate $x^{\mu}$, we have
\begin{align}
\label{twistorformofP}
	\slashed{P} = \begin{pmatrix}
		\slashed{x} & 1 \\ -x^2 & -\slashed{x}
	\end{pmatrix} = \begin{pmatrix}
		1 \\ -\slashed{x} 
	\end{pmatrix} \begin{pmatrix}
		\slashed{x} & 1 
	\end{pmatrix} \ .
\end{align}

Because of the broken rotational symmetry, it is convenient to work with projectors of definite helicity on the boundary. In real space, we use the normal $n^{\mu}$, and in the embedding we use $V^{A}=\delta^{A,d}$. The analog of the polarization tensor $Z$ is the polarization spinor $S$, which must satisfy a similar transversality condition
$\slashed{P} S = 0$.  This null eigenvalue condition is solved by
\begin{align}
	S &= \begin{pmatrix}
		1 \\ -\slashed{x}
	\end{pmatrix}s \ , & 
	\overline{S} &= \overline{s}\begin{pmatrix}
		\slashed{x} & 1
	\end{pmatrix} \ .
\end{align}
We use Dirac bra-ket notation to efficiently write the spinor-bilinear. The convention is that $\bra{i}\equiv \overline{S}_i$, $\ket{i}\equiv S_{i}$, and that vectors inside (unless otherwise specified) are contracted with $\Gamma$-matrices. Projection to real-space is performed by evaluating the bilinear using its expression in the tensor-product space, to reduce it to a real-space bilinear. The twistor form of $\slashed{P}$ (\ref{twistorformofP}) is particularly convenient for this purpose. For example, the structures entering the spinor-spinor-scalar three point function are
\begin{align}
	\bra{1}\ket{2} \rightarrow \bra{1}x_{12} \ket{2} \ , \; \; \; 
	\bra{1}P_3\ket{2} \rightarrow \bra{1}x_{13} x_{32} \ket{2} \ .
\end{align}
And these are manifestly real, as they map to themselves  under complex-conjugation and exchange $1\leftrightarrow 2$. 

To discuss conjugate spinors, we settle on a reality structure to impose on our correlators. This translates into reality conditions on the different spinorial structures and OPE coefficients. Although the space is Euclidean and so are the spinors, we impose the reality structure suitable for spinors of $\mathbb{R}^{2,d}$ in the embedding, $\mathbb{R}^{1,d-1}$ in real space. This means that the correlators we write will not be Hermitian as seen from the Euclidean viewpoint.

The breaking of rotational symmetry induced by the planar boundary is efficiently understood by the analogy with Weyl-spinors. A Dirac spinor in the bulk can be decomposed on the boundary into its two helicity parts. We encode boundary spinors as constrained Dirac spinors of the same type as the bulk ones. They are written $\rho_{\pm}$, and satisfy $\gamma\cdot n\rho_{\pm}=\pm \rho_{\pm}$. Their polarisations are written with a supplementary label, $s_{i}^{(\pm)} \equiv \ket{i,\mp}$ , $\overline{s}_{i}^{(\pm)} \equiv \bra{i,\pm}$. Naturally, one has $\bra{i,\pm}=\bra{i}\mathbb{P}_{\pm}$ and likewise for the kets. The perhaps counterintuitive notation is such that $\overline{s}_{+}\rho_{+}$ and $\overline{\rho}_{+}s_+$ are scalars, complex-conjugate to each other, and compatible with $\slashed{n}\rho_+ = \rho_+$ and $\overline{\rho}_{+}\slashed{n}=-\overline{\rho}_{+}$.

\section{Spinning Conformal Integral}
\label{app:conformalintegral}

In this appendix we collect useful results for the explicit evaluation of generic CPWs using the embedding space form of the shadow formalism. The idea is that partial waves can be computed through an integral representation using a conformally invariant sewing of simpler correlation functions. The resulting formulae take particularly simple form in the embedding space.

\subsection{Reduction to Standard Integral}

We first note the following useful formulae: 
\begin{align}
	\int \frac{D^{d}X}{(-2 Y\cdot X)^{d}}&=\frac{\sqrt{\pi}^{d}\Gamma(\frac{d}{2})}{\Gamma(d)} \frac{1}{(-Y\cdot Y)^{\frac{d}{2}}} \ , \\
\prod_{i=1}^{n}\frac{1}{A_i^{a_i}}&=\frac{\Gamma(\sum_{i=1}^{n} a_i)}{\prod_{i=1}^{n}\Gamma(a_i)}\int \left(\prod_{i=2}^{n}\frac{d\alpha_i}{\alpha_i}\alpha_i^{a_i}\right)\frac{1}{(A_1+\alpha_2 A_2 +\ldots \alpha_n A_n)^{\sum_{i=1}^{n} a_i}} \ ,  \\
\left(\sum_{i=1}^{n} A_{i}\right)^{l} &= \sum_{r_1+r_2 +\ldots r_n=l}\binom{l}{r_1 \ldots r_n} (A_1)^{r_1}\ldots (A_n)^{r_n} \ .
\end{align}
We can now evaluate the most general thing we will encounter 
\begin{align}
	\mathcal{I}^{(l)}_{(a_i)}(P_i,Z)&=\mathcal{P}\int D^{d}X (Z\cdot X)^{l}\prod_{i=1}^{n}\frac{1}{(-2P_i \cdot X)^{a_i}}\\
	&=\mathcal{P}\frac{\Gamma(d+l)}{\prod_{i=1}^{n}\Gamma(a_i)}\int \left(\prod_{i=2}^{n}\frac{d\alpha_i}{\alpha_i}\alpha_i^{a_i}\right)\frac{(Z\cdot X)^{l}}{\Big(-2\underbrace{(P_1+\alpha_2 P_2 +\ldots \alpha_n P_n)}_{Y}\cdot X)\Big)^{d+l}}\nonumber \\
	&=\mathcal{P}\frac{\Gamma(d+l)}{\prod_{i=1}^{n}\Gamma(a_i)}\frac{1}{2^{l}(d)_{l}}\int \left(\prod_{i=2}^{n}\frac{d\alpha_i}{\alpha_i}\alpha_i^{a_i}\right)\left(Z\cdot \pdv{}{Y} \right)^l \int D^{d}X\frac{1}{\left(-2Y\cdot X\right)^{d}}  \nonumber \\
	&=\mathcal{P}\frac{\sqrt{\pi}^{d}\Gamma(\frac{d}{2}+l)}{\prod_{i=1}^{n}\Gamma(a_i)}\int \left(\prod_{i=2}^{n}\frac{d\alpha_i}{\alpha_i}\alpha_i^{a_i}\right)\frac{(Y\cdot Z)^l}{(-Y\cdot Y)^{\frac{d}{2}+l}} \nonumber \\
	&=\frac{\sqrt{\pi}^{d}\Gamma(\frac{d}{2}+l)}{\prod_{i=1}^{n}\Gamma(a_i)} \sum_{r_1+\ldots+r_n =l}\binom{l}{r_1\ldots r_n}(Z\cdot P_1)^{r_1}\ldots (Z\cdot P_n)^{r_n}\mathcal{F}_{(a_i+r_i)}(P_i) \ ,
	\nonumber
	\end{align}
where in the last line we defined the object which we refer to as the ``standard integral'':
\begin{align}
	\mathcal{F}_{(a_i)}(P_i) = \mathcal{P} \int \left(\prod_{i=2}^{n}\frac{d\alpha_i}{\alpha_i}\alpha_i^{a_i}\right) \frac{1}{\Big(-(P_1 + \alpha_2 P_2+\ldots)\cdot (P_1 + \alpha_2 P_2+\ldots)\Big)^{\frac{\sum a_i}{2}}} \ .
\end{align}
The task of evaluating the most general correlator is then reduced to fixing $\mathcal{F}$. Note the integral still must be projected, via ${\mathcal P}$, on the right hypergeometric subspace.
	
\subsection{Evaluation of the bulk standard integral}

We consider 3 points, which are contracted with a full dot product $\cdot$, and one of them not on the lightcone, $P_3 = -V$. The standard integral becomes 

\begin{align}
	\mathcal{F}_{(a,b,c)} &= \mathcal{P} \int\frac{d\beta d \gamma}{\beta \gamma}\frac{\beta^{b}\gamma^{c} }{\Big(-(P_1 + \beta P_2-\gamma V)\cdot (P_1 + \beta P_2-\gamma V)\Big)^{\frac{a+b+c}{2}}} \\
	&= \mathcal{P}\int \frac{d\beta}{\beta}\frac{d\gamma}{\gamma}\frac{\beta^{b}\gamma^{c}}{\Big(\beta( P_{12}+\gamma 2P_2 \cdot V)+\gamma( 2P_{1}\cdot V  -\gamma)\Big)^{\frac{a+b+c}{2}}}  \nonumber \ .
\end{align}
After rescaling $\beta$ and $\gamma$, we can isolate an integral depending only on the cross-ratio $\xi$
\begin{align}
\label{Fabc}
	\mathcal{F}_{(a,b,c)}(P_1,P_2)&= \mathcal{P}\frac{1}{ (2P_1\cdot V)^{(a-b)}(P_{12})^{b}}\int \frac{d\beta}{\beta}\frac{d\gamma}{\gamma}\frac{\beta^{b}\gamma^{c}}{\Big(\beta(1+\frac{\gamma}{\xi}) +\gamma(1-\gamma)\Big)^{\frac{a+b+c}{2}}}  \nonumber \\
	&= \frac{1}{(2P_1\cdot V)^{a}(2P_2 \cdot V )^{b}}\frac{\pi}{\sin(\frac{a-b+c}{2}\pi)}\frac{\Gamma (a) \Gamma (b)(-1)^{\frac{1}{2} (-a+b-c)}}{\Gamma \left(1+\frac{a+b-c}{2}\right) \Gamma \left(\frac{a+b+c}{2}\right)} \nonumber \\
	&\times \, _2F_1\left(a,b;1+\frac{a+b-c}{2};-\xi\right)
\end{align}
where in the last line we used the monodromy projection to discard one of the hypergeometric functions from the integral in the line above.
In particular, we need a function which scales as a constant in the $\xi \to 0$ limit.  
This integral (\ref{Fabc}) is key to the derivation of the bulk CPWs (\ref{bulkCPWone}) and (\ref{bulkCPWtwo}).

\subsection{Evaluation of the boundary standard integral}

We look at two boundary points contracted with the reduced dot product $\bullet$.
Neither of the points have vanishing square under $\bullet$, 
\begin{align}
	\mathcal{F}_{(a,b)} &= \mathcal{P}\int\frac{d\beta}{\beta}\frac{\beta^{b}}{\Big(-(P_1 + \beta P_2)\bullet (P_1 + \beta P_2)\Big)^{\frac{a+b}{2}}} \nonumber \\
	&=\frac{\mathcal{P}}{(P_1\cdot V)^{a}(P_2\cdot V)^{b}}\int \frac{d\beta}{\beta}\frac{\beta^{b}}{\Big((1+\beta)^2+4\beta\xi\Big)^{\frac{a+b}{2}}} \ ,
\end{align}
where we used the definition of the cross-ratio $\xi$. Now, we can evaluate the integral by replacing the series expansion 
\begin{align}
\label{Fab}
	\mathcal{F}_{(a,b)} &=\frac{ \mathcal{P}}{(2P_1\cdot V)^{a}(2P_2\cdot V)^{b}}2^{a+b}\sum_{n=0}^{\infty} \frac{\left(\frac{a+b}{2}\right)_n}{n!}\left(-4\xi\right)^{n}\int \frac{d\beta}{\beta}\frac{\beta^{b+n}}{(1+\beta)^{a+b+2n}}  \nonumber \\
	 &=\frac{\Gamma (a) \Gamma \left(\frac{b-a}{2}\right)}{\Gamma \left(\frac{a+b}{2}\right)}\frac{2^{b-a}}{(P_1\cdot V)^{a}(P_2\cdot V)^{b}}\frac{1}{\xi^{a}} \, _2F_1\left(a,\frac{1+a-b}{2};1+a-b;-\frac{1}{\xi}\right) \ ,
\end{align}
where in the last line we used the hypergeometric identity 
\begin{align*}
	_2F_1\left(\alpha,\beta;\gamma;z\right)&=\frac{\Gamma(\gamma)\Gamma(\beta-\alpha)}{\Gamma(\beta)\Gamma(\gamma-\alpha)}(-z)^{-\alpha}\,_2F_1\left(\alpha,\alpha+1-\gamma;\alpha+1-\beta;\frac{1}{z}\right) \\&+\frac{\Gamma(\gamma)\Gamma(\alpha-\beta)}{\Gamma(\alpha)\Gamma(\gamma-\beta)}(-z)^{-\beta} \,_2F_1\left(\beta,\beta+1-\gamma;\beta+1-\alpha;\frac{1}{z}\right) \ .
\end{align*}
and applied the monodromy projection $\mathcal{P}$ which throws out the second term in the hypergeometric identity. 
In a nutshell, the monodromy projection should enforce that the CPW behaves as $\xi^{-a}$ in the boundary $\xi \to \infty$ limit, matching onto the
behavior expected from the OPE.
The integral (\ref{Fab}) is an essential ingredient in deriving (\ref{boundaryCPW}).

\section{Conformal Block through OPE resummation}
\label{app:OPEresummation}

\subsection{Fixing the boundary OPE}

This section focuses on the construction of the operator $\mathfrak{D}$ which encodes the boundary OPE of spinor operators. Our starting point is the definition (\ref{boundarypsiOPE}) given previously.
The operator $\mathfrak{D}$ is fixed by requiring the matching of the OPE with the form of correlators involving $\psi$ and $\rho$. Concretely, consider
\begin{align*}
	\expval{\psi_{\Delta_1}(x,z)\overline{\rho}_{\Delta,\alpha}(y)}&\propto \frac{\left(\gamma\cdot(x-y)+\gamma_n z\right)\mathbb{P}_{-\alpha}}{(2z)^{\Delta_1-\Delta}\left(\abs{x-y}^2+z^2\right)^{\Delta+\frac{1}{2}}}  \ ,\\
	\expval{\rho_{\Delta,\alpha}(x)\overline{\rho}_{\Delta,\alpha}(y)}&=\frac{\mathbb{P}_{\alpha}\gamma\cdot(x-y)}{\abs{x-y}^{2\Delta+1}}  \ .
\end{align*}
In this appendix, $\gamma\cdot x \equiv \gamma^{i}x_i$, and does not include the normal components. Likewise, $\Box = \partial_i \partial^i$. We have to match the expansion in $z$ in terms of a sum over descendants of a given primary $\rho$. Without loss of generality, we can bring $y\rightarrow 0$ to simplify expressions 
\begin{align*}
	\expval{\psi_{\Delta_1}(x,z)\overline{\rho}_{\Delta,\alpha}(0)}&\propto (2z)^{\Delta-\Delta_1}\sum_{n=0}^{\infty} \frac{(-1)^n(\Delta+\frac{1}{2})_{n}}{n!}
	\times \nonumber \\
	& \times \left( z^{2n}\mathbb{P}_{\alpha}\frac{\gamma\cdot x}{\abs{x}^{2\Delta+1+2n}}+z^{2n+1}\gamma_n\mathbb{P}_{-\alpha}\frac{1}{\abs{x}^{2\Delta+1+2n}}\right) \ .
\end{align*}

The overall power is that expected by the OPE, so we can ignore it. We rewrite each term as some derivatives acting on $\expval{\rho \overline{\rho}}$. Note the following useful identities 
\begin{align}
	\Box^{n} \expval{\rho_{\Delta,\alpha}\overline{\rho}_{\Delta,\alpha}}&=4^n \left(\Delta+\frac{1}{2}\right)_{n}\left(\Delta-\frac{p-1}{2}\right)_{n}\frac{\mathbb{P}_{\alpha} \gamma\cdot x}{\abs{x}^{2n+2\Delta+1}} \ , \nonumber \\
	\Box^{n} \gamma\cdot \partial\expval{\rho_{\Delta,\alpha}\overline{\rho}_{\Delta,\alpha}}&=-2^{2n+1}\left(\Delta+\frac{1}{2}\right)_{n}\left(\Delta-\frac{p-1}{2}\right)_{n+1}\mathbb{P}_{-\alpha}\frac{1}{\abs{x}^{2n+2\Delta+1}} \ . 
\end{align}
From these we can write 
\begin{align}
	\expval{\psi_{\Delta_1}(x,z)\overline{\rho}_{\Delta,\alpha}(0)}&\propto (2z)^{\Delta-\Delta_1}\sum_{n=0}^{\infty} \frac{1}{n!}\left(-\frac{z^2}{4}\Box\right)^{n}\left(\frac{1}{\left(\Delta-\frac{p-1}{2}\right)_{n}}\right. \nonumber \\
	&-\left.\frac{z \gamma_n\gamma \cdot\partial}{2}\frac{1}{\left(\Delta-\frac{p-1}{2}\right)_{n+1}}\right)\expval{\rho_{\Delta,\alpha}\overline{\rho}_{\Delta,\alpha}} \ ,
\end{align}
which fully specifies the form of the differential operator:
\begin{equation}
	\mathfrak{D}_{(\Delta)}(z,x,\partial_x)=\sum_{n=0}^{\infty}\frac{1}{n!}\left(-\frac{z^2}{4}\Box _x\right)^n \left(\frac{1}{\left(\Delta-\frac{p-1}{2}\right)_n}-\gamma_n\frac{z \gamma\cdot \partial_x}{2\left(\Delta-\frac{p-1}{2}\right)_{n+1}} \right) \ .
\end{equation}

To obtain the equivalent expression for the boundary OPE of $\overline \psi$, one should take the complex conjugate of the $\psi$ OPE:
\begin{align}
	\overline{\psi}(y,w) &= \sum_{\Delta,\alpha}\frac{\mu_{\Delta,\alpha}}{(2w)^{\Delta_1-\Delta}}\overline \rho_{\Delta,\alpha}(y)\overline{\mathfrak{D}}_{(\Delta)}(w,y,\overset{\leftarrow}{\partial}_y) \\
	\overline{\mathfrak{D}}_{(\Delta)}(w,y,\overset{\leftarrow}{\partial}_y)&=\sum_{k=0}^{\infty}\frac{1}{k!}\left(-\frac{w^2}{4}\overset{\leftarrow}{\Box}_y \right)^k \left(\frac{1}{\left(\Delta-\frac{p-1}{2}\right)_n}+\gamma_n\frac{w \gamma\cdot \overset{\leftarrow}{\partial}_y}{2\left(\Delta-\frac{p-1}{2}\right)_{n+1}} \right) \ , \nonumber
\end{align}
where the sign change in the second term comes from anticommuting the $\gamma_n$ and $\gamma_i$ after complex conjugation.

We now turn to the computation of the conformal blocks. These are given by inserting the boundary OPE expansion for a single primary field into both operators inside $\expval{\psi(x,z)\overline{\psi}(y,w)}$. 

We obtain the following expression for the sum of blocks times spinor structures 
\begin{align}
	\widehat{\mathcal{W}}_{\Delta,\alpha}(u)&=\frac{1}{(2z)^{\Delta_1-\Delta}(2w)^{\Delta_2-\Delta}} \mathfrak{D}_{(\Delta)}(z,x,\partial_x)\expval{\rho_{\Delta,\alpha}(x)\overline{\rho}_{\Delta,\alpha}(y)}\overline{\mathfrak{D}}_{(\Delta)}(w,y,\overset{\leftarrow}{\partial}_y)  \ .
\end{align}
This expression is complicated, and we will satisfy ourselves with only a partial check of the boundary conformal blocks, 
keeping only the terms proprtional to $\slashed{\partial}$.  To further simplify the expression, we set $y=0$ by translation symmetry.   

The check proceeds by brute force, evaluating the derivatives and making use of various Pochhammer identities.  The goal is to try to organize the sum
to match the sum (B.8) from \cite{Dolan:2001wg}\footnote{For an alternative reference see \cite{gradshtein}. This double sum defines an $F_4$ hypergeometric of 2 variables, formula (9.180). This reduction to a single Gauss hypergeometric then follows by formula (9.182), sub-equation 7.}:
\begin{align}
	\sum_{n,k}\frac{A^n B^k}{(A+B+C)^{\lambda+n+k}}\frac{(\lambda)_{n+k}(\rho)_{n+k}}{n!k!(\rho)_{n}(\rho)_k}&=\frac{1}{C^{\lambda}}\sum_{n}\frac{(\lambda)_{2n}}{n!(\rho)_n}\left(\frac{A B}{C} \right)^{n} \\
	&=\frac{1}{C^{\lambda}} \, _2F_1\left(\frac{\lambda }{2},\frac{\lambda +1}{2};\rho ;\frac{4 A B}{C^2}\right)  \ . \nonumber
\end{align}
In particular, one is able to identify $|x|^2 = A+B+C$, $A=-z^2$, and $B=-w^2$.  One finds hypergeometric functions which depend on 
$(1+2 \xi)^{-2}$, which after a quadratic transformation can be written in terms of hypergeometrics which depend on $-\frac{1}{\xi}$, and further simplify to the form (\ref{boundaryCPW}) previously given.

\subsection{Fixing the bulk OPE}

The goal of this section is to fully specify the differential operator encoding the bulk OPE of the two spinor fields, as given through the equation (\ref{bulkpsiOPE}). To do so, we will take as a starting point the differential operator generating the bulk OPE of scalar fields 
\begin{align}
	\phi_{1}(x)\phi_2(y) = \sum_{\Delta} \frac{C_{123}}{\abs{x-y}^{\Delta_1+\Delta_2-\Delta_3}}C_{(\Delta_1,\Delta_2,\Delta_3)}(s,\partial_y)\phi_{3}(y)\bigg\rvert_{s_\mu=(x-y)_\mu}+(\ell>0) \ .
\end{align}
The differential operator $C$ is given by \cite{Ferrara:1973yt,Billo:2016vm}.
\begin{align}
	C^{(\Delta_1,\Delta_2,\Delta_3)}(s,\partial_y)=& \int_{0}^{1}\frac{du}{B(a,b)}u^{a-1}(1-u)^{b-1}e^{u s\cdot \partial_y}\sum_{m=0}^{\infty}\left(-\frac{s^2 u(1-u)}{4}\Box_{y}\right)^m \times \nonumber \\
	& \frac{1}{m!\left(\Delta_3+1-\frac{d}{2}\right)_m} \ .
\end{align}
We leave $s$ free so that $[s,\partial_y]=0$, meaning one should not act with derivatives on any factor involving $s$. This is convenient to obtain this simple, factorised expression. The parameters $a$ and $b$ are given by  $a=\frac{\Delta_3+\Delta_1-\Delta_2}{2}$ and $b=\frac{\Delta_3+\Delta_2-\Delta_1}{2}$. 

This operator is fixed by requiring that its action on a scalar two point function reproduces the parts needed to generate the three point function 
\begin{align}
	\left.C_{(\Delta_1,\Delta_2,\Delta_3)}(s,\partial_y)\left[\frac{1}{\abs{y-z}^{2\Delta_3}}\right]\right\rvert_{s=(x-y)}=\frac{1}{\abs{x-z}^{\Delta_3+\Delta_1-\Delta_2}\abs{y-z}^{\Delta_3+\Delta_2-\Delta_1}} \ .
\end{align}

The general tactic is now to write the correlator involving spinors using derivatives and sums of correlators involving only scalars. Then, linear combinations of $C$'s with shifted dimension and derivatives can be used to construct the spinor $\mathfrak{C}$. Consider the first structure in the three point function 
\begin{align}
	\bev{\psi(x) \bar{\psi}(y)\phi(z)}^{(1)}&=\gamma\cdot(x-y)\expval{\phi_{\Delta_1+\frac{1}{2}}(x)\phi_{\Delta_2+\frac{1}{2}}(y)\phi_{\Delta_3}} \\
	&=\left.\frac{\gamma\cdot(x-y) C^{(\Delta_1+\frac{1}{2},\Delta_2+\frac{1}{2},\Delta_3)}(s,\partial_y)}{\abs{x-y}^{\Delta_1+\Delta_2+1-\Delta_3}}\bigg[\expval{\phi_{\Delta_3}(y)\phi_{\Delta_3}(z)}\bigg]\right\lvert_{s=x-y}  \ , \nonumber
\end{align}
from which we obtain that the bulk OPE of the first structure can be encapsulated efficiently using this $C$ differential operator, i.e.\ we find 
\begin{align}
	\mathfrak{C}_{(\Delta_1,\Delta_2,\Delta)}^{(1)}(x-y,\partial_y)[\bullet ] = \left.\gamma\cdot(x-y) C_{(\Delta_1+\frac{1}{2},\Delta_2+\frac{1}{2},\Delta)}(s,\partial_y)[\bullet]\right\lvert_{s=x-y} \ .
\end{align}

For the second structure, we have to be more creative. First note we can rewrite it as 
\begin{align}
	(\slashed{x}-\slashed{z})(\slashed{y}-\slashed{z})&=\slashed{s}(\slashed{y}-\slashed{z})+\abs{y-z}^2 \ .
\end{align}
Clearly, the second term can be reabsorbed into the power-laws to yield a scalar three point function with shifted weights. Meanwhile, the first term should be rewritten using factors of $s_{\mu}$ and $\partial_y$ acting on the three point function given out by the action of $C$. One can check the following identity 
\begin{align}
	\slashed{s}(\slashed{y}-\slashed{z})\expval{\phi_{\Delta_1}(x)\phi_{\Delta_2}(y)\phi_{\Delta_3}(z)}&=\frac{\left(\Delta_3-\Delta_2-\Delta_1+\slashed{s}\slashed{\partial}_{y}\right)}{2+\Delta_1-\Delta_2-\Delta_3}\expval{\phi_{\Delta_1}(x)\phi_{\Delta_2-1}(y)\phi_{\Delta_3-1}(z)}\\
	&=\frac{1}{\abs{x-y}^{\Delta_1+\Delta_2-\Delta_3}}\left.\frac{\slashed{s}\slashed{\partial}_y C_{(\Delta_1,\Delta_2-1,\Delta_3-1)}(s,\partial_y)}{2+\Delta_1-\Delta_2-\Delta_3}
	\frac{1}{|y-z|^{2\Delta_3-2}}
	\right\lvert_{s=x-y} \nonumber \ .
\end{align}

Meanwhile, the second piece can be incorporated using the same shifted dimension for $\Delta_2$ and $\Delta_3$. Taking into account the explicit form of the correlator $\bev{\psi \bar{\psi}\phi}^{(2)}$, and the powers entering it, we end up with the final expression 
\begin{align}
	&\bev{\psi(x) \bar{\psi}(y)\phi(z)}^{(2)}=(\slashed{x}-\slashed{z})(\slashed{z}-\slashed{y})\expval{\phi_{\Delta_1+\frac{1}{2}}(x)\phi_{\Delta_2+\frac{1}{2}}(y)\phi_{\Delta_3+1}} \\
	&=\left.\frac{1+\frac{\slashed{s}\slashed{\partial}_y}{1+\Delta_1-\Delta_2-\Delta_3}}{\abs{x-y}^{\Delta_1+\Delta_2-\Delta_3}}C_{(\Delta_1+\frac{1}{2},\Delta_2-\frac{1}{2},\Delta_3)}(s,\partial_y)
	\frac{1}{|y-z|^{2\Delta_3}}
	\right\lvert_{s=x-y} \nonumber
\end{align}
from which we gather that 
\begin{align}
	\mathfrak{C}_{(\Delta_1,\Delta_2,\Delta)}^{(2)}(x-y,\partial_y)[\bullet ] = \left.\abs{s}\left(1+\frac{\slashed{s}\slashed{\partial}_y}{1+\Delta_1-\Delta_2-\Delta}\right)C_{(\Delta_1+\frac{1}{2},\Delta_2-\frac{1}{2},\Delta)}(s,\partial_y)[\bullet]\right\lvert_{s=x-y} \ .
\end{align}
Using these results, we can compute the conformal blocks for the exchange of bulk scalars.

\subsection{Bulk OPE Resummation}

The bulk conformal partial waves are given by the action of $\mathfrak{C}^{(i)}_{(\Delta_1,\Delta_2,\Delta)}$ on the one point function of a scalar operator
\begin{align}
	\mathcal{W}_{\Delta,i}(x,y)=\frac{\mathfrak{C}^{(i)}_{(\Delta_1,\Delta_2,\Delta)}[\expval{\phi_{\Delta}}]}{\abs{x-y}^{\Delta_1+\Delta_2+1-\Delta_3}} \ .
\end{align}
Since the $\mathfrak{C}^{(i)}_{(\Delta_1,\Delta_2,\Delta)}$ are built from the $C$'s, one can first compute the result for $C$ to build the result for the spinors. The result for $C$ is of course the bulk block of a scalar field \cite{McAvity:1995tm}:
\begin{align}
	C_{(\Delta_1,\Delta_2,\Delta_3)}(s,\partial_y)& \frac{1}{(2 y \cdot n)^{\Delta_3}} 
	= \nonumber\\
	& \frac{1}{2^{\Delta_3}}\sum_{m=0}^\infty \int_{0}^{1}\frac{du}{B(a,b)}\frac{u^{a+m-1}(1-u)^{b+m-1}}{(y\cdot n+u s\cdot n)^{\Delta_3+2m}} 
	\left(-\frac{s^2}{4}\right)^m\frac{(\Delta_3)_{2m}}{m!\left(\Delta_3+1-\frac{d}{2}\right)_m} \ . \nonumber
	\end{align}
We used that the exponential $e^{u s \cdot \partial_y}$ generates a translation in the $y$ direction. The $u$ integral is an Euler parametrisation of a hypergeometric function with parameters including a shift by $m$. Inputting the explicit values for $a$ and $b$, the hypergeometric function becomes a rational function of $x\cdot n$ and $y \cdot n$.  The sum can then be done and gives yet another hypergeometric function
\begin{align}
	\left.C_{(\Delta_1,\Delta_2,\Delta_3)}(s,\partial_y)[\expval{\phi_{\Delta_3}(y)}]\right\lvert_{s=x-y}&=\frac{\, _2F_1\left(\frac{\Delta_3+\Delta_1-\Delta_2}{2},\frac{\Delta_3+\Delta_2-\Delta_1}{2};\Delta_3+1-\frac{d}{2};-\xi\right)}{(2y\cdot n)^{\frac{\Delta_2+\Delta_3-\Delta_1}{2}}(2x\cdot n)^{\frac{\Delta_1+\Delta_3-\Delta_2}{2}}} \ ,
\end{align}
and we made the cross-ratio explicit. Computing the block is now straightforward by acting with derivatives. One finds back the expressions 
(\ref{bulkCPWone}) and (\ref{bulkCPWtwo}) coming from the conformal integrals. In checking $\mathcal{W}_{\Delta,2}(x,y)$, the hypergeometric identity
\begin{align}
\left( 1 - \frac{2\xi}{1+\Delta_{12} - \Delta_3} \partial_\xi \right){}_2 F_1& \left( \frac{\Delta_3 + \Delta_{12}+1}{2}, \frac{\Delta_3 - \Delta_{12}-1}{2}, \Delta_3 - \frac{d}{2}+1, -\xi \right) =  \nonumber
\\
&{}_2 F_1 \left( \frac{\Delta_3 + \Delta_{12}+1}{2}, \frac{\Delta_3 - \Delta_{12}+1}{2}, \Delta_3 - \frac{d}{2}+1, -\xi \right) 
\end{align}
is useful.


\section{Renormalisation Group Analysis of Interacting Fermions in $d=3-\epsilon$} 
\label{app:perturbative}

We present the renormalisation group analysis of the system of fermions coupled to scalars in $d=3-\epsilon$. For completeness sake, we also consider a potential for the scalars. To the best of our knowledge, these results are new.

\subsection{Lagrangian and Feynman Rules}

The Lagrangian is given in eq.\ \eqref{eq:lag3dyuk}, and the conventions for the spinors are as before. We take $a,b,c,\ldots=1 \ldots N_f$ or $ 1 \ldots N_s$, depending on context. We will work in $d=3-\epsilon$, where both interactions are marginally relevant. We write the renormalized fields using brackets, $[\bullet]$. The various $Z$-factors from which we extract anomalous dimensions and that we use to renormalize the fields
are defined as follows:
\begin{equation}
	\begin{aligned}
		\psi &= \sqrt{Z_{\psi}} [\psi] \ , & \phi &=\sqrt{Z_{\phi}} [\phi] \ , & & Z_\psi [\overline{\psi}] [\psi] =Z_{\sigma}[\overline{\psi}\psi] \ , \\
	\gamma_\psi &= \frac{\mu}{2}\pdv{}{\mu}\log(Z_{\psi}) \ , & \gamma_\phi &= \frac{\mu}{2}\pdv{}{\mu}\log(Z_{\phi}) \ , & & \gamma_\sigma = \mu \pdv{}{\mu}\log(\frac{Z_\psi}{Z_\sigma}) \ .
	\end{aligned}
\end{equation}
The couplings are also renormalized, giving rise to beta functions
\begin{align}
	g &= \frac{Z_g}{Z_\phi Z_\psi}\mu^{\epsilon}[g] \ , &  \beta_g &= g\left( 2(\gamma_\psi + \gamma_\phi) -\mu \pdv{}{\mu}\log(Z_{g})-\epsilon\right)\ , \\
	h &=  \frac{Z_h}{Z_\phi^{3}}\mu^{2\epsilon}[h] \ , & \beta_h &=h \left( 6 \gamma_\phi-\mu \pdv{}{\mu} \log(Z_{h})-2\epsilon\right) \ .
\end{align}
For convenience later, we write $Z_{A} = 1 + \delta_A$. The anomalous dimensions $\gamma_\psi$ and $\gamma_\phi$ have to be positive by unitarity. We will use a renormalisation condition such that the loop corrections to the correlators vanish when all momenta (except one) are equal to some $p_\mu$ such that $p \cdot p = \mu^2$. The Feynman rules\footnote{The diagrams have all been drawn using TikZ-Feynman \cite{Ellis:2016jkw}.} are 
\begin{align}
\begin{tikzpicture}[baseline=-\the\dimexpr\fontdimen22\textfont2\relax] \begin{feynman}[inline=((a).base)]
\vertex (a) ;
\vertex [  right=of a](b);
\diagram* {
 (a) --[fermion] (b) 
}; \end{feynman} \end{tikzpicture} & = -\frac{\slashed{p}}{p^2} 
& \begin{tikzpicture}[baseline=-\the\dimexpr\fontdimen22\textfont2\relax] \begin{feynman}[inline=((a).base)]
\vertex [](a) {} ;
\vertex [  right=of a,crossed dot](b) {};
\vertex [right=of b] (c) {};
\diagram* {
 (a) --[fermion] (b) --[fermion] (c) }; \end{feynman} \end{tikzpicture} & = -\slashed{p}\delta_\psi \\
\begin{tikzpicture}[baseline=-\the\dimexpr\fontdimen22\textfont2\relax] \begin{feynman}[inline=((a).base)]
\vertex (a) ;
\vertex [  right=of a](b);
\diagram* {
 (a) --[scalar] (b) 
}; \end{feynman} \end{tikzpicture} & = \frac{-i}{p^2} 
& \begin{tikzpicture}[baseline=-\the\dimexpr\fontdimen22\textfont2\relax] \begin{feynman}[inline=((a).base)]
\vertex [](a) {} ;
\vertex [  right=of a,crossed dot](b) {};
\vertex [right=of b] (c) {};
\diagram* {
 (a) --[scalar] (b) --[scalar] (c) }; \end{feynman} \end{tikzpicture} & =i p^2\delta_\phi \\ 
\begin{tikzpicture}[baseline=-\the\dimexpr\fontdimen22\textfont2\relax] \begin{feynman}[inline=((a).base)]
\vertex (a) ;
\vertex [ above right=of a](b1) ;
\vertex [ above left =of a](b2) ;
\vertex [ below right=of a](c1) ;
\vertex [below left = of a] (c2) ;
\diagram* {
 (b2) --[fermion] (a) --[fermion] (b1), (c1) --[scalar] (a) --[scalar] (c2)
}; \end{feynman} \end{tikzpicture}
& = -ig & \begin{tikzpicture}[baseline=-\the\dimexpr\fontdimen22\textfont2\relax] \begin{feynman}[inline=((a).base)]
\vertex [crossed dot](a) {} ;
\vertex [ above right=of a](b1) ;
\vertex [ above left =of a](b2) ;
\vertex [ below right=of a](c1) ;
\vertex [below left = of a] (c2) ;
\diagram* {
 (b2) --[fermion] (a) --[fermion] (b1), (c1) --[scalar] (a) --[scalar] (c2)
}; \end{feynman} \end{tikzpicture}
& =-ig\delta_g \\
\begin{tikzpicture}[baseline=-\the\dimexpr\fontdimen22\textfont2\relax] \begin{feynman}[inline=((a).base)]
\vertex (a);
\vertex [ above=of a](c1) ;
\vertex [below= of a] (c2) ;
\vertex [above right= of a] (d1);
\vertex [above left= of a] (d2);
\vertex [below right= of a] (d3);
\vertex [below left= of a] (d4);
\diagram* {
(c1) --[scalar] (a),(c2) --[scalar] (a),(d1) --[scalar] (a),(d2) --[scalar] (a),(d3) --[scalar] (a),(d4) --[scalar] (a)
}; \end{feynman} \end{tikzpicture}&=-i h &\begin{tikzpicture}[baseline=-\the\dimexpr\fontdimen22\textfont2\relax] \begin{feynman}[inline=((a).base)]
\vertex [crossed dot](a) {};
\vertex [ above=of a](c1) ;
\vertex [below= of a] (c2) ;
\vertex [above right= of a] (d1);
\vertex [above left= of a] (d2);
\vertex [below right= of a] (d3);
\vertex [below left= of a] (d4);
\diagram* {
(c1) --[scalar] (a),(c2) --[scalar] (a),(d1) --[scalar] (a),(d2) --[scalar] (a),(d3) --[scalar] (a),(d4) --[scalar] (a)
}; \end{feynman} \end{tikzpicture} &=-ih\delta_\lambda
\end{align}
We omitted the $O(N_s)\times O(N_f)$ tensor structure of the vertices. When computing loop diagrams, one has to analytically continue each momentum integral to Euclidean signature.  Hence for each loop we have to add an $i$. We will also abbreviate the momentum integral 
\begin{align}
	\int dk \equiv \int \frac{d^{d}k}{(2\pi)^d} \ .
\end{align}

For the renormalisation group analysis, it is convenient to use a QFT normalisation of the coupling, as we just gave. In the rest of this paper, we used a CFT adapted normalisation for $g$: $\kappa_s g = 2 \lambda \sqrt{\epsilon}$ where $\kappa_s^{-1} = 4 \pi$.
To derive the epsilon scaling of the coupling, let us instead write $g = 8 \pi g_*$ and use $g_*$ for now instead of $\lambda$.
We find it convenient to introduce here also $h = 1920 \pi^2 h_*$.

\subsection{Field Strength Renormalisation}

The renormalisation of the fermion propagator is computed through 

\begin{align}
	0 &= \begin{tikzpicture}[baseline=-\the\dimexpr\fontdimen22\textfont2\relax] \begin{feynman}[small,baseline=(a)]
\vertex (a);
\vertex [ right=of a](b) ;
\vertex [right= of b] (c) ;
\vertex [right= of c] (d);
\diagram* {
 (a) --[fermion] (b) --[fermion] (c) --[fermion] (d), (b) --[scalar,half right, looseness=1.5] (c) --[scalar, half right, looseness=1.5] (b)
}; \end{feynman}  \end{tikzpicture} + \feynmandiagram [layered layout, inline=-\the\dimexpr\fontdimen22\textfont2\relax, horizontal=a to b] { a -- [fermion] b [crossed dot]--[fermion] c , };  \\
&=  -\frac{g^2 N_s}{2}\int dk \, dl \frac{\slashed{k}}{k^2l^2(l-p+k)^2}-\slashed{p}\delta_\psi \ .
\end{align}
The integral can be evaluated using Feynman parameters and dimensional regularization. Asking that the equality be achieved at a scale $p\cdot p = \mu^2$ extracting only the logarithmic piece, one finds
\begin{align}
	\delta_\psi	&= \frac{g^2 N_s }{96 \pi ^2}\log(\mu)+ \ldots 
\end{align}
which translates into an anomalous dimension 
\begin{align}
	\gamma_\psi = \frac{g_*^2 N_s}{3} + \ldots 
\end{align}
This result matches (\ref{Deltapsid3}) that we obtained from crossing symmetry.

To find the beta function, one needs also the renormalisation of the scalar sector. We look at two contributions, one from a $g^2$ interaction and one from  an $h^2$ diagram:

\begin{align}
	&\begin{tikzpicture}[baseline=-\the\dimexpr\fontdimen22\textfont2\relax] \begin{feynman}[small,baseline=(a)]
\vertex (a);
\vertex [ right=of a](b) ;
\vertex [right= of b] (c) ;
\vertex [right= of c] (d);
\diagram* {
 (a) --[scalar] (b) --[scalar] (c) --[scalar] (d), (b) --[scalar,half right, looseness=2] (c),(b) --[scalar,half right, looseness=1] (c),(c) --[scalar,half right, looseness=2] (b),(c) --[scalar,half right, looseness=1] (b)
}; \end{feynman} \end{tikzpicture} & 
	&\begin{tikzpicture}[baseline=-\the\dimexpr\fontdimen22\textfont2\relax] \begin{feynman}[small,baseline=(a)]
\vertex (a);
\vertex [ right=of a](b) ;
\vertex [right= of b] (c) ;
\vertex [right= of c] (d);
\diagram* {
 (a) --[scalar] (b) --[scalar] (c) --[scalar] (d), (b) --[fermion,half right, looseness=1.5] (c) --[fermion, half right, looseness=1.5] (b)
}; \end{feynman}  \end{tikzpicture} & &\feynmandiagram [layered layout, horizontal=a to b, inline=-\the\dimexpr\fontdimen22\textfont2\relax] { a -- [scalar] b [crossed dot]--[scalar] c , };
\end{align}

The renormalisation condition is then
\begin{equation}
\begin{aligned}
	\delta_\phi =& \overbrace{-\frac{h^2(N_s+2)(N_s+4)}{15(5!) p^2}\int\frac{ dk_1 \ldots dk_4 }{k_1^2 \ldots k_4^2 (p-k_1-k_2-k_3-k_4)^2}}^{\Sigma_h} \\ &\underbrace{-\frac{2g^2 N_f}{p^2} \int dk \, dl \frac{k\cdot l}{k^2l^2(k-p+l)^2}}_{\Sigma_g} \ .
\end{aligned}	
\end{equation}
The computation of $\Sigma_g$ is quite similar to the fermion case 
\begin{equation}
\begin{aligned}
	\Sigma_g &= \frac{4}{3}g_*^2 N_f\log(\mu)+\ldots 
\end{aligned}	
\end{equation}

The second piece, $\Sigma_h$ can be computed by chaining the loop integrals. There is no particular difficulty beyond keeping track of the different overall coefficients. We find 
\begin{equation}
\begin{aligned}
	\Sigma_h  &= \frac{(N_s+2) (N_s+4)}{6}h_*^2 \log (\mu )+\ldots 
\end{aligned}	
\end{equation}
	
As expected, these coefficients are positive, since $\Delta_{\phi}= \frac{d-2}{2}+\gamma_\phi\geq \frac{d-2}{2}$, by unitarity.  The end result is that 
\begin{align}
	\delta_\phi = \frac{1}{6}\left(8 N_f g_*^2 +(N_s+2)(N_s+4)h_*^2 \right)\log(\mu)+\ldots
\end{align}

\subsection{Scalar Vertex Renormalisation}

The diagrams that renormalise the $\phi^6$ vertex are at leading order  

\begin{align}
&\begin{tikzpicture}[baseline=-\the\dimexpr\fontdimen22\textfont2\relax] \begin{feynman}[baseline=(a)]
\vertex (a);
\vertex [right=of a](b) ;
\vertex [left=of a](c1) ;
\vertex [above left= of a] (c2);
\vertex [below left= of a] (c3);
\vertex [right=of b](d1) ;
\vertex [above right= of b] (d2);
\vertex [below right= of b] (d3);
\diagram* {
 (c1) --[scalar] (a),(c2) --[scalar] (a),(c3) --[scalar] (a),(d1) --[scalar] (b),(d2) --[scalar] (b),(d3) --[scalar] (b),(a) --[scalar,half right,looseness=1.5] (b)--[scalar,half right,looseness=1.5] (a) -- [scalar] (b)
}; \end{feynman} \end{tikzpicture} & 
&\begin{tikzpicture}[baseline=-\the\dimexpr\fontdimen22\textfont2\relax] \begin{feynman}[small,baseline=(a)]
\vertex (a);
\vertex [above left=of a](b1) ;
\vertex [below left=of a](b2) ;
\vertex [above left=of b1](bb1) ;
\vertex [below left=of b2](bb2) ;
\vertex [above right=of b1](c1) ;
\vertex [below right=of b2](c2) ;
\vertex [above right=of a](d1) ;
\vertex [below right=of a](d2) ;
\vertex [above right=of d1](dd1) ;
\vertex [below right=of d2](dd2) ;
\diagram* {
 (b1) --[scalar] (a),(b2) --[scalar] (a),(c1) --[scalar] (a),(c2) --[scalar] (a),(d1) --[scalar] (a),(d2) --[scalar] (a),(d2) --[fermion,bend right] (d1)--[fermion,bend right] (d2), (d1)--[scalar] (dd1), (d2)--[scalar] (dd2), (b1) --[scalar] (bb1),(b2) --[scalar] (bb2) 
}; \end{feynman} \end{tikzpicture}\end{align}
The multiplicity count of those diagram can be found by using double-line notation. As previously, we split the computation in two 
\begin{align}
	-i \delta_{h}h + i h(\Sigma_h^{'}+\Sigma_{g}^{'})=0 \ ,
\end{align}
with obvious form of the contributions from each diagram. The first piece is given by 
\begin{align}
	\Sigma_{h}^{'} &=h\frac{22+3N_s}{15} \int \frac{dk_1 dk_2 }{k_1^2k_2^2(p-k_1-k_2)^2} \\
		&= -8 (22+3N_s) h_* \log(\mu)+\ldots  \nonumber
\end{align}
while the second is
\begin{align}
	\Sigma_{g}^{'}&=-30g^2 N_f \int dk dl \frac{k\cdot(k+l-p)}{k^2 l^2 (l+k-p)^2(l-2p)^2} \\
	&= -60 N_f  g_*^2 \log (\mu )+\ldots  \nonumber
\end{align}

The end result is then 
\begin{align}
	\delta_h = -\left(8(3N_s+22)h_*+60 N_f g_*^2 \right)\log(\mu)+\ldots 
\end{align}
which implies a beta function 
\begin{align}
	\beta_{h}\propto h_*\left( 8 N_f g_*^2+(3N_s+22)h_*-\frac{\epsilon}{4}+\ldots \right)
\end{align}
As anticipated $g_*$ scales as $\sqrt{\epsilon}$, while $h_*$ scales as $\epsilon$.  
Thus the $h_*^2$ contribution to the anomalous dimension of $\phi$ will be higher order.

\subsection{Yukawa Vertex Renormalisation} 

The diagrams at order $g^2$ are all convergent; hence one must go to order $g^3$ to find a non-zero contribution. 
Indeed, $O(g^3)$ is consistent with $g^2 \sim \epsilon$ scaling at the fixed point. 
There are three families of divergent diagrams, which we consider separately
\begin{align}
	-ig \delta_{g} + i g( \Sigma_1+\Sigma_2+\Sigma_3)=0.
\end{align}

The first family is given by bug type diagrams, 
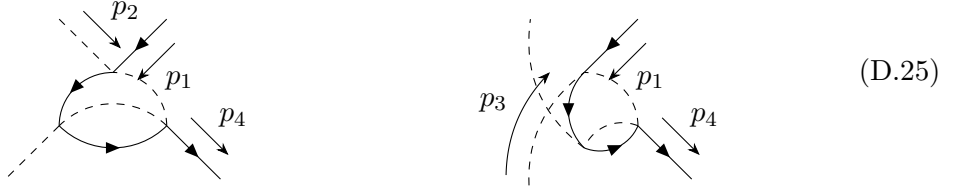
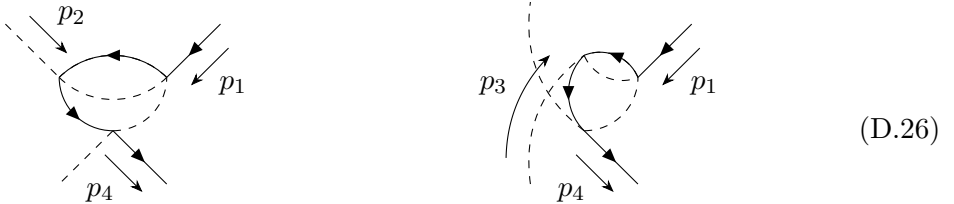
\begin{align}
&\begin{tikzpicture}[baseline=-\the\dimexpr\fontdimen22\textfont2\relax] \begin{feynman}[small,baseline=(a)]
\vertex (a);
\vertex [below right=of a](b) ;
\vertex [below left= of a] (c) ;
\vertex [above right= of a] (a1);
\vertex [above left= of a] (a2);
\vertex [below right= of b] (f1);
\vertex [below left= of c] (f2);
\diagram* {
 (a1) --[fermion,momentum=\(p_1\)] (a), (a2) -- [scalar,momentum=\(p_2\)] (a), (a) --[fermion, quarter right] (c) -- [fermion, quarter right] (b) --[scalar, quarter right] (a), (b) --[fermion, momentum =\(p_4\)] (f1), (c)--[scalar] (f2), (b) --[scalar, quarter right] (c)
}; \end{feynman} \end{tikzpicture}
 &
&\begin{tikzpicture}[baseline=-\the\dimexpr\fontdimen22\textfont2\relax] \begin{feynman}[small,baseline=(a)]
\vertex (a);
\vertex [below right=of a](b) ;
\vertex [below= of a] (c) ;
\vertex [above right= of a] (a1);
\vertex [above left= of a] (a2);
\vertex [below right= of b] (f1);
\vertex [below left= of c] (f2);
\diagram* {
 (a1) --[fermion,momentum=\(p_1\)] (a), (f2) -- [scalar,momentum=\(p_3\),bend left] (a), (a) --[fermion, quarter right] (c) -- [fermion, quarter right] (b) --[scalar, quarter right] (a), (b) --[fermion, momentum =\(p_4\)] (f1), (c)--[scalar, bend left] (a2), (b) --[scalar, quarter right] (c)
}; \end{feynman} \end{tikzpicture}
  \\ 
&\begin{tikzpicture}[baseline=-\the\dimexpr\fontdimen22\textfont2\relax] \begin{feynman}[small,baseline=(a)]
\vertex (a);
\vertex [above right=of a](b) ;
\vertex [above left= of a] (c) ;
\vertex [above right= of b] (a1);
\vertex [above left= of c] (a2);
\vertex [below right= of a] (f1);
\vertex [below left= of a] (f2);
\diagram* {
 (a1) --[fermion,momentum=\(p_1\)] (b), (a2) -- [scalar,momentum=\(p_2\)] (c), (b) --[fermion, quarter right] (c) --[scalar, quarter right] b, c -- [fermion, quarter right] (a) --[scalar, quarter right] (b), (a) --[fermion, momentum'=\(p_4\)] (f1), (a)--[scalar] (f2), (b) --[scalar, quarter right] (c)
}; \end{feynman} \end{tikzpicture}
& 
&\begin{tikzpicture}[baseline=-\the\dimexpr\fontdimen22\textfont2\relax] \begin{feynman}[small,baseline=(a)]
\vertex (a);
\vertex [above right=of a](b) ;
\vertex [above= of a] (c) ;
\vertex [above right= of b] (a1);
\vertex [above left= of c] (a2);
\vertex [below right= of a] (f1);
\vertex [below left= of a] (f2);
\diagram* {
 (a1) --[fermion,momentum=\(p_1\)] (b), (f2) -- [scalar,momentum=\(p_3\),bend left] (c), (b) --[fermion, quarter right] (c) --[scalar, quarter right] b, c -- [fermion, quarter right] (a) --[scalar, quarter right] (b), (a) --[fermion, momentum'=\(p_4\)] (f1), (a)--[scalar,bend left] (a2), (b) --[scalar, quarter right] (c)
}; \end{feynman} \end{tikzpicture}
\end{align}

We will impose the renormalisation condition at the point $p_1 = p_2 = p_3 = p$, such that $p^2 = \mu^2$. The sum of the four diagrams drawn evaluates in this simplified configuration to  
\begin{align}
	 \Sigma_1=2g^2 \int \frac{dk \, dl}{k^2 l^2(k-2p)^2} \left(\frac{\slashed{k}\slashed{l}}{(l-k-p)^2}+\frac{\slashed{l}\slashed{k}}{(l-k+p)^2}\right) \ .
\end{align}
Standard methods yield 
\begin{align}
	\Sigma_1 &= -8 g_*^2 \log(\mu)+ \ldots. 
\end{align}

We still have two diagrams to consider. The first one is made up of an internal fermion loop, the second of an internal scalar loop:
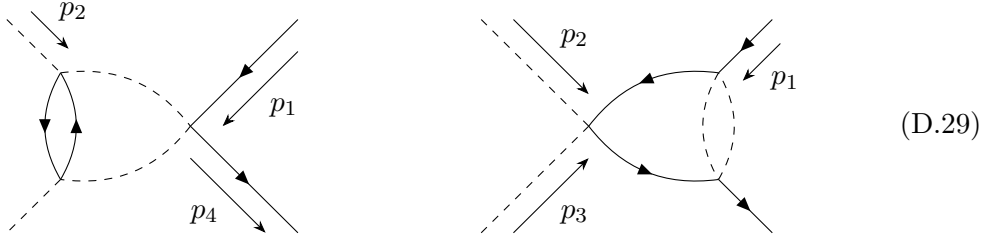
\begin{align}
&\begin{tikzpicture}[baseline=-\the\dimexpr\fontdimen22\textfont2\relax] \begin{feynman}[small,baseline=(a)]
\vertex (a);
\vertex [right= of a](aa);
\vertex [above right=of aa] (a1) ;
\vertex [below right=of aa] (a2) ;
\vertex [above right=of a1](aa1) ;
\vertex [below right=of a2](aa2) ;
\vertex [above left= of a] (b1);
\vertex [below left= of a] (b2);
\vertex [above left= of b1] (f1);
\vertex [below left= of b2] (f2);
\diagram* {
 (aa1) --[fermion,momentum=\(p_1\)] (aa) --[fermion,momentum'=\(p_4\)] (aa2), (f1) -- [scalar,momentum=\(p_2\)] (b1) --[scalar, bend left] (aa) --[scalar, bend left] (b2) --[scalar] (f2), (b1) --[fermion, bend right] (b2)--[fermion, bend right] (b1)
}; \end{feynman} \end{tikzpicture} & 
&\begin{tikzpicture}[baseline=-\the\dimexpr\fontdimen22\textfont2\relax] \begin{feynman}[small,baseline=(a)]
\vertex (a);
\vertex [left= of a](aa);
\vertex [above left=of aa] (a1) ;
\vertex [below left=of aa] (a2) ;
\vertex [above left=of a1](aa1) ;
\vertex [below left=of a2](aa2) ;
\vertex [above right= of a] (b1);
\vertex [below right= of a] (b2);
\vertex [above right= of b1] (f1);
\vertex [below right= of b2] (f2);
\diagram* {
 (aa1) --[scalar,momentum=\(p_2\)] (aa), (aa2) --[scalar,momentum'=\(p_3\)] (aa), (f1) -- [fermion,momentum=\(p_1\)] (b1) -- [fermion, bend right] (aa) --[fermion, bend right] (b2) --[fermion] (f2), (b1) --[scalar, bend right] (b2)--[scalar, bend right] (b1)
}; \end{feynman} \end{tikzpicture}
\end{align}

The diagram with the internal fermion loop gives
\begin{align}
	\Sigma_2 &= 2N_f g^2 \int dk \, dl \frac{k\cdot (k+p-l)}{k^2 l^2 (k+p-l)^2(l-2p)^2}
\end{align}
Extracting the log we find  
\begin{align}
	\Sigma_2 &= -4N_f g^2_* \log(\mu)+ \ldots
\end{align}
For the diagram with the internal scalar loop, we have instead 
\begin{align}
	\Sigma_3 &= \frac{g^2 N_s}{2} \int dk \, dl \frac{\slashed{k}(\slashed{k}-2\slashed{p})}{l^2 k^2(k-2p)^2(l-k+3p)^2} \\
	&=-2 N_s g_*^2 \log(\mu)+\ldots \nonumber
\end{align}

The end result of this computation is 
\begin{align}
	\delta_g = (-8-4N_f-2N_s)g_*^2 \log(\mu)+\ldots
\end{align}
from which 
we can compute the beta function 
\begin{align}
	\beta_g \propto g_*\left( g_*^2 (N_s+2N_f+3)-\frac{3}{8}\epsilon+\ldots\right)
\end{align}
%
The value of $g_*$ at the fixed point $3-\epsilon$ dimensions is then real.  

\subsection{Fermion Bilinear Renormalisation}

We are interested in the anomalous dimension of ${:}\overline{\psi}\psi{:}\equiv\sigma$. We must renormalise the divergences of the $\psi$ fields on their own as well as the new divergences that show up because of compositeness.  
We computed $Z_\psi$ to the requisite perturbative order already.  
The new divergences will be rescaled away by the $Z_\sigma$ factor we introduced
at the beginning of this appendix.
To fix $Z_\sigma$, we need to regularise diagrams with $\sigma$ insertions. The simplest situation is to consider a process in which $\sigma \rightarrow \psi \overline{\psi}$. 
We find possible contributions from two diagrams at $O(g^2)$:
\begin{align}
&\begin{tikzpicture}[baseline=-\the\dimexpr\fontdimen22\textfont2\relax] \begin{feynman}[baseline=(a)]
\vertex [blob] (a) {};
\vertex [left = of a](b);
\vertex [above right=of a] (a1) ;
\vertex [below right=of a] (a2) ;
\vertex [right=of a1](aa1) ;
\vertex [right=of a2](aa2) ;
\diagram*{
 (a2) --[fermion, bend left] (a) --[fermion,momentum=\(p_3\), bend left] (a1), (a1)--[scalar, bend right] (a2) --[scalar, bend right] (a1), (b) -- [ghost, momentum=\(p\)] (a), (aa2) --[fermion] (a2),(a1) --[fermion] (aa1)
}; \end{feynman} \end{tikzpicture} & 
 & \begin{tikzpicture}[baseline=-\the\dimexpr\fontdimen22\textfont2\relax] \begin{feynman}[baseline=(a)]
\vertex [blob] (a) {};
\vertex [left = of a](b);
\vertex [right = of a](c);
\vertex [right = of c](d);
\vertex [above right=of d] (a1) ;
\vertex [below right=of d] (a2) ;
\diagram*{
 (a2) --[fermion] (d) --[fermion] (a1), (b) -- [ghost, momentum=\(p\)] (a), (a) --[fermion,quarter left] (c) --[fermion, quarter left] (a), (c)--[scalar, half left] (d)--[scalar, half left] (c)
 }; \end{feynman} \end{tikzpicture}
\end{align}

It turns out that the second diagram is convergent; hence we can safely drop it. We then regularise the first diagram at a momentum scale $p^2 =\mu^2$ by adding a counterterm $\delta_\sigma$:
\begin{align}
	\delta_\sigma &= -\frac{g^2}{2}N_s \int dk \, dl \frac{(\slashed{k}-\slashed{p})\slashed{k}}{k^2 l^2(k-p)^2(l-k+p)^2} \ .
\end{align}
These integrals are straightforward. Isolating the logarithm, we extract 
\begin{align}
	\delta_\sigma &= -2 g_*^2 N_s \log(\mu) +\ldots 
\end{align}
The anomalous dimension of the composite operator is determined as usual
\begin{align}
	\gamma_\sigma &= \mu \pdv{}{\mu}\log(\frac{Z_\psi}{Z_\sigma})= \frac{8}{3}g_*^2 N _s \ .
\end{align}
This result matches (\ref{Deltasigmad3}) we computed using crossing symmetry and the equations of motion.


\section{Relation to EAdS$_{p+1}$ Amplitude}
\label{app:EAdS}

Intuitively it is understood that one can consider a bCFT and perform a Weyl transformation to obtain a QFT in AdS \cite{Paulos:2016fap}. It is useful 
to give an explicit form to this mapping in the embedding space picture. Embedding space offers a simple geometric and coordinate invariant view of the procedure. The goal of this section is to bridge the gap between our work, which we view along the bCFT perspective, and the various results coming from the study of QFT in AdS  \cite{Carmi:2018qzm,Giombi:2021cnr}.

Consider fields in a bCFT viewed in the embedding space, with points specified by $P^A P_A = 0$. Given a constant vector $V^{A}$, there is an AdS foliation of the lightcone along the direction $V$, i.e.\ there is a map from insertions on the lightcone in $\mathbb{R}^{1,d+1}$ to insertions on an AdS slice of $\mathbb{R}^{1,d}$, immersed in the higher space. The explicit map is given by a projection and rescaling 
\begin{align}
	X^{A}&=\left(\eta^{AB}-V^{A}V^{B}\right)\frac{P_{B}}{P\cdot V} \rightarrow X^{A}X_{A} = -1 \ .
\end{align}
The orthogonal projection kills off one component, while the rescaling brings us to a canonical AdS slice. Because the vector is transverse, the inner product on the AdS slice is given by the reduced inner product $\bullet$. Intrinsic coordinates on the manifold spanned by the $X^{A}$ see an $EAdS_{p+1}$ metric.
Viewed from the bCFT embedding, this is simply a change of section, whose effect is to rescale the fields by their Weyl factors. This parametrisation is singular at boundary points, $P\cdot V \rightarrow  0$, which are sent to infinity. Effectively, this maps insertions of fields in a bCFT to those of fields in AdS. This map is invariant under the residual conformal symmetry of the boundary, which is kept manifest. The case of $dS_{p+1}$ follows through, at the kinematic level, by considering $V^2=-1$ and working on the lightcone of the space $\mathbb{R}^{2,d}$.

Let us now see how generic fields behave under this mapping. First, from the scaling property   
\begin{align}
	\phi_{AdS}(X^{A})=2^{\Delta}\phi_{bCFT}\left(\frac{P^{A}}{P\cdot V}\right)=(2P\cdot V)^{\Delta}\phi_{bCFT}(P) \ .
\end{align}
Hence for bulk insertions, the equivalent expression in $AdS$ follows straightforwardly by taking power laws out of correlation functions. For boundary insertions, this rescaling is not well defined since $P\cdot V=0$, and one must retain the scaling law. We find back that bulk insertions at $P^{A}$ are mapped to bulk fields at coordinate $X^{A}$ on the AdS hyperboloid, $X^{A}X_{A}=-1$. Boundary insertions are mapped to insertions on the lightcone of $R^{1,d}$, the conformal boundary of AdS. The fate of tensorial objects is similarly dealt with. In the bCFT embedding space, tensors satisfy 
\begin{align}
	P^{A}T_{A\ldots} &= 0 \ ,  & T_{A\ldots}  &\sim T_{A\ldots}+\alpha P_{A}\  T'_{\ldots}  \ .
\end{align}
Both conditions are needed to kill two components for each index. Meanwhile, tensors in the embedding space of AdS have one component killed by the transversality relation, $X^{A}T_{A\ldots} =0$. This last relation is not true of general bulk bCFT tensors. However, one is free to use the gauge freedom to select the representative tensor out of the gauge-equivalence class which is transverse to both $P^{A}$ and $X^{A}$, equivalently, to both $P^{A}$ and $V^{A}$. In the index-free picture, the argument follows through by applying this logic to the polarisation vector $Z^{A}$. The bulk bCFT polarisation $Z^{A}$ is gauge-fixed to  $W^{A}= Z^{A}-\frac{Z\cdot V}{P\cdot V}P^{A}$. $X^{A}W_A=0=V^{A}W_A$. The renaming into $W$ is to match with both the usual notation in AdS, and the notation for the boundary polarisation vectors which satisfy generally $W^{A}V_A=0$. 

The rule to translate tensorial insertions from $bCFT$ to $AdS$ is then to take the correlation function and perform the rewriting:
\begin{equation}
\begin{aligned}
	Z_i^{A} &\rightarrow W_i^{A} \ , & W_i^{A}V_A &= 0= W^{A}W_A \ , \\
	P_{i}^{A}& \rightarrow P_i\cdot V(X_{i}^{A}+V^{A}) \ , & X_i^{A}V_A &= 0 \ , \\
	A\cdot B &\rightarrow A\bullet B + (A\cdot V)(B\cdot V) \ .
\end{aligned}
\end{equation} 

One can also take an expression written using the AdS variables and rewrite them using bCFT variables. One should simply give arbitrary distance $P_{i}\cdot V$ to each individual insertion. 
As an example, consider the bCFT two-point function cross-ratio. Under this transformation, it becomes the usual AdS geodesic distance 
\begin{align}
	\xi &= -\frac{P_1\cdot P_2}{4P_1\cdot VP_2\cdot V}=-\frac{\left(-X_1\bullet X_2-1\right)}{4}=\frac{(X_1-X_2)\bullet (X_1-X_2)}{2} \ .
\end{align}
We can also consider the tensor structures entering a bCFT bulk-bulk correlation function of symmetric traceless operators. These are usually constructed starting from the basis
\begin{equation}
	\begin{aligned}
	{\mathcal S}_{1} &= \frac{P_2\cdot P_1 Z_1\cdot Z_2- P_2\cdot Z_1 P_1\cdot Z_2}{P_1\cdot P_2} \rightarrow W_1 \bullet W_2 +\frac{4}{\xi}W_1\bullet X_2 W_2\bullet X_1 \ , \\
	{\mathcal S}_{2} &= \left(Z_1\cdot V-\frac{P_2\cdot Z_1 P_1\cdot V}{P_1\cdot P_2}\right)\left(Z_2\cdot V-\frac{P_1\cdot Z_2 P_2\cdot V}{P_1\cdot P_2}\right) \rightarrow \frac{16}{\xi^2}W_1\bullet X_2 W_2\bullet X_1 \ ,	
		\end{aligned}
\end{equation}
where we performed the substitution described and rewrote the final results using the AdS variables. Looking now at the whole correlator, we have found a mapping between the correlator in the bCFT and in AdS 
\begin{equation}
	\begin{aligned}
		\expval{F_{L}(P_1,Z_1)F_{L}(P_2,Z_2)} &= \sum_{k=0}^l\frac{({\mathcal S}_1)^{k}({\mathcal S}_2)^{L-k}}{(2P_1\cdot V)^{\Delta_1}(2P_2\cdot V)^{\Delta_2}}g_k(\xi) \ , \\ 
	\expval{F_{L}(X_1,W_1)F_{L}(X_2,W_2)}_{AdS}&=\sum_{k=0}^{L}(W_1\bullet W_2)^{k}(W_1\bullet X_2 W_2 \bullet X_1)^{L-k}f_k(z) \ , \\
	f_{k}(z)&=\sum_{n=k}^{L}\binom{n}{k}\left(\frac{\xi}{4}\right)^{n+k-2L}g_{n}\left(\xi\right) \ .
	\end{aligned}
\end{equation}
The functions of the cross-ratio are the same up to a change of basis. The physical content is that the blocks are left unchanged, as is expected by conformal invariance. The functions $g_{k}$ or $f_k$ admit an expansion in terms of bulk and boundary conformal blocks. On the bCFT side, the bulk blocks are all known. They can all be extrapolated from the knowledge of the seed block for a scalar and by acting with weight shifting operators \cite{Karateev:2018uk}. The seed conformal blocks \cite{Costa:2011vf} consist of the exchange of a boundary spin-$L$ operator between two spin-$L$ bulk operators, and configurations with higher bulk spin can be reached by acting with another weight shifting operator \cite{Billo:2016vm,Herzog:2021spv}. 

The bulk expansion is convergent whenever the bulk is conformal, which corresponds to a CFT in AdS. The boundary expansion however is always present. The seed boundary conformal blocks can be found by solving the conformal Casimir equation, which in the AdS picture is equivalent to considering the propagator for massive tensorial fields with spin-$L$ and some specific mass-squared. This has been done and the boundary blocks are known \cite{Costa:2014kfa}. It follows that the seed blocks for tensorial operators in a bCFT two point function are all known.

Now that the situation for tensors is clearly laid out, we can turn to our main interest, spinors. These are defined in the bCFT embedding space through the eigenvalue equation 
\begin{align}
	P^{A}\Gamma_{A}\Psi = 0 \rightarrow X^{A}\Gamma_{A}\psi = \left(-V\cdot \Gamma \right)\Psi \ .
\end{align}
We can define $\Gamma_{\circ} = V \cdot \Gamma$.  We then find fermions in AdS which satisfy the eigenvalue equation outlined in \cite{Nishida:2018opl}. Note, this clearly holds in any dimension, since the embedding of spinors on the lightcone is insensitive to $d$. Consider the representation of $\Psi$. For a bCFT$_{d}\sim $AdS$_{p+1}$, we use embedding spinors of $\mathbb{R}^{1,d+1}$. These always take the form of a doublet of Dirac spinors of two dimensions lower. Consider $d$ odd; then $\Psi$ is a Dirac spinor of $\mathbb{R}^{1,d+1}$, which is also the Dirac spinor of $\mathbb{R}^{1,d}$. This is an irreducible representation in even total dimension; a $\Gamma_{\circ}$ matrix is available in the embedding of $EAdS_{d}$. If $d$ is even, $\Psi$ is a Dirac spinor of $\mathbb{R}^{1,d+1}$, and it is made up of a symplectic doublet of Dirac spinors of $\mathbb{R}^{1,d}$. As seen from the embedding space of $AdS_{d}$, this forms a reducible representation, whose symplectic form gives rise to a $\Gamma_{\circ}$-type matrix. This construction is precisely the one highlighted in \cite{Pethybridge:2021rwf}, where this structure was constructed explicitly to study the embedding of fermions in dS. The output of this discussion is that one can use the same embedding formalism in (A)dS in both even and odd dimensions. To further make contact with the eigenvalue equation used there, note that one can redefine the $\Gamma$-matrix in the AdS picture as $\Gamma'_{A}=-i \Gamma\cdot V(\Gamma^{B}-V^{B}\Gamma\cdot V)$. Then, one finds the eigenvalue equation:
\begin{align}
	X^{A}\Gamma'_{A}\Psi = i \Psi  \ .
\end{align}
The factor of $i$ is needed to preserve the signs in the Clifford Algebra. Where $V^{A}$ timelike, one would not need the $i$, and we would find back the constraint for dS. Note that taking $V^{A}V_{A}=-1$ yields a projection to $EdS_{d} \equiv S_{d}$. To obtain Lorentzian $dS_{p+1}$ one needs to consider a Lorentzian bCFT$_{d}$, which lifts to an embedding space $\mathbb{R}^{2,p+1}$. It is quite interesting to note that, although the embedding picture for AdS$_{p+1}$, $CFT_{p}$ and dS$_{p+1}$ live in the same space, $\mathbb{R}^{1,p+1}$, the map from bCFT$_{d}$ to these is signature dependent. Regarding the index-free formalism for spinors, no modification is needed beyond the identification $\slashed{V}\leftrightarrow \Gamma_{\circ}$. As we did for the tensor, we can relate the bulk-bulk correlators in both pictures, by simply translating the structures in the dictionary:
\begin{equation}
	\begin{aligned}
		\expval{\Psi(P_1,\overline{S}_1)\overline{\Psi}(P_2,S_2)}&=\frac{\overline{S}_1S_2 F(\xi)+ \overline{S}_1 \slashed{V}S_2 G(\xi)}{(2P_1\cdot V)^{\Delta_1+\frac{1}{2}}(2P_2\cdot V)^{\Delta_2+\frac{1}{2}}}  \ ,\\
	\expval{\Psi(X_1,\overline{S}_1)\overline{\Psi}(X_2,S_2)}_{AdS}&=\overline{S}_1S_2 F\left(\xi\right)+\overline{S}_1\Gamma_{\circ}S_2 G\left(\xi\right) \ .
	\end{aligned}
\end{equation}
Using these equations, one can easily make contact between our computations and the ones done in AdS, in a coordinate invariant manner.

\bibliographystyle{jhep}
\bibliography{paper.bib}

\end{document}